\documentclass[floats,floatfix,showpacs,amssymb,prd,superscriptaddress,nofootinbib,nolongbibliography,reprint]{revtex4-2}

\usepackage{amssymb,amsmath,verbatim,mathtools,needspace,enumitem,etoolbox,graphicx,physics,microtype,afterpage,bigints,gensymb,booktabs}

\usepackage[dvipsnames, usenames]{xcolor}
\definecolor{linkcolor}{rgb}{0.0,0.3,0.5}
\definecolor{dodgerblue}{HTML}{1E90FF}
\usepackage[unicode, colorlinks=true, linkcolor=linkcolor, citecolor=linkcolor, filecolor=linkcolor,urlcolor=linkcolor, pdfusetitle]{hyperref}
\usepackage[all]{hypcap}
\usepackage[T1]{fontenc}
\usepackage[utf8]{inputenc}
\usepackage{orcidlink}
\usepackage[htt]{hyphenat}
\usepackage{graphicx}

\usepackage{rotating}
\usepackage{tikz}


\usepackage{lipsum}

\newcommand{\mrs}{\affiliation{Aix-Marseille Universit\'e, Universit\'e de Toulon, CNRS, CPT, Marseille, France}}

\def\pymcpop{\texttt{pymcpop-gw}}

\begin{document}

\title{
A weakly modelled view of the joint compact-binary mass plane: population structure and spectral-siren cosmology
}

\author{Viola De Renzis\texorpdfstring{$\,$}{ }\orcidlink{0000-0001-7038-735X}}
\email{viola.DE-RENZIS@univ-amu.fr}
 \mrs

\author{Alessandro Agapito$\,$\orcidlink{0009-0005-9004-3163}
}
\email{alessandro.AGAPITO@univ-amu.fr}
\mrs

 \author{Michele Mancarella\texorpdfstring{$\,$}{ }\orcidlink{0000-0002-0675-508X}}
 \mrs


\date{\today}

\begin{abstract}

We introduce a flexible mixture model for the joint source-frame component-mass distribution of compact binaries, inferred simultaneously with cosmology, together with a %
framework that identifies structures %
in the reconstructed density, quantifying their posterior stability, local prominence, and cosmological relevance. %
Applied to GWTC--5.0, the workflow clarifies several familiar and emergent %
structures. 
The ${\sim}35\,M_\odot$ feature is classified as a clear
near-equal-mass peak despite an ambiguous %
interpretation in the primary-mass projection. %
A weaker unequal-mass concentration near $(21,12)\,M_\odot$ is recovered  with high recurrence but lower prominence.
A tentative joint transition near \(60\,M_\odot\) appears at mildly unequal masses, with a more robust manifestation in the primary-mass projection.
For cosmology, the two--dimensional mass model constraint on $H_0$ is $33\%$ narrower than a model that flexibly reconstructs only the primary mass, 
and only \({\sim}10\%\) wider than %
analyses based on strongly parameterized one-dimensional spectral %
structure.
At the feature level, we show that the ${\sim}10\,M_\odot$ and ${\sim}35\,M_\odot$
concentrations depend on $H_0$ mainly through a common %
scale which captures the entire ${\sim}(1+z)$ cosmological rescaling of both masses; the intermediate unequal-mass feature instead retains residual dependence in the orthogonal mass-ratio direction.
More generally, we argue that population prominence, formal inclusion in an astrophysical feature catalogue, and cosmological relevance need not coincide.
Together, these results suggest that a genuinely two-dimensional view may be required to fully exploit spectral-siren information, with weakly modelled approaches retaining precision while reducing sensitivity to mass-model systematics.
Our workflow provides an end-to-end, agnostic, reproducible way to detect and interpret population structure and identify which aspects of its geometry are associated with spectral-siren constraints. The feature finder is directly applicable to other population reconstructions and theoretical predictions.

\end{abstract}

\maketitle

\section{Introduction}
\label{sec:introduction}

The rapid growth of the LIGO--Virgo--KAGRA
(LVK)~\cite{LIGOScientific:2014pky,VIRGO:2014yos,KAGRA:2020tym}
catalogues~\cite{LIGOScientific:2025hdt,LIGOScientific:2025slb,
LIGOScientific:2026wfs} is turning gravitational-wave (GW) astronomy into an
increasingly large-sample domain~\cite{LIGOScientific:2025pvj,
LIGOScientific:2026ctl}. Population inference is correspondingly moving toward
more flexible and general descriptions. A broad class of weakly
modelled methods allows population structure to be inferred without assigning
it in advance to a small set of fixed functional components
\cite{DelPozzo:2018dpu,Tiwari:2020vym,Edelman:2021zkw,Rinaldi:2021bhm,
Edelman:2022ydv,Callister:2023tgi,Ray:2023upk,Farah:2024xub,Heinzel:2024jlc,
Heinzel:2024hva,Alvarez-Lopez:2025ltt,Tenorio:2025nyt,Tiwari:2026bvx}.
At the same time, multidimensional analyses show that correlations and structure in the joint population cannot in general be represented
by one-dimensional marginals alone
\cite{Heinzel:2024jlc,Tenorio:2025nyt,LIGOScientific:2026ctl}.

These developments are particularly important for spectral-siren
cosmology~\citep{Taylor:2011fs,Farr:2019twy,Mastrogiovanni:2021wsd,
Ezquiaga:2022zkx}. GWs measure luminosity distance, while the masses entering
the waveform are redshifted relative to their source-frame values, so
structure in the source-frame mass distribution can provide an internal
redshift reference. The assumed mass distribution is therefore also a leading
source of systematic uncertainty in spectral-siren inference
\cite{Mastrogiovanni:2021wsd,Pierra:2023deu,LIGOScientific:2026uyd}.
Standard strongly parameterized descriptions prescribe localized
spectral structure through a one-dimensional mass model together
with a relatively simple secondary-mass prescription~\cite{LIGOScientific:2026uyd}.

Recent spectral-siren analyses have begun relaxing parametric assumptions, often
through one-dimensional spline representations
\cite{Tagliazucchi:2026gxn,Bertheas:2026odj,Gennari:2026dfy,Li:2026amt}.
These approaches provide substantially greater flexibility, although choices
such as the number and spacing of spline nodes can still affect the inferred
$H_0$ posterior. 
 At fixed cosmology, adaptive methods
such as PixelPop~\cite{Heinzel:2024jlc} or Delaunay triangulation~\cite{Tenorio:2025nyt} already address this
problem more directly and in higher dimensions.  
A natural next step for spectral sirens is therefore
to let the data determine where local structure is needed and how much complexity is supported, while retaining the multidimensional
correlations that population studies are increasingly resolving.  %

In this work, we address these questions introducing a weakly modelled distribution
for the source-frame component masses that is inferred jointly with
cosmology.  The model uses correlated
Gaussian components directly in the two-dimensional mass plane, with their
locations, widths, correlations, and weights inferred from the data, thereby providing a further significant step towards the reduction of systematic uncertainty related to modelling the mass spectrum.  This
construction builds on earlier Gaussian-mixture and Dirichlet-process
approaches to adaptive density estimation and compact-binary populations~\cite{DelPozzo:2018dpu,Rinaldi:2021bhm}.
A parallel
effort to this work has recently pursued this
program using one- and two-dimensional Gaussian-process mass models~\cite{Leyde:2026bat}. 

Such flexibility, however, raise an interpretational question:
\emph{what constitutes a population feature?} In a strongly parametrized
model, a peak location or cutoff may be an explicit parameter; in a weakly
modeled reconstruction, the inferred object is instead a posterior over
population densities. A structure can shift, broaden, split, or
disappear across posterior realizations, while a one-dimensional feature may
be the projection of a qualitatively different structure in the full parameter
space. Previous work has addressed parts of this problem in one-dimension~\cite{Callister:2023tgi,Farah:2023vsc,Fabbri:2025faf,Corelli:2026thw}.
For spectral sirens, this raises an additional and deeper question: \emph{which structures in a weakly modelled reconstruction act as cosmological anchors and how can they be identified?}

We address this by defining features directly from the morphology of
the reconstructed density. An automated finder identifies localized
concentrations, under-densities, and changes of slope, and propagates
uncertainty in their recovery, location, geometry, and strength. Because these
structures are defined directly from the density, their identification does not depend on how a feature is
represented internally by the population model. 
This provides a reproducible
interpretation of multidimensional structure beyond visual inspection, while allowing to explicitly track their correlation with cosmological parameters.
In this sense, the feature--cosmology analysis generalizes feature-level spectral-siren diagnostics developed for one-dimensional parametric descriptions~\cite{Mali:2024wpq} to weakly modelled multidimensional reconstructions.
The same analysis
can be applied directly to other multidimensional reconstructions and theoretical
predictions.
The corresponding Python implementation is available open source at Ref.~\cite{plandscape}.

We apply the framework to GWTC--5.0 while jointly inferring cosmology. We first
characterize the recovered two-dimensional structures and relate them to their
one-dimensional projections. We then compare a model that flexibly
reconstructs only the primary-mass spectrum with the full joint
component-mass reconstruction, testing how much additional spectral-siren
information is retained in the second mass dimension. Finally, we identify the
features associated with $H_0$ and decompose their geometry into a common
two-component mass scale, following the cosmological mass--redshift rescaling,
and an orthogonal mass-ratio direction. This allows us to test whether the
cosmological dependence of a feature is exhausted by coherent rescaling of
both component masses or retains genuinely two-dimensional structure.

\section{Methods}\label{sec:methods}

\subsection{Population model}
\label{sec:population_model}

We describe the source-frame compact-binary population through the differential event rate
\begin{equation}
\frac{dN}{dm_1\,dm_2\,dz}
=
R_0\,
p_m(m_1,m_2\mid\Lambda_m)\,
\psi(z\mid\Lambda_z)\,
\frac{dV_c}{dz}\,
\frac{1}{1+z}.
\label{eq:differential_rate_reference}
\end{equation}
Here $p_m$ describes the distribution of the source-frame component masses, while
$\psi(z\mid\Lambda_z)$ describes the evolution of the merger-rate density with
redshift with a
Madau--Dickinson-like function (see App.~\ref{app:redshift_evolution}). The remaining factors account for the comoving volume and
for the conversion between source-frame and detector-frame time. $R_0$ is an
overall rate amplitude. The shape of the mass
distribution does not evolve with redshift, and we assume spins are isotropic and distributed uniformly in magnitude, as in the parameter estimation prior.

We model the mass distribution flexibly, allowing the data to determine the number, location, and shape of the population features that are supported. Specifically, we represent it as a finite mixture of components in the
$(m_1,m_2)$ plane. The full distribution is %
a weighted mixture of these components, with the mixture weights inferred from the data.
The mixture is defined in logarithmic source-frame masses,
\begin{equation}
x_1 \equiv \log m_1,
\qquad
x_2 \equiv \log m_2.
\label{eq:mass_log_coordinates}
\end{equation}
Logarithmic coordinates are convenient because compact-binary masses span a
broad range. %

Each component $k$ is a correlated two-dimensional Gaussian. It is useful to
write it as a Gaussian for the primary mass multiplied by a Gaussian for the
secondary mass conditional on the primary,
\begin{equation}
g_k(x_1,x_2)
=
\mathcal{N}\!\big(
x_1\mid\mu_{1k},\sigma_{1k}^{2}
\big)
\mathcal{N}\!\big(
x_2\mid\mu_{2\mid 1,k}(x_1),
\sigma_{2\mid 1,k}^{2}
\big),
\label{eq:conditional_component_kernel}
\end{equation}
where
\begin{align}
\mu_{2\mid 1,k}(x_1)
&=
\mu_{2k}
+
\rho_k
\frac{\sigma_{2k}}{\sigma_{1k}}
\left(x_1-\mu_{1k}\right),
\label{eq:conditional_mean}
\\
\sigma_{2\mid 1,k}
&=
\sigma_{2k}\sqrt{1-\rho_k^{2}}.
\label{eq:conditional_width}
\end{align}

The parameters $(\mu_{1k},\mu_{2k})$ are the center of the two-dimensional Gaussian in the mass
plane, while $\sigma_{1k}$ and $\sigma_{2k}$ set its extent in the two
directions. 
The correlation coefficient $\rho_k$ controls its tilt: %
for nonzero $\rho_k$, the component can follow an inclined feature in the
mass plane.

We parameterize the secondary coordinate of the component center relative to
the primary one,
\begin{equation}
\mu_{2k}
=
\mu_{1k}
+
\log q_{\mu,k},
\qquad
0<q_{\mu,k}<1.
\label{eq:q_ratio_location_parameterization}
\end{equation}
The parameter $q_{\mu,k}$ is the mass ratio at the center of component $k$.
This parameterization guarantees that the component center lies in the ordered
region $m_2<m_1$. Away from the center, the correlation term in
Eq.~\eqref{eq:conditional_mean} allows the preferred mass ratio to vary with
the primary mass.

In physical masses, the model is
\begin{equation}
\begin{split}
& p_m(m_1,m_2\mid\Lambda_m)
\propto {}
S_{\mathrm{low}}(m_1)\,
S_{\mathrm{high}}(m_1)\,
S_{\mathrm{low}}(m_2)
\\
&\times \frac{\mathbf{1}\!\left(m_2<m_1\right)}{m_1m_2}
\times \sum_{k=1}^{K_{\max}}
w_k\,
g_k(\log m_1,\log m_2).
\end{split}
\label{eq:reference_mass_model}
\end{equation}
The factor $(m_1m_2)^{-1}$ is the Jacobian for transforming a density in
$(\log m_1,\log m_2)$ into a density in $(m_1,m_2)$. The indicator function
enforces the physical ordering of the component masses.
The low-mass tapers $S_{\mathrm{low}}$ smoothly suppress the distributions of both $m_1$ and
$m_2$ near their respective lower boundaries, while the high-mass taper $S_{\mathrm{high}}$ acts at the high--mass end of the distribution. The boundary locations and transition widths are inferred jointly with the mixture parameters. The taper functions and the sampling details of their parameters are defined explicitly in App.~\ref{app:tapers}. %

Note that mixture weights $w_k$ are defined before applying the mass tapers and the
restriction $m_2<m_1$, and therefore do not correspond directly to physical
population fractions. Moreover, the Gaussian components represent auxiliary basis
functions rather than distinct astrophysical features: the same structure in the reconstructed density can be represented
by different combinations of components across posterior realizations. We
therefore base the physical interpretation on the complete population density rather than on individual mixture components; corresponding
physical-fraction diagnostics are defined in
Appendix~\ref{app:mixture_complexity}.

For comparison, we also consider a model in which only the primary-mass
distribution is reconstructed flexibly, while the secondary mass is described
through a single global power-law envelope $\propto q^\beta$. We refer to this
as the flexible one-dimensional comparator, in contrast to the reference model
in which both component-mass dimensions and their correlations are reconstructed
flexibly.
Details are given in App.~\ref{app:alternative_parameterizations}.

\subsection{Prior construction for the mass mixture}
\label{sec:mass_mixture_priors}

We outline here the construction of the mixture prior, with full mathematical details given in Appendix~\ref{app:prior-maths}; see~\cite{DelPozzo:2018dpu,Rinaldi:2021bhm} for earlier use of similar constructions in the context of GW populations.\\

\paragraph{Stick-breaking weights.}
\label{sec:stick_breaking_weights}

We adopt a finite stick-breaking construction for the mixture weights,\footnote{ This can be understood by imagining a unit-length
stick that represents the total mixture probability. The first component
takes a fraction $v_1$ of the stick. The second takes a fraction $v_2$ of what
remains, and so on. Components that are not required by the data naturally receive very little weight. Because the construction is stopped after a finite number of components, we normalize the retained weights to unity.}
\begin{equation}
\begin{split}
    &v_k\sim{\rm Beta}(1,\alpha_{\rm DP}),
    \\
    &w_k\propto
    v_k\prod_{\ell<k}(1-v_\ell),
    \label{eq:raw_stick_weights}
    \end{split}
\end{equation}
with $k \in [1 , \dots K_{\max} ]$ with fixed and finite $K_{\max}$.
The concentration parameter $\alpha_{\mathrm{DP}}$ controls how fragmented
the mixture tends to be. %
We infer this
complexity from the data by assigning a log-normal prior to its inverse,
\begin{align}
\ell_\alpha
&\sim
\mathrm{Normal}\!\left(
\mu_{\log\alpha^{-1}},
\sigma_{\log\alpha^{-1}}
\right),
\\
\alpha_{\mathrm{DP}}^{-1}
&=
\exp(\ell_\alpha).
\label{eq:dp_concentration_prior}
\end{align}
Thus $K_{\max}$ specifies the maximum available complexity rather than
the number of structures imposed on the population. %

\paragraph{Component locations.}
The Gaussian centers are allowed to move within the region supported by
the event-level posterior information. We determine this region by propagating
posterior samples to source-frame masses while varying cosmology over its prior.
The primary-mass centers span the resulting source-frame $m_1$ range, while the
allowed range of the characteristic mass ratio $q_{\mu,k}$ is based on
posterior quantiles to reduce sensitivity to the broader tails of the
mass-ratio posteriors. The construction is detailed in
Appendix~\ref{app:data_adaptive_bounds}.

\paragraph{Component widths and resolved scale.}
For coordinate $j\in\{1,2\}$ and component $k$, we write
\begin{equation}
    \sigma_{jk}
    =
    L_j+\tau_j\exp(\epsilon_{jk}),
    \qquad
    \epsilon_{jk}\sim{\rm Normal}(0,s_j),
    \label{eq:component_width_hierarchy}
\end{equation}
where $L_j$ sets the minimum resolved logarithmic mass scale, $\tau_j$ sets a
shared characteristic scale above this floor, and $\epsilon_{jk}$ allows
component-to-component variation. The floors $L_j$ are calibrated from the
event-level mass uncertainties, preventing the population model from resolving
structure below the scale supported by individual observations.
The prior range of $\tau_j$ spans the catalog support.
Their constructions are given in
Appendix~\ref{app:width_hierarchy}.

\paragraph{Correlations and component shape.}
Correlations between the two logarithmic masses are inferred for each
component with a symmetric prior.  %
The prior prevents collapse onto an arbitrarily thin line,
while a separate soft regularization discourages very large disparities
between the two marginal component widths unless supported by the data. The
corresponding priors are defined in
Appendix~\ref{app:aspect_regularization}.

\paragraph{Reference configuration.}
The complete numerical specification of the reference model and its induced
mass prior is given in Appendix~\ref{app:reference_configuration}. We use at
most $K_{\max}=15$ components and explicitly test variations of the mixture
complexity, resolution scale, component-shape regularization, mixture-weight
prior, and adaptive bounds, for a total of 12 alternative configurations. These robustness tests are summarized in
Appendix~\ref{app:baseline_variations}.

\subsection{Inference details}
\label{sec:inference}

We infer the population and cosmological hyperparameters within a hierarchical Bayesian framework~\cite{Mandel:2018mve,Vitale:2020aaz}. The analysis combines the information from all detected events, accounts for the priors adopted in the single-event analyses, and corrects for the fact that only a fraction of the underlying population is detected.

We use the likelihood-evaluation strategy introduced in Ref.~\cite{Mancarella:2025uat}, where the parameters of all events are sampled jointly with the population and cosmological hyperparameters.
Selection effects are evaluated by reweighted Monte Carlo integration over simulated injections~\cite{Tiwari:2017ndi,Farr:2019rap,Essick:2022ojx}. %
Following Refs.~\cite{Talbot:2023pex,Heinzel:2025ogf}, we monitor the Monte Carlo uncertainty of the selection estimate and require the resulting variance of the estimated log likelihood to remain below unity.

The full hierarchical likelihood, the treatment of the single-event priors, the selection correction, and the bounded posterior approximation are described in App.~\ref{app:stats}. Posterior sampling is performed 
with an extension of the \texttt{NumPyro} implementation of the package \pymcpop~\cite{pymcpop,Mancarella:2025uat}, using the No-U-Turn Sampler~\cite{JMLR:v15:hoffman14a,Brooks_2011} and running on a GPU. 
We verify convergence by ensuring that the Gelman-Rubin $\hat{r}$ statistics~\cite{10.1214/ss/1177011136} over four independent chains satisfies $\hat{r}\approx1.0$ for all variables.

\subsection{Identification and characterization of population features}
\label{sec:feature_framework}

We reconstruct the normalized source-frame density per logarithmic component
mass,  
\begin{equation}
   p_{\ln}(m_1,m_2 \mid\Lambda_m)
    \equiv
    m_1 m_2\ p_m(m_1,m_2\mid\Lambda_m)\, .
    \label{eq:joint_log_mass_density}
\end{equation}

To study structures in the mass plane, their interpretation, and their
relation to cosmology in a quantitative way, it is useful to adopt an explicit
and reproducible definition of what is meant by a population feature.
To this end, we use a custom feature-finding framework developed to identify
two-dimensional structures in a deterministic and reproducible way. The
framework identifies localized concentrations (``peaks''), extended regions
of suppressed density (``valleys''), and changes of logarithmic slope, and
measures their locations, geometries, strengths, and posterior recovery.
Reference structures are first identified on the posterior-median density and
then matched across posterior realizations.

To compare the genuinely two-dimensional information with that retained by
one-dimensional projections, and to relate the joint structures to known
features in the component-mass distributions, we also construct independent
one-dimensional catalogues of peaks, dips, and slope changes in the $m_1$
and $m_2$ marginals. The one- and two-dimensional catalogues are constructed
independently and compared only after both searches have been completed.

The framework is implemented in the open-source package
\texttt{posterior-landscape}~\cite{plandscape}; a detailed description and
validation are given in a companion paper~\cite{plandscape-paper}.
Appendix~\ref{app:feature-finder} summarizes the feature-identification
procedure. %

\subsection{Validation on simulations}\label{sec:simulations}

We validate the reconstruction on a simulated full-CBC population constructed
to reproduce the main features of the GWTC--5 population constraints
\cite{LIGOScientific:2026ctl}. We generate an O5-like catalogue of
$1000$ detections with $\mathrm{SNR}>8$ and analyze it with the same inference settings used for the real data.
The workflow recovers
the injected joint component-mass distribution and its principal features
within posterior uncertainty, without evidence for additional structures, as well as
the injected cosmological parameters. We use a large catalog so that statistical uncertainty is reduced, making systematic reconstruction
biases easier to expose. The complete injected population,
simulation setup, and recovery diagnostics are given
in Appendix~\ref{app:mass_model_simulations}.

\section{Application to the GWTC-5.0 catalog}

In this section we present results on the GWTC-5.0 catalog.
The section is organized as follows.
Section~\ref{sec:data} summarizes data choices, overall model design and performance, and robustness tests. 
Section~\ref{sec:two_dimensional_mass_features} characterizes the structures
identified in the joint mass plane and relates them to their one-dimensional
manifestations. Section~\ref{sec:h0_model_comparison} compares the cosmological
constraints obtained with the flexible one- and two-dimensional population
models. Section~\ref{sec:joint_features_h0} then examines which joint
population features act as spectral-siren anchors and how their cosmological
dependence is distributed between the common-mass and mass-ratio directions.
Finally, Section~\ref{sec:population_literature_comparison} compares the
population and cosmological results with previous studies.

\subsection{Data and model}\label{sec:data}
We analyze the subset of GWTC-5.0 events used in the LVK population and cosmology analyses~\cite{LIGOScientific:2026ctl,LIGOScientific:2026uyd},
including all CBC detected with false-alarm rate $\mathrm{FAR}<0.25\,\mathrm{yr}^{-1}$.
This yields a total of $N_{\rm obs}=237$ detections spanning the LVK observing
runs considered in the GWTC-5.0 release.\footnote{Note that the LVK population paper~\cite{LIGOScientific:2026ctl} includes additional O4a events whose FAR
estimates were revised from above to below the chosen threshold between the GWTC-4.0 and GWTC-5.0 data releases. We do not include those events here. The LVK cosmology paper~\cite{LIGOScientific:2026uyd} instead excludes the BBH event GW231123. The event-level parameter-estimation (PE) data products are available at~\cite{ligo_scientific_collaboration_and_virgo_2022_6513631,ligo_scientific_collaboration_and_virgo_2023_8177023,ligo_scientific_collaboration_and_virgo_2025_17014085,ligo_scientific_collaboration_and_virgo_2026_20348005,ligo_scientific_collaboration_and_virgo_2026_20348006}.} %
For BBH--only events, we use samples obtained with the \texttt{IMRPhenomXPHM} waveform model for events in O1 to O3, and \texttt{IMRPhenomXPHM\_SpinTaylor} for O4 events. The only exception is the heavy binary GW231123~\cite{LIGOScientific:2025rsn} for which we use the \texttt{NRSur7dq4} waveform samples.
For NS-containing events we use the \texttt{IMRPhenomPv2\_NRTidal} samples. %
For selection effects, we use the publicly released injection sets accompanying the GWTC-5.0 population
products, available at~\cite{ligo_scientific_collaboration_2026_19500052}; see~\cite{Essick:2025zed,Essick:2023toz} for additional details.

Throughout the results below, we adopt the baseline configuration summarized
in
Appendix~\ref{app:reference_configuration}.
We infer the cosmological parameters $\{H_0, \Omega_{\rm m} \}$ within a $\Lambda$CDM scenario, and the Madau-Dickinson rate parameters alongside. Prior ranges are given in Apppendix~\ref{app:reference_configuration}. 

We first examine how much of the available model complexity is used and which characteristic component widths are selected by the data.
The posterior
corresponds to a median effective number of about ${\sim}3.6$ appreciably contributing
mixture components, compared with a prior median of ${\sim}5.0$, despite
$K_{\max}=15$ components being available. The characteristic widths selected
by the data correspond to fractional mass scales of approximately $23\%$ and
$27\%$ along the two component-mass directions, or roughly
$2$--$3\,M_\odot$ at $10\,M_\odot$ and $7$--$8\,M_\odot$ at
$30\,M_\odot$. These scales remain above the event-resolution floor of
about $5\%$ ($\sim0.5\,M_\odot$ at $10\,M_\odot$ and
$\sim1.5\,M_\odot$ at $30\,M_\odot$).
Details are reported in Appendix~\ref{app:mixture_complexity}.

We additionally assess the stability of
the inference under a broad set of variations of the mixture complexity,
minimum resolved scale, component-shape regularization, mixture-weight prior,
and data-adaptive bounds, for a total of 12 variations. %
The complete set of
robustness variations is presented in
Appendix~\ref{app:baseline_variations}.

\subsection{Joint mass-plane morphology and its marginal manifestations}
\label{sec:two_dimensional_mass_features}

\begin{figure*}[t]
    \centering
    \includegraphics[width=0.75\textwidth]
    {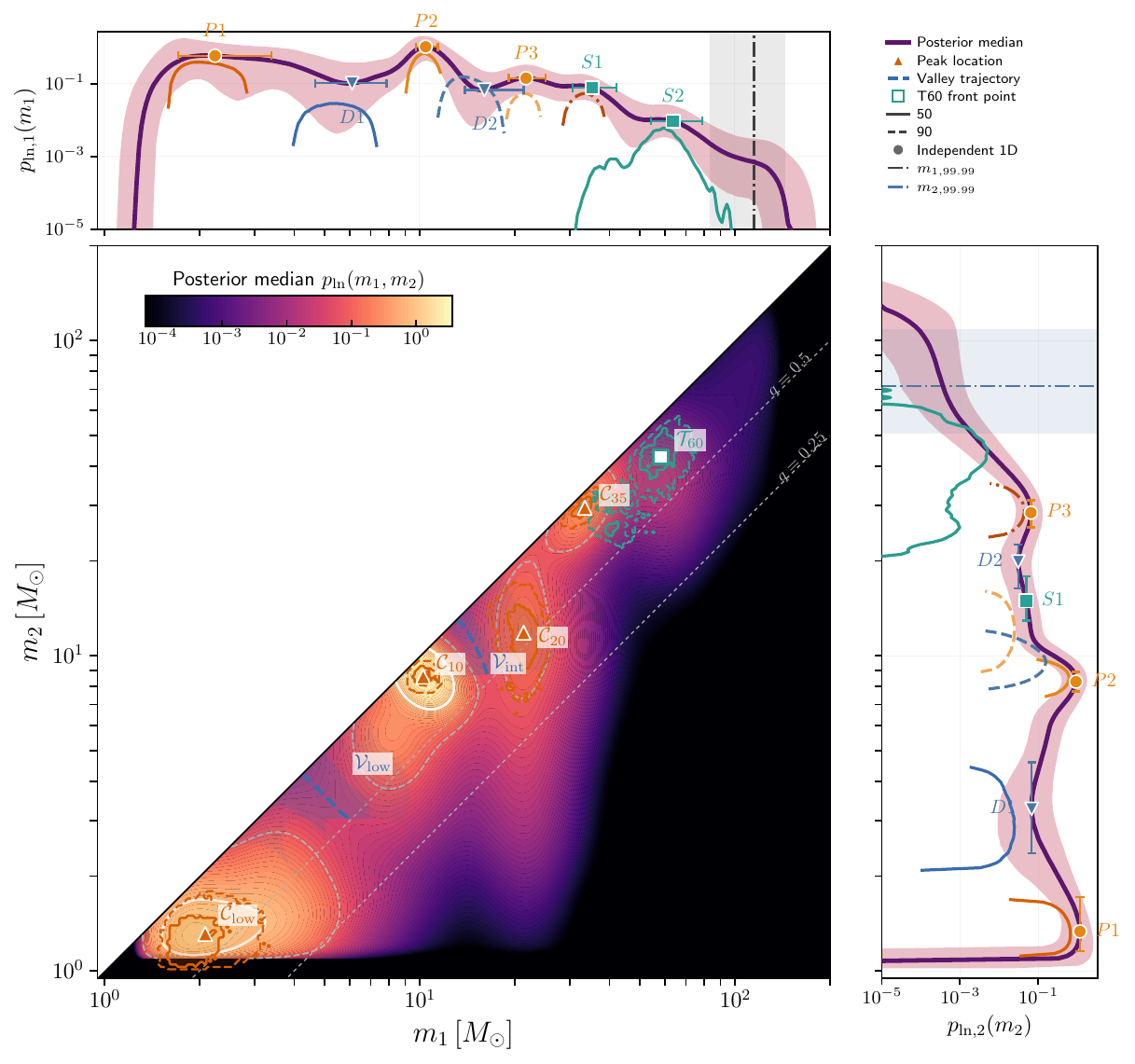}
    \caption{Joint component-mass landscape reconstructed from the GWTC-5.0 catalog and
its one-dimensional projections. The central panel shows the posterior
median of \(p_{\ln}(m_1,m_2)\). Orange markers and contours summarize
concentration maxima; blue dashed curves and shaded corridors summarize
valley trajectories; the turquoise square and contours locate a slope transition front. The location summaries are conditional on recovery
and an available measurement, with \(50\%\) and \(90\%\) contours or corridors.
The upper and right panels show posterior medians and central \(90\%\)
intervals of the one--dimensional marginals. Coloured curves show projected
feature measures; the turquoise front-position distributions have amplitudes
rescaled for display. Independently detected one-dimensional features have
central \(90\%\) location intervals. Grey and blue bands, with dash-dotted
median lines, give central \(90\%\) intervals of \(m_{1,99.99}\) and
\(m_{2,99.99}\). Diagonal dotted lines mark fixed mass ratios.}
    \label{fig:joint_mass_feature_origins}
\end{figure*}

\begin{table*}[t]
    \centering
    \small
    \renewcommand{\arraystretch}{1.20}
    \caption{Principal structures in the joint mass plane and their
    one-dimensional manifestations. Reference geometries are in
    $M_\odot$ and are shown at their posterior medians.
These geometries describe concentration maxima, valley endpoints, and the slope-change front.
    $P_{\rm morph}$ is the unconditional morphology-recovery probability.
    Marginal location uncertainties are central $90\%$ posterior intervals
    conditional on morphological recovery.
    Strengths are posterior medians and central $68.3\%$ intervals conditional
    on a valid draw-level strength measurement. For concentrations, the
    strength is the relative prominence
    $A_{\rm P}=1-p_{\ln,\mathrm{saddle}}/p_{\ln,\max}$; for valleys it is the
    relative depth $A_{\rm V}$ defined in
    Eq.~\eqref{eq:relative_valley_depth}; for $\mathcal T_{60}$ it is the
    logarithmic-slope contrast $\Delta\alpha$. Strength values characterize
    different notions of contrast for the different morphology classes and
    should therefore be compared quantitatively only within a given class.
    Marginal recovery probabilities and classifications refer to the
    independently constructed one-dimensional catalogues.}
    \label{tab:joint_feature_census}
    \resizebox{\textwidth}{!}{%
    \begin{tabular}{@{}llcccp{0.21\textwidth}p{0.21\textwidth}@{}}
        \hline
        Feature & Type & Reference geometry & $P_{\rm morph}$ & Strength
        & $m_1$ manifestation & $m_2$ manifestation \\
        \hline

        $\mathcal C_{\rm low}$ & concentration
        & $(2.09,1.31)$
        & $99.1\%$
        & $A_{\rm P}=0.985^{+0.014}_{-0.100}$
        & peak at $2.24^{+1.14}_{-0.53}$ ($99.3\%$)
        & peak at $1.33^{+0.38}_{-0.18}$ ($>99.9\%$)
        \\

        $\mathcal V_{\rm low}$ & valley
        & $(5.9,3.0)\rightarrow(4.3,4.2)$
        & $52.2\%$
        & $A_{\rm V}=0.886^{+0.078}_{-0.182}$
        & dip at $6.11^{+1.74}_{-1.44}$ ($99.2\%$)
        & dip at $3.26^{+1.32}_{-0.90}$ ($>99.9\%$)
        \\

        $\mathcal C_{10}$ & concentration
        & $(10.26,8.58)$
        & $99.7\%$
        & $A_{\rm P}=0.977^{+0.021}_{-0.059}$
        & peak at $10.45^{+0.98}_{-0.72}$ ($100\%$)
        & peak at $8.28^{+0.61}_{-0.57}$ ($100\%$)
        \\

        $\mathcal V_{\rm int}$ & valley
        & $(16.3,8.7)\rightarrow(13.2,12.9)$
        & $73.5\%$
        & $A_{\rm V}=0.933^{+0.036}_{-0.094}$
        & dip at $16.05^{+5.32}_{-2.14}$ ($92.2\%$);
          paired with $P_3$ quoted below
        & no unique marginal counterpart
        \\

        $\mathcal C_{20}$ & concentration
        & $(21.4,11.8)$
        & $95.4\%$
        & $A_{\rm P}=0.551^{+0.239}_{-0.275}$
        & $P_3$ at $21.75^{+3.35}_{-2.56}$ ($93.6\%$);
          $89.9\%$ peak, $10.1\%$ slope change;
          $A_{\rm P}^{1{\rm D}}=0.649^{+0.248}_{-0.418}$ when a valid peak
        & feature at $14.94^{+2.93}_{-1.99}$ ($84.8\%$);
          $86.0\%$ slope change, $14.0\%$ peak
        \\

        $\mathcal C_{35}$ & concentration
        & $(33.4,29.5)$
        & $99.9\%$
        & $A_{\rm P}=0.857^{+0.092}_{-0.134}$
        & feature at $35.26^{+6.91}_{-4.70}$ ($90.2\%$);
          $50.5\%$ peak, $49.5\%$ slope change
        & peak at $28.45^{+2.66}_{-2.90}$ ($99.9\%$);
          $99.1\%$ peak
        \\

        $\mathcal T_{60}$ & slope change
        & $(58.2,42.8)$
        & $36.7\%$
        & $\Delta\alpha=4.92^{+3.13}_{-2.12}$
        & high-mass transition at
          $63.64^{+15.24}_{-9.47}$ ($86.4\%$)
        & no secure direct counterpart
        \\

        \hline
    \end{tabular}}
\end{table*}

Figure~\ref{fig:joint_mass_feature_origins} shows the reconstructed joint mass
distribution and its $m_1$ and $m_2$ marginals, together with the posterior
geometry of the identified two-dimensional structures, their projections, and
the independently detected one-dimensional features.

We describe concentrations by their maxima and valleys
by their trajectories. For slope changes, a curve
traces where the density locally flattens before steepening again. A representative \emph{front point}
is constructed in each posterior draw from the separate medians of
the logarithmic common mass \(\ln\sqrt{m_1m_2}\) and logarithmic mass
asymmetry \(\tfrac12\ln(m_1/m_2)\) along this curve.
The construction is detailed in Appendix~\ref{app:feature-finder}.

We characterize each joint feature through two complementary quantities:
\begin{itemize}
\item \emph{Morphology stability} quantifies how frequently a compatible
structure is recovered across posterior realizations,
\begin{equation}
P_{{\rm morph},f}
=
\frac{1}{N_s}\sum_{s=1}^{N_s} Z_f^{(s)},
\label{eq:joint_feature_morphology_probability}
\end{equation}
where $Z_f^{(s)}=1$ when feature $f$ is recovered in realization $s$.

\item \emph{Feature strength} quantifies how strongly the recovered structure
stands out from its local surroundings. For a concentration, we use
\begin{equation}
A_{\rm P}
=
1-\frac{p_{\ln,\mathrm{saddle}}}{p_{\ln,\max}},
\label{eq:relative_peak_prominence}
\end{equation}
where $p_{\ln,\max}$ is the density at the peak and $p_{\ln,\mathrm{saddle}}$ the one at the neighboring saddle.
For a valley we use
\begin{equation}
A_{\rm V}
=
1-\frac{p_{\rm V}}{\sqrt{p_{\rm L}p_{\rm R}}},
\label{eq:relative_valley_depth}
\end{equation}
where $p_{\rm V}$ is the density along the trough and $p_{\rm L}$ and
$p_{\rm R}$ are the local maxima reached on its two sides. For a slope change,
the strength is
\begin{equation}
\Delta\alpha=\alpha_{\rm pre}-\alpha_{\rm post},
\label{eq:slope_change_strength}
\end{equation}
with $\alpha_{\rm pre}$ and
$\alpha_{\rm post}$ being the logarithmic slopes measured over the intervals immediately preceding and
following the flattening.
The precise constructions are given in Appendix~\ref{app:feature-finder}.
\end{itemize}

We test the catalogue against variations of both the population model and the
feature-finder settings; details are given in
Appendices~\ref{app:baseline_variations} and
\ref{app:feature_finder_variations}. Unless stated otherwise, the structures
discussed below are stable under these variations. Their principal properties
and one-dimensional manifestations are summarized in
Table~\ref{tab:joint_feature_census}.

To provide a compact census, the principal structures are summarized
below. %
We retain in the discussion structures with posterior occurrence probability at least $50\%$, together with lower-probability structures having a primary independently detected one-dimensional counterpart.
Quoted location uncertainties are central \(90\%\) credible intervals; strength uncertainties are central \(68.3\%\) intervals.
\\

\paragraph{Familiar population structures and their joint interpretation.}

Widely know features are robustly recovered by the algorithm, with the two--dimensional view providing additional insight. We discuss them first.

\begin{itemize}

\item \textbf{$\boldsymbol{\mathcal C_{\rm low}}$ at
$\boldsymbol{(2.09^{+1.01}_{-0.43},\,1.31^{+0.37}_{-0.23})\,M_\odot}$:
a stable and highly prominent low-mass concentration.}

This is the familiar over-density corresponding to BNS systems~\cite{LIGOScientific:2026ctl,LIGOScientific:2025pvj}, recovered
with high $P_{\rm morph}$ and prominence.

\item \textbf{$\boldsymbol{\mathcal V_{\rm low}}$ from
$\boldsymbol{(5.9^{+3.0}_{-2.0},\,3.0^{+2.1}_{-1.2})\,M_\odot}$
to
$\boldsymbol{(4.3^{+2.4}_{-1.7},\,4.2^{+2.4}_{-1.7})\,M_\odot}$,:
a moderately stable but deep oblique deficit whose marginal dips are much more
persistent than its full joint geometry.
}

This underdensity is the expected transition between NS and BH systems~\cite{Farah:2021qom,LIGOScientific:2026ctl,LIGOScientific:2025pvj}.
Its relatively low $P_{\rm morph}$ indicates that the full joint valley
morphology is not recovered consistently across posterior realizations; nevertheless, when measurable, the suppression is pronounced, with a
median valley density of only ${\sim}11\%$ of the characteristic density on
the two sides of the trough (Table~\ref{tab:joint_feature_census}). %
The independently reconstructed marginals
show much more persistent dips, indicating that marginalization compresses the oblique deficit into apparently simpler and more stable
one-dimensional underdensities.

\item \textbf{$\boldsymbol{\mathcal C_{10}}$ at
$\boldsymbol{(10.26^{+0.96}_{-0.88},\,8.58^{+0.80}_{-0.87})\,M_\odot}$:
a stable, highly prominent concentration with a clear peak in each marginal.}

This peak corresponds to the familiar population scale near $10\,M_\odot$~\cite{Tiwari:2020otp,LIGOScientific:2020kqk, KAGRA:2021duu, Edelman:2022ydv, Farah:2023vsc, Callister:2023tgi, Gennari:2025nho, Legred:2026oiz, Bertheas:2026odj}.
It is robustly recovered with high prominence and projects onto independently
recovered peaks.

\item \textbf{$\boldsymbol{\mathcal C_{35}}$ at
$\boldsymbol{(33.4^{+3.1}_{-3.9},\,29.5^{+3.3}_{-3.0})\,M_\odot}$:
a stable and prominent near-equal-mass peak whose one-dimensional
classification depends on the projection.}

This peak corresponds to the familiar feature around $35\,M_\odot$~\cite{KAGRA:2021duu,LIGOScientific:2025pvj, Mould:2026sww, Sadiq:2025vly,LIGOScientific:2026ctl}. The
two-dimensional reconstruction identifies it as a stable, localized, and
prominent concentration close to the equal-mass diagonal.
Its primary-mass manifestation, however, is classified almost
equally often as a peak ($50.5\%$) and a slope change ($49.5\%$) (whereas the
secondary-mass counterpart is classified as
a peak in $99.1\%$ of its recoveries).
This analysis therefore clarifies a recent ambiguity in the interpretation of this feature in the one dimensional marginal as either a peak or a slope-change~\cite{LIGOScientific:2026ctl}: this ambiguity is a property of the
one-dimensional projection rather than of the joint population morphology, where the underlying structure is unambiguously a peak.

\end{itemize}

\paragraph{Emergent intermediate-mass structure.}

Next, we identify some emergent, weak structure that is not universally found in existing analyses (see Sec.~\ref{sec:population_literature_comparison} below for an extended discussion). 

\begin{itemize}

\item \textbf{$\boldsymbol{\mathcal V_{\rm int}}$ from
$\boldsymbol{(16.3^{+7.5}_{-3.2},\,8.7^{+9.8}_{-2.6})\,M_\odot}$
to
$\boldsymbol{(13.2^{+4.7}_{-1.8},\,12.9^{+4.6}_{-1.7})\,M_\odot}$.:
a moderately stable but deep oblique under-density. %
}

Similarly to $\mathcal V_{\rm low}$, this intermediate under-density is not recovered in part of the posterior ($P_{\rm morph}=0.735$).
However, when measurable it is deep, with
a median trough density of only ${\simeq}6.7\%$ of the characteristic density
on its two sides (Table~\ref{tab:joint_feature_census}). Its primary-mass
projection overlaps the independently recovered dip $D_2$, which forms a
recurrent ordered dip--excess pattern with the higher-mass feature $P_3$ near
$20\,M_\odot$: these two features coexist in $87.6\%$ of posterior
realizations. Thus, marginalization compresses two geometrically distinct
joint structures---the extended oblique deficit $\mathcal V_{\rm int}$ and
the localized off-diagonal concentration $\mathcal C_{20}$---into a single
dip-followed-by-excess pattern.

\item \textbf{$\boldsymbol{\mathcal C_{20}}$ at
$\boldsymbol{(21.4^{+3.3}_{-3.5},\,11.8^{+3.9}_{-3.3})\,M_\odot}$:
a recurrent but weak unequal-mass concentration whose morphology is clarified
only in the joint mass plane.}

In the baseline posterior, $\mathcal C_{20}$ is recovered with high
$P_{\rm morph}$ but has conditional relative prominence
$A_{\rm P}=0.551^{+0.239}_{-0.275}$, substantially below the other principal
concentrations; at its median strength, the surrounding pass remains at about
$45\%$ of the peak density. %
Its median location,
$m_2/m_1\simeq0.55$, identifies a specific off-diagonal excess of
unequal-component-mass systems.

Its one-dimensional manifestations are less definite. The primary-mass
counterpart is recovered in $93.6\%$ of posterior realizations and is
classified as a peak in $89.9\%$ of its recoveries and as a slope change in
$10.1\%$. The secondary-mass counterpart is recovered in $84.8\%$ of
realizations and is predominantly classified as a slope change ($86.0\%$),
rather than a peak ($14.0\%$). The joint reconstruction therefore gives the
clearest interpretation of this region as a localized, comparatively weak
unequal-mass concentration whose appearance depends strongly on projection.

\end{itemize}

Under variations of the default feature-finder settings, we find
a common exception for three of the weaker structures. $\mathcal C_{20}$ is
matched in 11 of the 12 population-model variations, while
$\mathcal V_{\rm low}$ and $\mathcal V_{\rm int}$ likewise lack compatible
counterparts in the same remaining variation. This configuration combines a
weaker aspect prior with a lower minimum component-width scale and is therefore
among the most permissive configurations considered.
Repeating the catalogue extraction with alternative feature-finder settings
shows that these omissions are also threshold dependent: a calibration yielding a
lower effective persistence cutoff for including a structure in the catalogued features recovers compatible counterparts of all
three structures, including $\mathcal C_{20}$ with
$P_{\rm morph}\simeq0.933$ and conditional median prominence
$A_{\rm P}\simeq0.538$. Details are given in
Appendix~\ref{app:feature_finder_variations}.

Taken together, these results support weak intermediate-mass structure while
warranting caution in its interpretation. Threshold sensitivity does not by
itself establish whether the underlying structure is astrophysical: a genuine
low-contrast feature may fall below a catalogue cut, while apparent weak
structure may also arise from finite-catalogue fluctuations. Additional data
or simulation-based investigations~\cite{Farah:2023vsc} will therefore be needed to determine how
robustly these structures can be resolved.

Importantly, however, the global $H_0$ posterior in the population-model configuration
where these intermediate structures are not retained by the default catalogue
is essentially unchanged. At the present sensitivity, their formal catalogue
inclusion therefore affects primarily the astrophysical interpretation of the
population morphology rather than the global cosmological constraint (see Sec.~\ref{sec:h0_model_comparison} below and Appendix~\ref{app:baseline_variations}).\\

\paragraph{High-mass transition and population extent.}

Finally, we discuss tentative structure at the high--end of the mass spectrum and the population extent.

\begin{itemize}

\item \textbf{\(\mathcal T_{60}\) at the front location
\((58.2^{+12.6}_{-18.1},\,42.8^{+9.6}_{-17.0})\,M_\odot\):
a tentative unequal-mass slope change with a more persistent primary-mass transition.}

The joint morphology is recovered in only $36.7\%$ of posterior realizations.
With the default feature-finder settings, a compatible family occurs in all
12 population-model configurations, but always with low morphology stability,
$0.249\leq P_{\rm morph}\leq0.411$. 
An independently identified primary-mass feature at
\(m_1=63.64^{+15.24}_{-9.47}\,M_\odot\) is recovered in \(86.4\%\) of draws.
The high-mass scale is therefore more persistent in the primary marginal.
We therefore retain
$\mathcal T_{60}$ as a tentative joint morphology and regard the marginal
high-mass transition as the more persistent observational signature.

\end{itemize}

Finally, the localized catalogue is supplemented by a strongly asymmetric upper tail.
The $99.99$th percentiles of the two marginals are
\begin{equation}
    \begin{split}
        m_{1,99.99} &=115.2^{+29.4}_{-32.0}\,M_\odot,\\
        m_{2,99.99} &= 71.7^{+37.2}_{-20.9}\,M_\odot.
    \end{split}
    \label{eq:component_upper_tail_results}
\end{equation}
The larger extent of the primary-mass tail reflects the asymmetry of the joint
high-mass population. These quantiles characterize the global extent of the
distribution rather than localized features or sharp physical edges.

\subsection{Cosmological precision from flexible primary-mass and joint component-mass models}
\label{sec:h0_model_comparison}

\begin{figure}[t]
    \centering
    \includegraphics[width=0.4\textwidth]
    {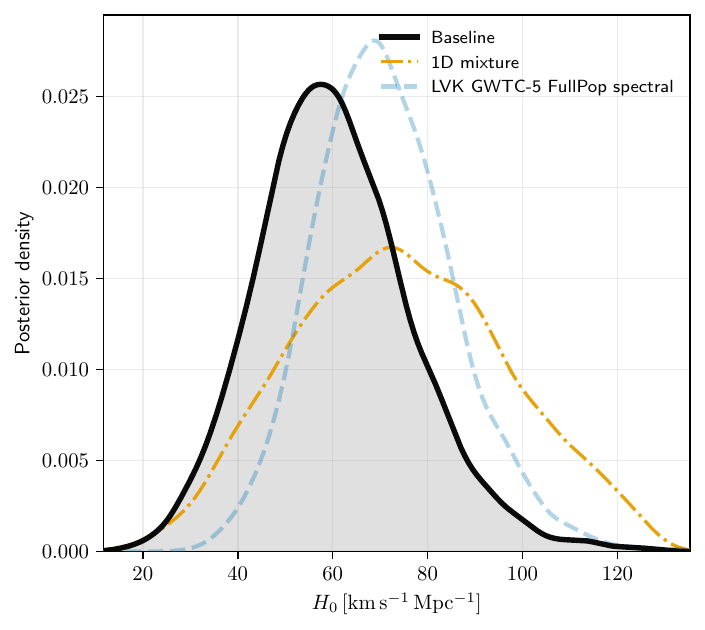}
    \caption{ Marginal posterior density of $H_0$ from spectral sirens in the GWTC-5.0 catalog. The thick black curve shows
    the result with the baseline joint flexible mass model of this wok, the ochre dash-dotted curve the flexible
    one-dimensional mixture comparator, and the blue dashed curve the LVK GWTC--5.0 \textsc{FullPop} spectral siren result for comparison.
    }
    \label{fig:H0_model_comparison}
\end{figure}

We next isolate the cosmological value of retaining the joint component-mass
structure and investigate the cosmological importance of the population
features identified above. We first compare the global $H_0$ constraints
obtained with the one- and two-dimensional population models, and subsequently
use the draw-resolved feature catalogue to determine which structures carry the
corresponding posterior information.

The baseline analysis yields
\begin{equation}
    H_0
    =59.5^{+16.5}_{-14.6}\,
    {\rm km\,s^{-1}\,Mpc^{-1}}
    \qquad (68.3\%).
    \label{eq:h0_baseline_2d_result}
\end{equation}
Importantly, the result is essentially unchanged under all the considered
variations around the baseline, indicating a remarkable stability of the
inferred cosmology---see Appendix~\ref{app:baseline_variations} for details.

For comparison, the one-dimensional mixture model described in
Sec.~\ref{sec:population_model} gives
$H_0=73.6^{+23.9}_{-22.8}\,{\rm km\,s^{-1}\,Mpc^{-1}}$.
Both posterior distributions are shown in
Fig.~\ref{fig:H0_model_comparison}. Relative to this flexible primary-mass
comparator, the widths of the $68.3\%$ and $90\%$ credible intervals decrease
by $33.4\%$ and $28.3\%$, respectively.
This reduction highlights the added value of flexibly reconstructing the full
joint component-mass distribution. In the one-dimensional mixture, only the
primary-mass structure is reconstructed flexibly, while the secondary mass
follows a single global ${\sim}q^\beta$ relation. The joint model instead
constrains structures through their paired $m_1$ and $m_2$ locations and
geometries, reducing the freedom to trade changes in $H_0$ against changes in
the source-frame mass distribution and thereby adding substantial information.

Figure~\ref{fig:H0_model_comparison} also shows, for reference, the LVK
GWTC--5 spectral-siren result obtained with the strongly parameterized
\textsc{FullPop} model~\cite{LIGOScientific:2026uyd},
$H_0=70.1^{+15.1}_{-13.3}\,{\rm km\,s^{-1}\,Mpc^{-1}}$.
The LVK and our flexible one-dimensional posteriors have similar central
locations, but the latter is substantially broader, reflecting the greater
freedom allowed in the reconstructed mass spectrum.
The joint reconstruction retains this weakly modelled treatment of population
structure while recovering much of the lost precision: its $68.3\%$ interval
is only about $10\%$ wider than the LVK result.
This is particularly significant because our model allows the number,
locations, widths, and joint geometry of localized structures to emerge
directly in the $(m_1,m_2)$ plane rather than prescribing a structured
one-dimensional spectrum. The near-LVK precision therefore shows that the
additional information carried by the two-dimensional mass morphology can
compensate for relaxing prescribed spectral structure, reducing sensitivity to
mass-model systematics while retaining comparable cosmological precision.
Further comparison with other results is given in
Sec.~\ref{sec:population_literature_comparison}.

\begin{figure*}[t]
    \centering
    \includegraphics[width=\textwidth]
    {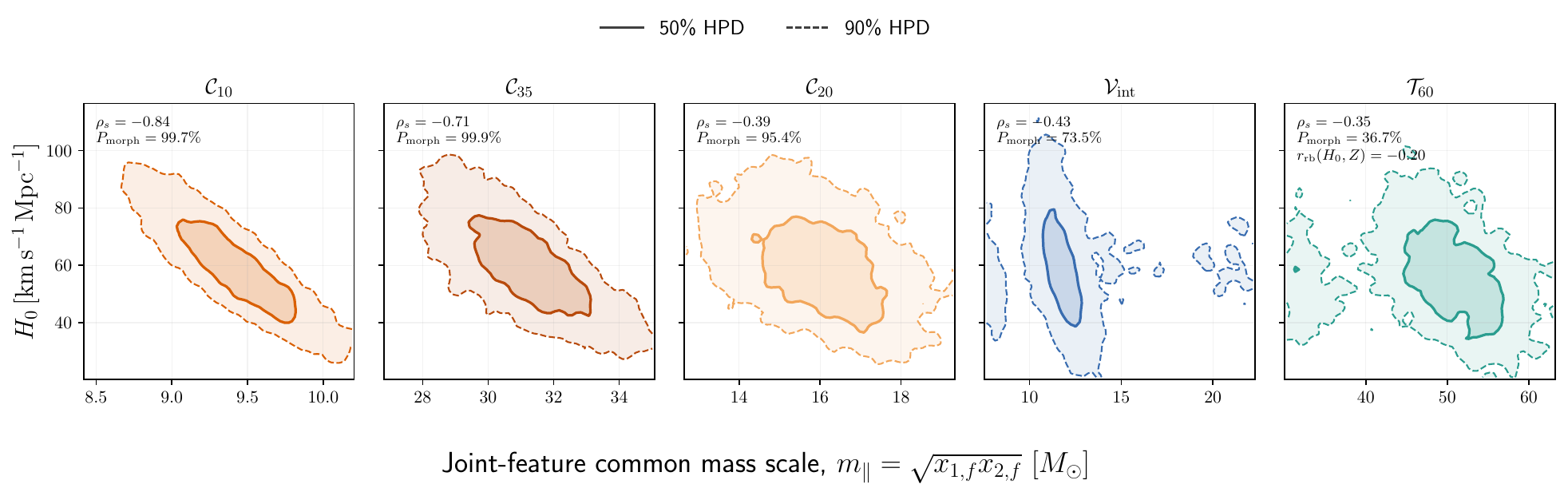}
    \caption{ Posterior association between \(H_0\) and the common mass scale
\(m_\parallel=\sqrt{x_{1,f}x_{2,f}}\) of five joint-density structures.
The position is the centroid for concentrations and valleys, and the front
point for \(\mathcal T_{60}\). Each panel and its Spearman coefficient
\(\rho_s\) use all draws with a valid position; \(P_{\rm morph}\) is the
unconditional morphology-recovery probability. The disconnected support of
\(\mathcal V_{\rm int}\) reflects its less frequent secondary location
configuration. The \(\mathcal T_{60}\) front scale is moderately
anticorrelated with \(H_0\); its rank-biserial coefficient separately
quantifies the association between \(H_0\) and morphology recovery.}
    \label{fig:H0_joint_feature_anchors}
\end{figure*}

The change in population model, however, also shifts the posterior location: the joint
reconstruction favors lower $H_0$ than either of the one-dimensional results.
A plausible origin of this shift is the different reconstruction of the
dominant $\mathcal C_{10}$ peak. In the one-dimensional model, this feature
is coupled to the global $q^\beta$ envelope, which favors nearly equal masses,
whereas the joint model infers its full two-dimensional geometry and places the
concentration slightly farther from the equal-mass boundary. This is accompanied by a
small ($\sim 4\%$) displacement of the characteristic mass scale associated with the feature toward higher mass.
Since the $\mathcal C_{10}$ mass scale is strongly anticorrelated with
$H_0$, such shift can naturally move the
$H_0$ posterior toward lower values. More targeted studies of population
parameterizations in the two-dimensional mass plane will be needed to
determine whether such shifts could become a source of systematic uncertainty
as statistical errors shrink.\footnote{Note, however, that the injection--recovery test in
Appendix~\ref{app:mass_model_simulations} recovers the injected
$H_0$ within posterior uncertainty for the simulated population
considered, suggesting this is not the case.}

\subsection{Joint mass-plane features as spectral-siren anchors}
\label{sec:joint_features_h0}
We now examine how the locations and geometries of the reconstructed population
features covary with $H_0$. 
We follow an approach introduced by Ref.~\cite{Mali:2024wpq} in a parametric one--dimensional case, generalized here to more dimensions and weakly modelled approaches.
Our aim is to identify which structures act as
spectral-siren anchors and whether their cosmological dependence is adequately
described by a single mass scale or requires their full two-dimensional
geometry.

For each posterior realization in which a concentration or valley \(f\) has valid geometry, we define its centroid as
\begin{equation}
    \mu_{a,f}^{(s)}
    =
    \frac{
        \displaystyle
        \int_{\mathcal R_f^{(s)}}
        m_a\,W_f^{(s)}\,d^2\ln m
    }{
        \displaystyle
        \int_{\mathcal R_f^{(s)}}
        W_f^{(s)}\,d^2\ln m
    },
    \qquad a\in\{1,2\},
    \label{eq:joint_feature_centroid_h0}
\end{equation}
where $W_f=p_{\ln}$ for a concentration and is the reconstructed
missing-density field for a valley. The feature regions $\mathcal R_f^{(s)}$
are defined in Appendix~\ref{app:feature-finder}.

We denote the component-mass position used for the
\(H_0\) analysis by \((x_{1,f},x_{2,f})\). For concentrations and
valleys, we use the centroids, \(x_{a,f}=\mu_{a,f}\); for
\(\mathcal T_{60}\), we use the front point defined in
Sec.~\ref{sec:two_dimensional_mass_features}.

We decompose each position into a common mass scale and a mass ratio,
\begin{equation}
 m_{\parallel,f}=\sqrt{x_{1,f}x_{2,f}},\qquad
 q_f=\frac{x_{2,f}}{x_{1,f}}.
 \label{eq:joint_feature_parallel_and_ratio}
\end{equation}

Under the spectral-siren mapping \(m_{\rm src}=m_{\rm det}/(1+z)\), both
source-frame masses rescale coherently. This shifts
\(\ln m_\parallel=(\ln x_1+\ln x_2)/2\), while
\(\ln q_f=\ln x_2-\ln x_1\) is unchanged.
These coordinates describe orthogonal directions in the log-mass plane:
the common scale traces the direct mass--redshift degeneracy, whereas
dependence of \(q_f\) on \(H_0\) at fixed \(m_\parallel\) probes changes
in the relative placement of the two component masses.

The clearest signature of a spectral-siren anchor is therefore a visible correlation of $H_0$ with the common mass scale $m_\parallel$.
Figure~\ref{fig:H0_joint_feature_anchors} shows the joint posterior of $H_0$
and $m_\parallel$ for the structures with the clearest potential cosmological
relevance, conditioned on realizations in which their two-dimensional
locations are measurable. The Spearman coefficient $\rho_s$ is reported to
quantify the direction and monotonicity of this association.
We do not show the low--mass structures as it shows mild correlation with $H_0$: the common mass
scales of $\mathcal C_{\rm low}$ and $\mathcal V_{\rm low}$ have
Spearman correlations with $H_0$ of $\rho_s=-0.125$ and
$\rho_s=+0.018$, respectively.

To determine instead how information is distributed across the two mass coordinates in Eq.~\ref{eq:joint_feature_parallel_and_ratio}, 
we compare three representations: the log location
\(X_{1,f}=\ln m_{1,f}^{\rm (1D)}\) of the independently reconstructed
primary-mass counterpart, the common scale \(\ln m_{\parallel,f}\), and
the full position \((\ln m_{\parallel,f},\ln q_f)\).
The first two are alternative, non-nested summaries.\footnote{The primary-mass feature
$X_{1,f}$ is identified from the $m_1$ marginal of the baseline joint
reconstruction, independently of the two-dimensional catalogue. This
feature-level comparison is distinct from the model-level reduction in the
$H_0$ interval obtained when replacing the one-dimensional population model
with the joint reconstruction.}
For each feature, all three representations are evaluated on the same
posterior realizations for which the required one- and two-dimensional
measurements are simultaneously available.

\begin{figure}[t]
    \centering
    \includegraphics[width=0.5\textwidth]
    {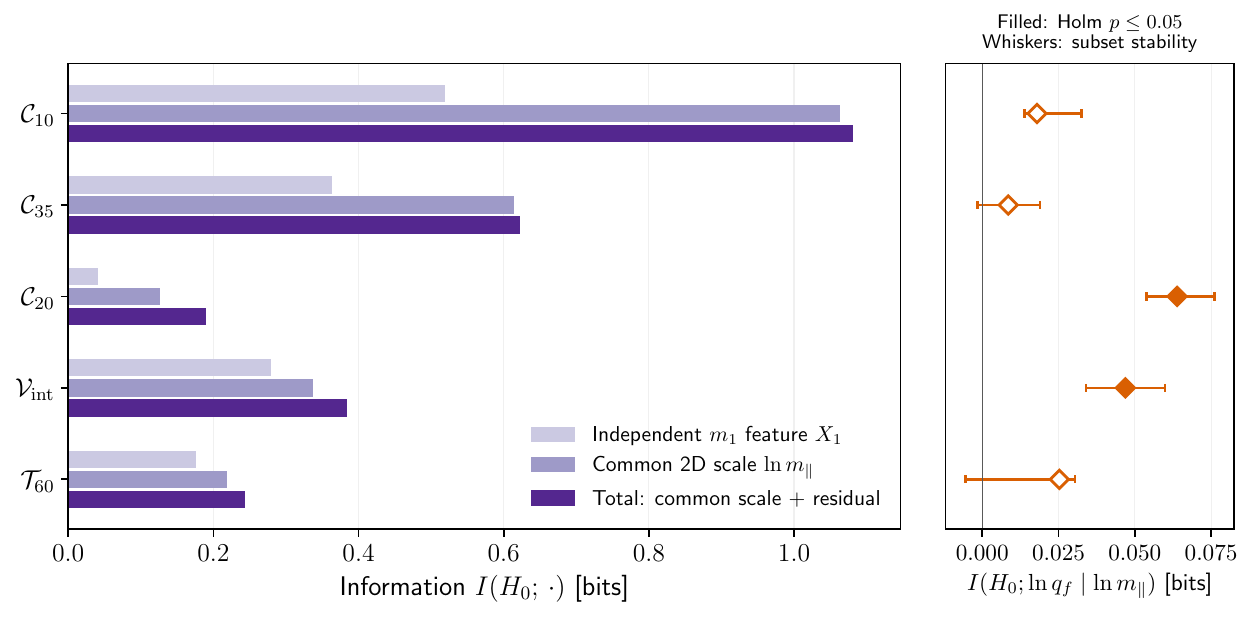}
    \caption{Posterior information about \(H_0\) in alternative feature representations.
Light bars show the independent primary-mass scale \(X_1\), intermediate
bars the joint common scale \(\ln m_\parallel\), and dark bars the total
for \((\ln m_\parallel,\ln q_f)\).
Values are in bits and use the same
common-valid draws within each feature. The right panel shows
\(I(H_0;\ln q_f\mid\ln m_\parallel)\). Whiskers span the \(5\)th--\(95\)th
percentiles of estimates from \(200\) random \(80\%\) subsets.
Filled diamonds denote residuals with \(p_{\rm Holm}<0.05\) across the five features in this test family. Appendix~\ref{app:h0_information_calibration}
describes estimation and conditional-randomization calibration.}
    \label{fig:H0_information_by_representation}
\end{figure}

We quantify their posterior dependence on $H_0$ using mutual information
$I(H_0;X)$, expressed in bits.
This measures how much knowing $X$ reduces the uncertainty about $H_0$, averaged over their joint posterior distribution.
Unlike a rank correlation, mutual information
also captures nonlinear or non-monotonic dependence. We use the
estimator described in
Appendix~\ref{app:h0_information_calibration}.  %
To isolate information specific to the mass-ratio direction, we further
evaluate
  
\begin{equation}
 I\!\left(H_0;\ln q_f\mid\ln m_{\parallel,f}\right).
 \label{eq:h0_conditional_mass_ratio_information}
\end{equation}

This conditional information measures the dependence of the mass ratio
on \(H_0\) after fixing the common scale. We report the full-position estimate as the common-scale
estimate plus this directly estimated residual. %
Note that these quantities should not be interpreted as additive or causal
contributions of individual features to the global $H_0$ constraint; rather,
they quantify the posterior dependence between $H_0$ and a given feature
representation.

The two low-mass structures,
\(\mathcal C_{\rm low}\) and \(\mathcal V_{\rm low}\), show negligible
dependence on \(H_0\) and are not carried forward to this analysis.
For each of the five remaining features,
\(\mathcal C_{10}\), \(\mathcal C_{35}\), \(\mathcal C_{20}\),
\(\mathcal V_{\rm int}\), and \(\mathcal T_{60}\), we ask whether
the mass ratio retains an association with \(H_0\) among posterior
draws with approximately the same common mass scale.

We assess this by keeping each draw's \(H_0\) and \(m_\parallel\)
fixed and replacing its mass ratio with one from a draw having a
nearby \(m_\parallel\). This leaves the observed
\(H_0\)--common-scale relation unchanged and approximately retains
the mass-ratio--common-scale relation, while disrupting the additional
pairing between \(H_0\) and mass ratio. Repeating this procedure,
called local conditional randomization, provides comparison datasets
representing the absence of residual dependence~\cite{H0UpdateRunge2018}. The resulting
\(p\)-value quantifies how often they produce residual information
at least as large as observed.

A small \(p\)-value can occur by chance even when a feature has no
residual dependence on \(H_0\). Testing five features gives multiple
opportunities for such a false positive. Holm's
correction~\cite{H0UpdateHolm1979} adjusts the five individual
\(p\)-values upward to account for these multiple comparisons.
We denote the adjusted values by \(p_{\rm Holm}\) and declare a
residual significant only when \(p_{\rm Holm}<0.05\).
This controls the probability of making even one false claim of
residual dependence across the five features at no more than \(5\%\).
These tests concern residual association with \(H_0\), not the
physical existence of the features.
Details are given in Appendix~\ref{app:h0_information_calibration}.

Figure~\ref{fig:H0_information_by_representation} compares the mutual
information retained by the three feature representations and isolates the
conditional mass-ratio contribution. The corresponding numerical values,
together with $P_{\rm morph}$, the common-valid fraction, and $\rho_s$, are
reported in Table~\ref{tab:H0_feature_information}.

\begin{table*}[t]
    \centering
    \small
    \caption{Posterior dependence between \(H_0\) and the principal cosmologically
relevant structures. \(P_{\rm morph}\) is unconditional morphology recovery
and \(f_{\rm cv}\) is the fraction of draws valid for all three representations.
Here \(\rho_s\) uses this common-valid sample, whereas
Fig.~\ref{fig:H0_joint_feature_anchors} uses all valid positions.
\(I_{X_1}\) and \(I_\parallel\) are raw mutual-information estimates in bits;
\(I_{q\mid\parallel}\) is the directly estimated conditional information.
The reported total is \(I_{\rm 2D}=I_\parallel+I_{q\mid\parallel}\). \(p_{\rm Holm}\) corrects the
conditional-randomization \(p\)-values across the five residual tests.
All continuous quantities are conditional on the required measurements
being available.}
    \label{tab:H0_feature_information}
    \resizebox{\textwidth}{!}{%
    \begin{tabular}{@{}lccccccccl@{}}
        \hline
        Feature & $P_{\rm morph}$ & $f_{\rm cv}$
        & $\rho_s(H_0,m_\parallel)$
        & $I_{X_1}$ & $I_\parallel$ & $I_{\rm 2D}$
        & $I_{q\mid\parallel}$ & $p_{\rm Holm}$ & Cosmological role \\
        \hline
        $\mathcal C_{10}$ & $0.997$ & $0.834$ & $-0.84$
        & $0.520$ & $1.064$ & $1.082$ & $0.018$ & $0.132$
        & dominant common-scale anchor \\
        $\mathcal C_{35}$ & $0.999$ & $0.899$ &  $-0.73$
        & $0.364$ &  $0.614$ &  $0.622$ &  $0.009$ &  $0.143$
        & secondary common-scale anchor \\
        $\mathcal C_{20}$ & $0.954$ & $0.897$ &  $-0.38$
        & $0.041$ & $0.126$ &  $0.190$ & $0.064$ &  $0.0025$
        & feature with residual mass-ratio dependence \\
        $\mathcal V_{\rm int}$ & $0.735$ & $0.604$ & $-0.43$
        &  $0.279$ &  $0.337$ &  $0.384$ & $0.047$ &  $0.0025$
        & feature with residual mass-ratio dependence \\
        $\mathcal T_{60}$ & $0.367$ &  $0.308$ &  $-0.38$
        &  $0.175$ &  $0.219$ &  $0.244$ &  $0.025$ &  $0.132$
        &  conditional common-scale anchor \\
        \hline
    \end{tabular}}
\end{table*}

The resulting hierarchy of spectral-siren structures is as follows.

\begin{itemize}

\item \textbf{$\boldsymbol{\mathcal C_{10}}$ at
$\boldsymbol{(10.3,8.6)\,M_\odot}$ is the dominant spectral-siren anchor,
with its cosmological dependence carried almost entirely by a coherent
two-component mass scale.}

Its common scale shows the strongest monotonic association with $H_0$,
$\rho_s=-0.84$. On the same posterior realizations, the mutual information
increases from  $0.520$ bits for the independently reconstructed primary-mass
feature to $1.064$ bits for the joint common scale. Including the mass-ratio
coordinate adds little further dependence, with no significant residual after
conditioning on $m_\parallel$. Thus, the two marginal mass scales are most
informative when recognized as a single joint structure translating coherently
with cosmology.

\item \textbf{$\boldsymbol{\mathcal C_{35}}$ at
$\boldsymbol{(33.4,29.5)\,M_\odot}$ is the second main spectral-siren anchor,
again dominated by the common joint mass scale.}

Its common scale is strongly anticorrelated with $H_0$,  \(\rho_s=-0.73\) on the common-valid sample, and
retains more dependence than the independently reconstructed primary-mass
feature ( $0.614$ versus $0.364$ bits). As for $\mathcal C_{10}$, adding the
mass-ratio coordinate yields no significant residual information. The joint
reconstruction therefore gives this otherwise projection-dependent population
feature a well-defined role as a coherent two-component cosmological scale.

\item \textbf{$\boldsymbol{\mathcal C_{20}}$ at
$\boldsymbol{(21.4,11.8)\,M_\odot}$ is a weaker anchor overall, but the
clearest concentration for which two-dimensional geometry matters.}

Its total dependence on $H_0$ is smaller than for the two dominant anchors, but
increases markedly from the primary-mass representation to the common scale
and full centroid ($0.041$, $0.126$, and  $0.190$ bits, respectively).
In the baseline reconstruction, after fixing the common mass scale, the mass-ratio coordinate
retains $=0.064~{\rm bits}$ of conditional information, 
with  \(p_{\rm Holm}=0.0025\). The cosmological dependence of this weak unequal-mass concentration therefore contains information
that cannot be reduced to a coherent rescaling of its two component masses.

\item \textbf{$\boldsymbol{\mathcal V_{\rm int}}$ provides a second, less stable example of residual dependence beyond the common mass scale, carried
by an oblique population deficit.}

In the baseline reconstruction, its residual mass-ratio information is $0.047$ bits and survives the
family-wise calibration ( \(p_{\rm Holm}=0.0025\)). This shows that extended
under-densities can also retain cosmological dependence beyond their common
mass scale, although the lower morphology stability and less uniform
$H_0$--$q_f$ relation make $\mathcal V_{\rm int}$ a less clean anchor than
the localized concentrations.

\item  \textbf{\(\mathcal T_{60}\) is a non-negligible conditional common-scale
anchor, despite its low joint morphology recovery.}

Its front scale is moderately anticorrelated with \(H_0\)
(\(\rho_s=-0.38\) on the common-valid sample), with \(0.219\) bits of
common-scale information and \(0.244\) bits in the full position.
The residual mass-ratio information is not significant
(\(p_{\rm Holm}=0.132\)). This conditional scale association is distinct
from the \(36.7\%\) recovery of the joint morphology; the independent
primary-mass feature is recovered much more often, in \(86.4\%\) of draws.

 Recovery also depends on \(H_0\): realizations in which the feature is recovered have median
$H_0=55.9\,{\rm km\,s^{-1}\,Mpc^{-1}}$, compared with
$61.5\,{\rm km\,s^{-1}\,Mpc^{-1}}$ when it is absent, with
$p=5.0\times10^{-4}$ under the circular-shift calibration
(Appendix~\ref{app:h0_recovery_dependence}).  The continuous front-scale association and the recovery dependence are complementary aspects of the high-mass feature's cosmological relevance.

\end{itemize}

The centroid positions of the low-mass NS concentration and
low-mass under-density show little posterior association with
$H_0$, despite their prominent population signatures.
Their absence from the hierarchy is therefore
itself informative: population prominence and cosmological relevance are
distinct properties.

 Taken together, the joint mass plane exhibits complementary forms of
posterior association with \(H_0\): coherent common scales, dominated by
\(\mathcal C_{10}\) and \(\mathcal C_{35}\) but also relevant for
\(\mathcal T_{60}\) when recovered; residual mass-ratio dependence for
\(\mathcal C_{20}\) and \(\mathcal V_{\rm int}\); and cosmology-dependent
recovery, separately demonstrated for \(\mathcal T_{60}\).

\subsection{Comparison with previous studies}
\label{sec:population_literature_comparison}

The principal structures identified here have established or
tentative counterparts in previous studies. We focus here on the
additional interpretation supplied by their joint morphology
and cosmological associations.

\paragraph{Established structures.}

The low-mass concentration $\mathcal C_{\rm low}$ and suppression
$\mathcal V_{\rm low}$ correspond to the established neutron-star population
and NS--BH transition
\cite{KAGRA:2021duu,Farah:2021qom,LIGOScientific:2025pvj,LIGOScientific:2026ctl,
vanSon:2022myr}, while the $\sim10\,M_\odot$ peak is a persistent feature
across LVK catalogues and independent reconstructions,  with several astrophysical interpretations proposed
\cite{Tiwari:2020otp,LIGOScientific:2020kqk,KAGRA:2021duu,Edelman:2022ydv,
Farah:2023vsc,Callister:2023tgi,LIGOScientific:2025pvj,
LIGOScientific:2026ctl,Godfrey:2026pbc}.
The $\sim30$--$40\,M_\odot$ structure is likewise well established
\cite{Edelman:2022ydv,LIGOScientific:2025pvj,Mould:2026sww,
Sadiq:2025vly,LIGOScientific:2026ctl} and has been discussed in connection
with pulsational pair instability
\cite{Farmer:2019jed,Hendriks:2023yrw,Golomb:2023vxm,
Croon:2023kct,Ulrich:2024nez}. The latest LVK analysis also finds preferential
near-equal-mass pairing around $35\,M_\odot$~\cite{LIGOScientific:2026ctl},
consistent with the geometry of $\mathcal C_{35}$ which is found to be a localized near-equal-mass concentration.

\paragraph{Emergent intermediate-mass structure.}
The situation around $15$--$20\,M_\odot$ is less settled. The GWTC--4
population analysis reported a possible third mass-spectrum feature near
$20\,M_\odot$~\cite{LIGOScientific:2025pvj}, not confirmed in the latest GWTC--5
analysis~\cite{LIGOScientific:2026ctl}. Nevertheless, several independent analyses have
reported structure in this intermediate region, including tentative
overdensities around $18$--$20\,M_\odot$
\cite{Edelman:2022ydv,Toubiana:2023egi,Tiwari:2023xff,Tiwari:2026bvx,
Tagliazucchi:2026gxn,Leyde:2026bat}, a $\sim20\,M_\odot$ overdensity
preceded by a dip~\cite{Gennari:2025nho,Mould:2026hfh}, a pronounced suppression around
$12$--$16\,M_\odot$ followed by a recovery at higher masses~\cite{Legred:2026oiz}, and
a low-significance intermediate-mass mode in a joint
mass--spin reconstruction~\cite{Heinzel:2024hva}. Astrophysically motivated population decompositions
also find structure in this mass range~\cite{Galaudage:2026opk}. These results
provide natural counterparts to the
$\mathcal V_{\rm int}$--$\mathcal C_{20}$ sequence found here. 
Recent analyses
of GWTC--5 additionally find a transition in the mass-ratio and spin
properties over approximately $18$--$30\,M_\odot$, including a substantially
flatter mass-ratio distribution than at lower masses
\cite{Flanagan:2026ayy}, which is particularly relevant to our identification
of $\mathcal C_{20}$ as an unequal-mass concentration. %
Importantly, our morphology framework turns the question of tentative structure
from a binary ``feature/no-feature'' assessment into a quantitative one: the relevant question is not simply whether the structure exists, but how strongly it stands out from the surrounding population. Specifically,
$\mathcal C_{20}$ is recurrent but only moderately prominent. Its measured
prominence provides an observable that can be compared directly with
the strength of intermediate-mass structure predicted by astrophysical models or observed across alternative analyses.

\paragraph{High-mass structure.}

Tentative structure around $60$--$70\,M_\odot$ has also been reported in
weakly modelled and dedicated analyses
\cite{LIGOScientific:2025pvj,Pierra:2026ffj}, with evidence for a distinct
population extending to still higher masses
\cite{Plunkett:2026pxt,Li:2025iux,LIGOScientific:2026ctl}.
 These results provide context for the tentative
joint morphology $\mathcal T_{60}$ and its more persistent
primary-mass manifestation.

\paragraph{Cosmology.}
Comparison of our headline result with the LVK GWTC--5 spectral-siren analysis has been discussed in Sec.~\ref{sec:h0_model_comparison}.
The LVK GWTC--5 analysis also reports a $26\%$ narrower $68.3\%$
$H_0$ interval with \textsc{FullPop} than with the BBH-only
\textsc{MLTP} model~\cite{LIGOScientific:2026uyd}.
The LVK analysis associates this improvement with the larger sample and
additional NS-related mass features, although the models also differ in their
treatment of the secondary mass. Our finding that the low-mass NS structures
contribute little to the $H_0$ constraint suggests that the gain need not be
driven predominantly by these additional scales, and that joint-mass modelling
may also play an important role (see also Ref.~\cite{Farah:2023swu} for dedicated parametric studies at fixed cosmology).

A complementary comparison is provided by the recent Ref.~\cite{Leyde:2026bat}, which
finds $H_0=66^{+17}_{-20}$ and
$57^{+20}_{-15}\,{\rm km\,s^{-1}\,Mpc^{-1}}$ with flexible one- and
two-dimensional Gaussian-process models of GWTC--5 BBHs. The shift toward
lower $H_0$ is in the same direction as in our analysis, while their
two-dimensional $68.3\%$ interval, $35\,{\rm km\,s^{-1}\,Mpc^{-1}}$, is only
about $11\%$ wider than our full-CBC result. Although not directly comparable,
the two analyses independently indicate that flexible joint BBH modelling can
retain substantial spectral-siren information compared to the inclusion of NS scales.

Finally, using a strongly parametrized model, Ref.~\cite{Pierra:2026ffj} identifies a
$63.3^{+4.8}_{-4.8}\,M_\odot$ scale in GWTC--4 whose inclusion tightens the
spectral-siren constraint by about $30\%$. 
The GWTC--5 LVK analysis likewise
considers a \textsc{FullPop + 3 Peaks} model containing an additional
high-mass feature~\cite{LIGOScientific:2026uyd}.
 Our corresponding primary-mass feature is recovered in \(86.4\%\) of draws, compared with $36.7\%$ for its joint counterpart. When present, its front position carries non-negligible common-scale
dependence on \(H_0\); recovery itself also depends on \(H_0\).
Thus, the present data support sensitivity of $H_0$ to the high-mass population, while the interpretation of that structure remains model dependent.

\section{Conclusions}
\label{sec:conclusion}

We have introduced a flexible mixture model for the joint source-frame
component-mass distribution of compact binaries, inferred simultaneously with
cosmology, together with a morphology-based framework that identifies
structures directly from the reconstructed density and quantifies their
posterior stability, strength, geometry, and cosmological dependence. Applied
to GWTC--5, the framework shows that the joint mass plane is useful not only
for interpreting population structure, but also contains spectral-siren
information that is not retained when the secondary-mass dependence is
restricted to a simple global mass-ratio prescription.

Our main conclusions are:

\begin{itemize}

\item \textbf{Combining a joint mass plane view with automated morphology
identification provides a reproducible interpretation of familiar and emerging
population structure.}

Rather than identifying structures by visual inspection of the reconstructed density, or the one–dimensional marginals, we argue that morphology should be assigned quantitatively in two dimensions. The clearest example is the feature near  ${\sim}35\,M_\odot$, identified here as a stable, prominent,
near-equal-mass concentration even though its primary-mass projection is
ambiguous between a peak and a slope change. At intermediate masses we recover
a tentative oblique deficit and a recurrent but less prominent unequal-mass
concentration near $(21,12)\,M_\odot$, whose projections can take different
one-dimensional forms. 
 A slope change near \(60\,M_\odot\) is recovered in \(86.4\%\) of
primary-mass marginals, compared with \(36.7\%\) for its joint unequal-mass
slope-change counterpart.
More generally, our framework separates how
consistently a morphology is recovered, through $P_{\rm morph}$, from how
strongly it stands out when present, through its prominence or depth.
A compelling population feature will in practice reflect a combination of
these two properties, rather than either criterion alone.
This
moves the interpretation of tentative structure beyond a binary
``feature/no-feature'' classification: a structure may be
real but only weakly prominent.

\item \textbf{A two-dimensional population description can be necessary to fully capture spectral-siren information.}

 For each feature we decompose its position into a common mass scale
\(m_\parallel=\sqrt{x_{1,f}x_{2,f}}\) and the orthogonal mass-ratio
coordinate \(q_f=x_{2,f}/x_{1,f}\). %
The \(10\,M_\odot\) and \(35\,M_\odot\) concentrations, and
\(\mathcal T_{60}\) when recovered, depend on \(H_0\) mainly through
the common scale. The intermediate concentration and valley instead retain residual mass-ratio dependence after conditioning on
\(m_\parallel\). Thus, coherent rescaling of the
two component masses might not exhaust the cosmological dependence of every
population structure, and information can be present in the joint mass plane
that cannot be represented by a structured primary-mass spectrum combined
with a global $q^\beta$ prescription.

\item \textbf{A weakly modelled reconstruction in the mass plane retains much of the
spectral-siren precision of strongly parameterized population models}

The baseline reconstruction gives
$H_0=59.5^{+16.5}_{-14.6}\,{\rm km\,s^{-1}\,Mpc^{-1}}$ and narrows the
$68.3\%$ interval by about $33\%$ relative to the comparator that flexibly
reconstructs only the primary-mass spectrum. Its interval is also only about
$10\%$ wider than the strongly parameterized LVK result. Unlike approaches in
which localized spectral structure is prescribed through a parametric one-dimensional
mass spectrum, the joint reconstruction allows the number and joint geometry of localized structures to emerge from data. %
The near-LVK precision therefore shows that the additional information carried by
the two-dimensional morphology can compensate for relaxing prescribed
spectral structure, reducing sensitivity to mass-spectrum model choice while
retaining comparable cosmological precision.

\item \textbf{A hierarchy of spectral-siren anchors.}

At the feature level, $\mathcal C_{10}$ and $\mathcal C_{35}$ provide the
strongest coherent mass-scale dependence. 
The less prominent $\mathcal C_{20}$ and $\mathcal V_{\rm int}$ carry weaker total dependence on $H_0$ but the clearest residual mass-ratio dependence.
Note that formal inclusion in an astrophysical feature catalogue is not a prerequisite for cosmological relevance: weak structure can fall below the morphology-selection threshold while leaving the global spectral-siren constraint essentially unchanged.
The prominent low-mass NS structures instead contribute little to the $H_0$ constraint; our results suggest that the additional NS mass
scales are not the dominant source of the precision gained in full-population
spectral-siren analyses.
Population prominence and cosmological relevance are therefore distinct
properties. 
 Finally, the high-mass transition is a non-negligible common-scale
anchor when recovered, %
with currently no significant residual mass-ratio dependence. Its
primary-mass manifestation is much more persistent than its joint
morphology. Separately, lower \(H_0\) values favour recovery of that
joint morphology.

\end{itemize}

The weaker intermediate structures require greater caution. Their formal
catalogue inclusion is more sensitive to population-model freedom and to the
feature-finder threshold, which can reflect either a genuine low-contrast
astrophysical structure or finite-catalogue fluctuations. Additional observations or
catalogue-sized simulations will be important for distinguishing these
possibilities~\cite{Farah:2023vsc,Mould:2026sww}.
At the same time, the measured
prominence provides a quantitative observable that can be compared directly
with population-synthesis predictions, enabling quantitative comparison between observed and predicted morphology rather than relying on visual similarity alone.

Finally, the cosmological results are conditional on assuming that the source-frame mass
distribution is independent of redshift. Allowing the mass spectrum to evolve
introduces an additional degeneracy with the cosmological mapping and is beyond
the scope of this work; extending the joint reconstruction and
feature--cosmology framework in this direction will be important as larger
catalogues make population evolution measurable~\cite{Agapito:2026kak}.

In conclusion, we stress that the explicit morphology classification turns structures in any reconstructed density into reproducible objects whose astrophysical and cosmological properties can be quantified across the posterior. The framework can be applied directly to other reconstructions and population predictions for quantitative comparison.

For spectral-siren cosmology, we allowed the number, locations, widths, and relative weights of population structures to be inferred from the data, taking a significant step forward in addressing systematics related to mass feature representations while exploiting the full two-dimensional mass information. 
Despite this flexibility, the resulting inference retains nearly the precision of strongly parameterized analyses while reducing sensitivity to mass-model systematics.
At the same time, some emerging structures exhibit genuinely two-dimensional cosmological dependence that cannot be reduced to a single coherent mass rescaling.
Overall, fully flexible two-dimensional mass models may therefore represent an important direction for future spectral-siren cosmology.

\section*{Data availability}
The population inference code used for this work is available at~Ref.~\cite{pymcpop}. 
The version used for this paper will be archived on Zenodo. %
The feature-finder package \texttt{posterior-landscape} is available at~\cite{plandscape}.
Data products for events in the GWTC-5.0 catalog are available at~\cite{ligo_scientific_collaboration_and_virgo_2025_17014085, ligo_scientific_collaboration_and_virgo_2023_8177023, ligo_scientific_collaboration_and_virgo_2022_6513631}.
Sensitivity estimates are available at~\cite{ligo_scientific_collaboration_2026_19500052}.

\acknowledgements

We thank Amanda Farah, Gr\'egoire Pierra, Simone Mastrogiovanni, Jack Heinzel, Matthew Mould, Jose Ezquiaga, for valuable discussions. %
This work is supported by the French government under the France 2030 investment plan, as part of the Initiative d'Excellence d'Aix-Marseille Universit\'e -- A*MIDEX AMX-22-CEI-02.
We acknowledge use of computational resources of the MesoCentre of Aix-Marseille University through allocation b435-GWCROSS.
This document has LIGO DCC document number LIGO-P2600539.

{\small This research has made use of data or software obtained from the Gravitational Wave Open Science Center (gwosc.org), a service of the LIGO Scientific Collaboration, the Virgo Collaboration, and KAGRA. This material is based upon work supported by NSF's LIGO Laboratory which is a major facility fully funded by the National Science Foundation, as well as the Science and Technology Facilities Council (STFC) of the United Kingdom, the Max-Planck-Society (MPS), and the State of Niedersachsen/Germany for support of the construction of Advanced LIGO and construction and operation of the GEO600 detector. Additional support for Advanced LIGO was provided by the Australian Research Council. Virgo is funded, through the European Gravitational Observatory (EGO), by the French Centre National de Recherche Scientifique (CNRS), the Italian Istituto Nazionale di Fisica Nucleare (INFN) and the Dutch Nikhef, with contributions by institutions from Belgium, Germany, Greece, Hungary, Ireland, Japan, Monaco, Poland, Portugal, Spain. KAGRA is supported by Ministry of Education, Culture, Sports, Science and Technology (MEXT), Japan Society for the Promotion of Science (JSPS) in Japan; National Research Foundation (NRF) and Ministry of Science and ICT (MSIT) in Korea; Academia Sinica (AS) and National Science and Technology Council (NSTC) in Taiwan.
}

\bibliography{correlations}

\appendix

\section{Redshift evolution}
\label{app:redshift_evolution}

We model the source-frame merger-rate evolution with a
Madau--Dickinson-like function,
\begin{equation}
\psi(z\mid\Lambda_z)
\propto
\frac{(1+z)^{\gamma}}
{
1+
\left[
\dfrac{1+z}{1+z_p}
\right]^{\gamma+\kappa}
},
\qquad
\Lambda_z=\{\gamma,\kappa,z_p\}.
\label{eq:md_redshift_evolution}
\end{equation}
At low redshift, $\gamma$ controls the rise of the rate. The parameter $z_p$
sets the approximate turnover redshift, and $\kappa$ controls the decline
beyond the turnover.

\section{Mathematical details of the mass-mixture prior}
\label{app:prior-maths}

This appendix provides the mathematical details of the mass model and its prior construction. We also present diagnostics
of the prior on the mixture weights and the complete prior induced over the physical mass distribution.

\subsection{Mass tapers.}
\label{app:tapers}

We model the mass boundaries with sigmoid tapers, with $\sigma(x)=(1+e^{-x})^{-1}$ and $\operatorname{logit}(p)=\log[p/(1-p)]$. For the low-mass taper, let $a_{\rm l}=\operatorname{logit}(\epsilon_{\rm low})$, $b_{\rm l}=\operatorname{logit}(1-\epsilon_{\rm on})$, $s_{\rm l}=\Delta m/(b_{\rm l}-a_{\rm l})$, and $c_{\rm l}=m_{\rm low}-s_{\rm l}a_{\rm l}$. For the high-mass taper, let $a_{\rm h}=\operatorname{logit}(p_{\rm high})$, $b_{\rm h}=\operatorname{logit}(\epsilon_{\rm high})$, $s_{\rm h}=\Delta m_{\rm high}/(a_{\rm h}-b_{\rm h})$, and $c_{\rm h}=m_{\rm high}+s_{\rm h}a_{\rm h}$. Then
\begin{equation}
S_{\rm low}(m)=\sigma \left(\frac{m-c_{\rm l}}{s_{\rm l}}\right),\qquad
S_{\rm high}(m)=\sigma \left(\frac{c_{\rm h}-m}{s_{\rm h}}\right),
\end{equation}
giving $S_{\rm low}(m_{\rm low})=\epsilon_{\rm low}$, $S_{\rm low}(m_{\rm low}+\Delta m)=1-\epsilon_{\rm on}$, $S_{\rm high}(m_{\rm high})=p_{\rm high}$, and $S_{\rm high}(m_{\rm high}+\Delta m_{\rm high})=\epsilon_{\rm high}$. We use $\epsilon_{\rm low}=\epsilon_{\rm high}=10^{-6}$, $\epsilon_{\rm on}=10^{-3}$, and $p_{\rm high}=0.99$, applying low-mass tapers separately to $m_1$ and $m_2$ and the high-mass taper to $m_1$.

Taper locations and widths are inferred jointly with the mixture parameters through transformed independent standard-normal variables:
\begin{equation}
\begin{aligned}
m_{1,\rm low}&=a_1+(b_1-a_1)\sigma(u_{\rm raw})^{3/2}, \\
m_{2,\rm low}&=a_2+\sigma(v_{\rm raw})(m_{1,\rm low}-a_2),\\
m_{\rm high}&=a_{\rm high}+(b_{\rm high}-a_{\rm high})\sigma(r_{\rm high}),\\
\Delta m_i&=\exp \left[\log a_{\Delta i}+
\bigl(\log b_{\Delta i}-\log a_{\Delta i}\bigr)\sigma(r_{\Delta i})\right], \, i\in{1,2},\\
\Delta m_{\rm high}&=a_{\Delta{\rm h}}+
(b_{\Delta{\rm h}}-a_{\Delta{\rm h}})\sigma(r_{\Delta{\rm h}}),
\end{aligned}
\end{equation}
where $u_{\rm raw},v_{\rm raw},r_{\rm high},r_{\Delta1},r_{\Delta2},r_{\Delta{\rm h}}\sim\mathcal N(0,1)$. The first construction enforces $m_{2,\rm low}<m_{1,\rm low}$. All numerical bounds $a_1,b_1,a_2,a_{\rm high},b_{\rm high},a_{\Delta i},b_{\Delta i},a_{\Delta{\rm h}},b_{\Delta{\rm h}}$ are given in Table~\ref{tab:reference_parameter_ranges}.

\subsection{Data-adaptive bounds on component locations}
\label{app:data_adaptive_bounds}

For the determination of the component locations, we use a Monte Carlo procedure in which, for each trial $t$, one
sample is drawn for every event from its posterior distribution. The sampled
luminosity distance is converted to redshift through the inverse
distance--redshift relation $z=z(d_L;\Lambda_c)$, where the cosmological
parameters $\Lambda_c$ are drawn uniformly from their prior intervals. The
detector-frame component masses are then converted to the source frame using
this redshift, while the mass ratio $q=m_2/m_1$ is unchanged.

Each trial therefore provides one pair
\begin{equation}
\left\{
\log m_{1,i}^{(t)},\,q_i^{(t)}
\right\},
\qquad
i=1,\ldots,N_{\rm obs},
\end{equation}
for every event in the catalogue. We define the prior bounds on the primary
component centers from lower and upper values of such draws, while the bounds on the characteristic mass ratios are obtained from quantiles of the trial-wise
mass-ratio distributions,

\begin{align}
q_{\mu}^{\rm low}
&=
\min_t
Q_{q_{q,\rm low}}
\left(
\left\{
q_i^{(t)}
\right\}_{i=1}^{N_{\rm obs}}
\right),
\\
q_{\mu}^{\rm up}
&=
\max_t
Q_{1-q_{q,\rm up}}
\left(
\left\{
q_i^{(t)}
\right\}_{i=1}^{N_{\rm obs}}
\right).
\end{align}
Here $Q_p$ denotes the $p$th quantile, and the lower- and upper-tail
probabilities may be chosen independently. This procedure allows the bounds to account for
both event-level posterior uncertainty and uncertainty in the
distance--redshift relation. These intervals define the prior support for the
component parameters $\mu_{1k}$ and $q_{\mu,k}$, subject to the overall
physical bounds of the model.

The component locations are sampled through bounded logistic-normal
transformations,
\begin{equation}
\theta_k
=
\theta^{\rm low}
+
\left(
\theta^{\rm up}-\theta^{\rm low}
\right)
\sigma(u_{\theta,k}),
\qquad
u_{\theta,k}\sim\mathcal{N}(0,1),
\label{eq:app_bounded_location_prior}
\end{equation}
where $\sigma(u)=(1+e^{-u})^{-1}$ and
$\theta\in\{\mu_1,q_{\mu}\}$. This parameterization keeps every component
inside the allowed range while avoiding hard boundaries in the sampling
coordinates. 

The purpose of this construction is only to automate the choice of broad
prior support over the mass range represented by the catalogue; it does not
place prior preference at the locations of individual population features or artificially restrict the allowed mass range.
Thus the procedure plays the role that would
otherwise be served by manually fixing sufficiently broad catalogue-covering
bounds. Sensitivity to the mass-ratio-tail choice is tested explicitly in
Appendix~\ref{app:baseline_variations}.

\subsection{Component widths and the minimum resolved scale}
\label{app:width_hierarchy}

For coordinate $j\in\{1,2\}$ and component $k$, the width is given by Eq.~\ref{eq:component_width_hierarchy}.

We estimate the minimum resolved scale $L_j$ from the resolution of the single-event posteriors in the
same coordinates used by the population mixture. For event $i$, define
\begin{equation}
s_{i,1}
=
\mathrm{Std}\!\left[\log m_1\right]_i,
\qquad
s_{i,2}
=
\mathrm{Std}\!\left[\log m_2\right]_i.
\label{eq:app_event_mass_resolutions}
\end{equation}
The floor is then obtained from a chosen percentile of the event-by-event
resolutions,
\begin{equation}
L_j
=
f_j\,
Q_{p_j}
\left(
\{s_{i,j}\}_{i=1}^{N_{\mathrm{obs}}}
\right),
\qquad
j\in\{1,2\}.
\label{eq:app_pe_resolution_floor}
\end{equation}
Here $Q_{p_j}$ denotes the $p_j$ percentile across events and $f_j$ is a
calibration factor. Using a catalogue percentile rather than the single best
or worst measured event makes the floor robust to outliers. The calibration
factors and percentiles are varied in the robustness analysis.

The shared excess-width scale $\tau_j$ is bounded using a data-adaptive upper
scale $U_j$ derived from the logarithmic span of the reconstructed source-frame
masses in the same coordinate. We sample
\begin{equation}
    0<\tau_j<U_j-L_j
\end{equation}
through a bounded-sigmoid parameterization. The individual widths
$\sigma_{jk}$ are not themselves capped at $U_j$, because the local factors
$\exp(\epsilon_{jk})$ remain free to vary between components.

Since the mixture coordinates are logarithmic,
\begin{equation}
\frac{\Delta m_j}{m_j}
\simeq
\Delta\log m_j
\end{equation}
for small displacements. The floors can therefore be interpreted as
approximate fractional mass resolutions. They prevent the model from
claiming structure on scales that the individual observations cannot resolve.

\subsection{Correlations and aspect regularization}
\label{app:aspect_regularization}

The correlation coefficient allows each Gaussian component to follow an
inclined feature in the mass plane. We use
\begin{equation}
    r_k\sim{\rm Beta}(\eta_\rho,\eta_\rho),
    \qquad
    \rho_k=
    \sqrt{1-f_{\rm minor}^2}\,(2r_k-1),
    \label{eq:correlation_prior}
\end{equation}
which is symmetric about $\rho_k=0$. The parameter $f_{\rm minor}$ bounds the
correlation away from $|\rho_k|=1$ and therefore prevents the conditional
Gaussian width from collapsing to zero.
The
parameter $f_{\mathrm{minor}}$ keeps the conditional width from becoming
arbitrarily small:
\begin{equation}
\sigma_{2\mid 1,k}
\geq
f_{\mathrm{minor}}\,\sigma_{2k}.
\end{equation}

The reference model also uses a soft regularization against a very large
disparity between the marginal widths in the two mass directions. We measure
this disparity through
\begin{equation}
a_k
=
\sqrt{
\left[
\log\!\left(
\frac{\sigma_{1k}}{\sigma_{2k}}
\right)
\right]^2
+
\epsilon_a^2
} \,, \quad \epsilon_a = 10^{-6} \, .
\label{eq:app_component_anisotropy}
\end{equation}
This quantity is close to zero when the two marginal widths are comparable
and grows when one is much larger than the other. We add the smooth
one-sided penalty
\begin{equation}
\Delta\log\pi_{\mathrm{aspect}}
=
-\lambda_a
\sum_{k=1}^{K_{\max}}
w_k
\left[
s_a\,
\mathrm{softplus}\!\left(
\frac{a_k-a_{\max}}{s_a}
\right)
\right]^2.
\label{eq:app_aspect_ratio_penalty}
\end{equation}
The penalty is negligible below the characteristic scale $a_{\max}$ and
turns on gradually above it. It is weighted by $w_k$, so a component carrying
negligible population probability is not strongly constrained. Physically,
this regularization discourages the model from explaining the data with
needle-like components unless such a shape is genuinely required. We vary
its strength and transition width as part of the robustness analysis.

\subsection{Reference configuration and induced prior}
\label{app:reference_configuration}

The preceding subsections define a class of flexible mass-mixture models. A
concrete inference additionally requires numerical choices for the maximum
number of available basis functions, the stick-breaking hyperprior, the
data-adaptive support of the component locations and widths, the minimum
resolved scale, the mass tapers, and the shape regularization. We collect here
the choices defining the reference configuration used throughout the main
analysis.

\begin{table*}[t]
\centering
\small
\renewcommand{\arraystretch}{1.18}
\caption{
Reference configuration of the two-dimensional mass model. The table
collects the numerical and structural choices required to instantiate the
general mixture construction defined in the main text and in the preceding
Appendix subsections. Data-adaptive quantities are determined from the
event-level posterior information before the population inference.}
\label{tab:reference_population_settings}
\begin{tabular}{@{}lll@{}}
\hline
\parbox[t]{0.19\textwidth}{\raggedright Setting\strut} &
\parbox[t]{0.27\textwidth}{\raggedright Reference choice\strut} &
\parbox[t]{0.46\textwidth}{\raggedright Interpretation\strut} \\
\hline

\parbox[t]{0.19\textwidth}{\raggedright Mass coordinates\strut} &
\parbox[t]{0.27\textwidth}{\raggedright $(\log m_1,\log m_2)$\strut} &
\parbox[t]{0.46\textwidth}{\raggedright Logarithmic coordinates make
component widths approximately fractional mass scales.\strut}
\\[0.6ex]

\parbox[t]{0.19\textwidth}{\raggedright Mixture structure\strut} &
\parbox[t]{0.27\textwidth}{\raggedright Shared-weight conditional Gaussian
mixture\strut} &
\parbox[t]{0.46\textwidth}{\raggedright A common weight labels the primary
kernel and its conditional-secondary counterpart, defining a coherent
two-dimensional basis function.\strut}
\\[0.6ex]

\parbox[t]{0.19\textwidth}{\raggedright Maximum components\strut} &
\parbox[t]{0.27\textwidth}{\raggedright $K_{\max}=15$\strut} &
\parbox[t]{0.46\textwidth}{\raggedright Upper bound on the number of available
basis functions; the effective complexity is determined by the
stick-breaking weights and is diagnosed below.\strut}
\\[0.6ex]

\parbox[t]{0.19\textwidth}{\raggedright Component centers\strut} &
\parbox[t]{0.27\textwidth}{\raggedright
$\mu_{2k}=\mu_{1k}+\log q_{\mu,k}$\strut} &
\parbox[t]{0.46\textwidth}{\raggedright The characteristic mass ratio
$q_{\mu,k}<1$ specifies the secondary coordinate of each component center.\strut}
\\[0.6ex]

\parbox[t]{0.19\textwidth}{\raggedright Mixture-weight prior\strut} &
\parbox[t]{0.27\textwidth}{\raggedright Finite stick breaking;
$\log\alpha_{\rm DP}^{-1}\sim
\mathcal N(-1.5,0.5^2)$\strut} &
\parbox[t]{0.46\textwidth}{\raggedright Controls how concentrated the mixture
probability is among the available basis functions.\strut}
\\[0.6ex]

\parbox[t]{0.19\textwidth}{\raggedright Primary-center support\strut} &
\parbox[t]{0.27\textwidth}{\raggedright Data-adaptive source-frame
$m_1$ span\strut} &
\parbox[t]{0.46\textwidth}{\raggedright Allows component centers to explore
the primary-mass range spanned by the catalogue realizations.\strut}
\\[0.6ex]

\parbox[t]{0.19\textwidth}{\raggedright Mass-ratio-center support\strut} &
\parbox[t]{0.27\textwidth}{\raggedright Data-adaptive $q_\mu$ interval;
$5\%$ lower and upper quantiles\strut} &
\parbox[t]{0.46\textwidth}{\raggedright Reduces sensitivity of the allowed
component-center range to the broad tails of the event-level mass-ratio
posteriors.\strut}
\\[0.6ex]

\parbox[t]{0.19\textwidth}{\raggedright Minimum component widths\strut} &
\parbox[t]{0.27\textwidth}{\raggedright
$(p_1,p_2)=(25,25)\%$,
$(f_1,f_2)=(0.25,0.25)$\strut} &
\parbox[t]{0.46\textwidth}{\raggedright The minimum logarithmic widths are
calibrated from the event-level posterior resolutions in the same
coordinates as the population mixture.\strut}
\\[0.6ex]

\parbox[t]{0.19\textwidth}{\raggedright Local width scatter\strut} &
\parbox[t]{0.27\textwidth}{\raggedright $(s_1,s_2)=(0.5,0.5)$\strut} &
\parbox[t]{0.46\textwidth}{\raggedright Controls component-to-component
fluctuations around the global excess-width scales.\strut}
\\[0.6ex]

\parbox[t]{0.19\textwidth}{\raggedright Global width scales\strut} &
\parbox[t]{0.27\textwidth}{\raggedright Bounded-sigmoid
parameterization\strut} &
\parbox[t]{0.46\textwidth}{\raggedright Samples the typical excess widths
smoothly over the data-adaptive range above the resolution floor.\strut}
\\[0.6ex]

\parbox[t]{0.19\textwidth}{\raggedright Mass correlation\strut} &
\parbox[t]{0.27\textwidth}{\raggedright
$r_k\sim\mathrm{Beta}(3,3)$,
$\rho_k=\sqrt{1-0.25^2}\,(2r_k-1)$\strut} &
\parbox[t]{0.46\textwidth}{\raggedright Allows inclined two-dimensional
structures while preventing collapse onto an arbitrarily thin line.\strut}
\\[0.6ex]

\parbox[t]{0.19\textwidth}{\raggedright Low-mass behavior\strut} &
\parbox[t]{0.27\textwidth}{\raggedright Separate inferred sigmoid tapers for
$m_1$ and $m_2$\strut} &
\parbox[t]{0.46\textwidth}{\raggedright Allows independent low-mass onset
locations and transition widths for the two component masses.\strut}
\\[0.6ex]

\parbox[t]{0.19\textwidth}{\raggedright High-mass behavior\strut} &
\parbox[t]{0.27\textwidth}{\raggedright Inferred sigmoid taper in
$m_1$\strut} &
\parbox[t]{0.46\textwidth}{\raggedright Smoothly suppresses the distribution
at high primary mass; the ordered domain restricts the secondary mass.\strut}
\\[0.6ex]

\parbox[t]{0.19\textwidth}{\raggedright Aspect regularization\strut} &
\parbox[t]{0.27\textwidth}{\raggedright
$a_{\max}=1$, $s_a=0.5$, $\lambda_a=0.5$\strut} &
\parbox[t]{0.46\textwidth}{\raggedright Softly suppresses components with a
very large disparity between their two marginal widths.\strut}
\\[0.6ex]

\hline
\end{tabular}
\end{table*}

Table~\ref{tab:reference_population_settings} summarizes the reference
mass-model settings. The mixture is defined in
$(\log m_1,\log m_2)$ and contains at most $K_{\max}=15$ shared-weight
conditional Gaussian components. 

The allowed range of $\mu_{1k}$ follows the source-frame primary-mass span obtained from the event-level posterior realizations described in Appendix~\ref{app:data_adaptive_bounds}, while the allowed interval of $q_{\mu,k}$ is constructed from the corresponding posterior-sample quantiles. The reference choice removes the lower and upper $5\%$ tails when constructing the mass-ratio interval. %
Variations of these choices are considered explicitly in
Appendix~\ref{app:baseline_variations}.

The component widths follow the hierarchical construction described in
Appendix~\ref{app:width_hierarchy}. Their minimum scales are tied to the
event-level mass resolution $L_j$.
with reference choices
$(p_1,p_2)=(25,25)\%$ and
$(f_1,f_2)=(0.25,0.25)$. The local component-to-component scatter is
$(s_1,s_2)=(0.5,0.5)$. Correlations between the two logarithmic mass
coordinates are allowed, while a soft aspect-ratio regularization is used. %

For completeness, the prior ranges of the inferred mass boundaries, taper
widths, cosmological parameters, and merger-rate evolution are collected in
Table~\ref{tab:reference_parameter_ranges}. The reference cosmological model
infers $H_0$ and $\Omega_m$, and fixes the dark-energy equation of state to
$w_0=-1$. %

\begin{figure}[t]
    \centering
    \includegraphics[width=0.45\textwidth]
    {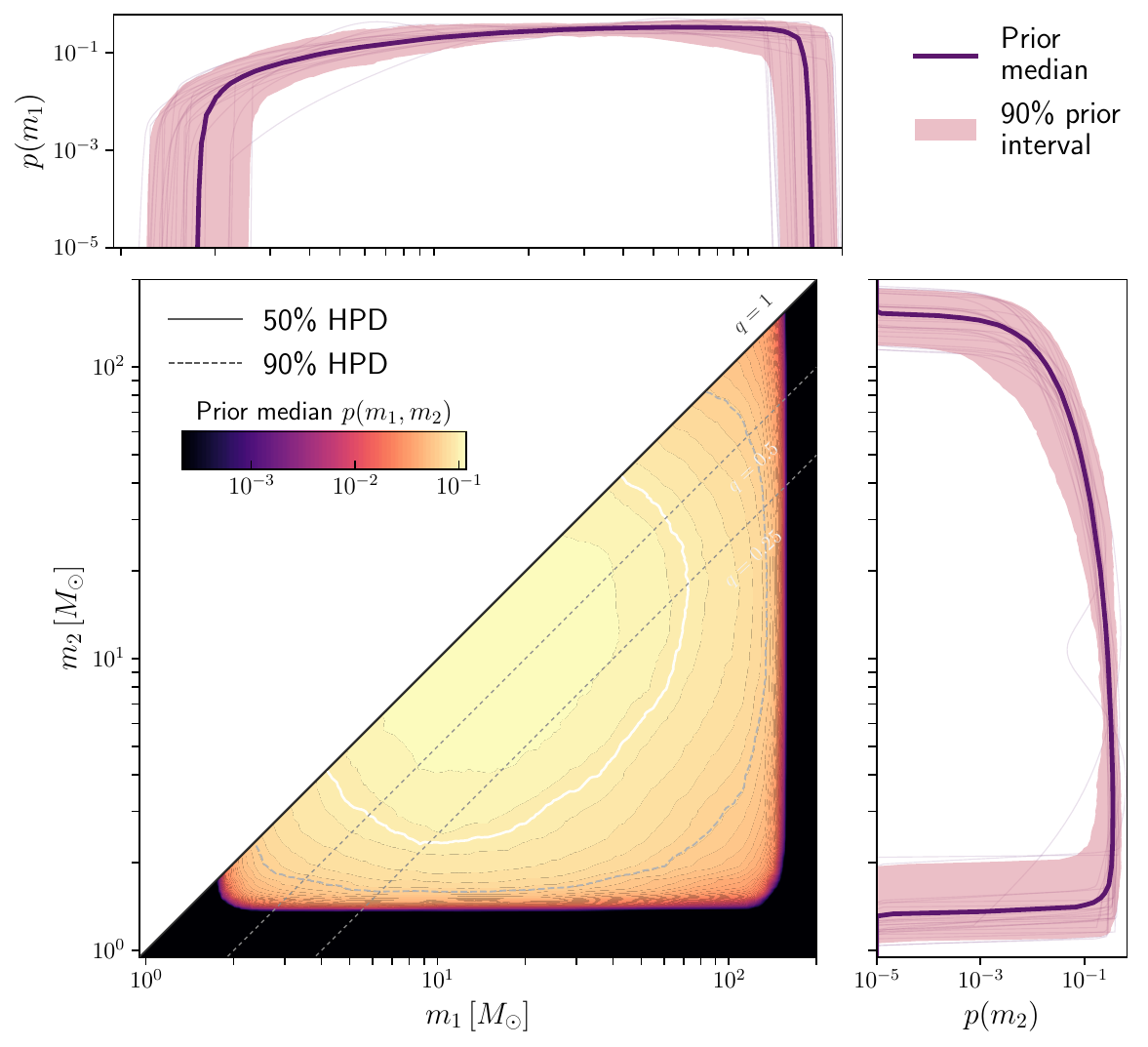}
    \caption{Induced prior over the normalized source-frame component-mass
    distribution for the reference configuration. The central panel shows the
    pointwise prior median of the joint mass density over the physical domain
    $m_2<m_1$. The upper and right panels show the corresponding marginal
    distributions $dP/d\ln m_1$ and $dP/d\ln m_2$. Thick curves denote
    pointwise medians, shaded regions the central $90\%$ pointwise prior
    intervals, and thin curves individual prior realizations. Each realization
    is normalized over the ordered physical mass domain. Dotted guides indicate
    selected constant mass ratios.}
    \label{fig:reference_mass_prior}
\end{figure}

It is useful to inspect the distribution over mass spectra induced by this
complete reference prior before conditioning on the GWTC--5 observations.
Figure~\ref{fig:reference_mass_prior} shows the pointwise prior distribution
of the normalized joint mass density and of its two marginals. Although the
mixture provides up to $K_{\max}=15$ Gaussian basis functions, the induced
prior is broad and largely smooth. The stick-breaking hierarchy typically
assigns appreciable weight to only a subset of the available components, while
the hierarchical width prior favors relatively broad kernels. Individual
prior realizations can nevertheless contain localized one- and
two-dimensional structure at different locations, so the construction does
not impose a fixed set of population peaks or characteristic masses.

\begin{figure}[t]
    \centering
    \includegraphics[width=0.45\textwidth]
    {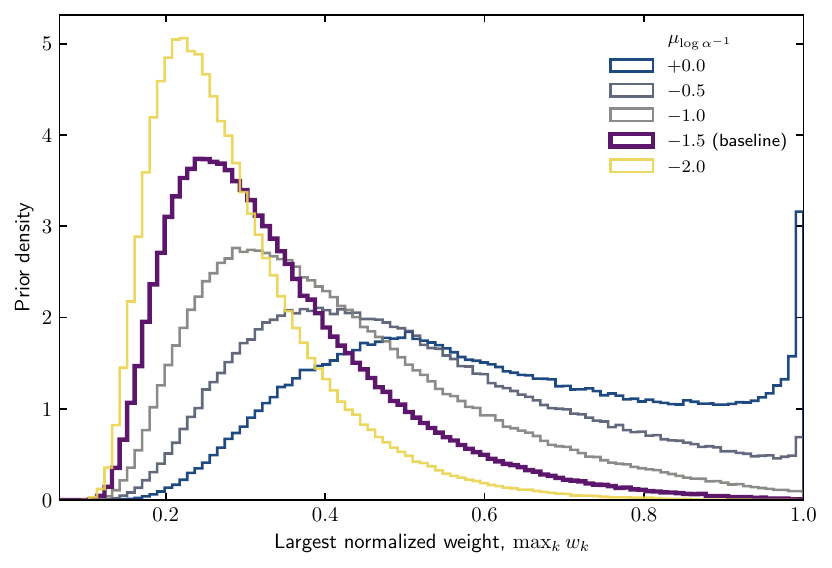}
    \caption{Prior distribution of the largest normalized stick-breaking
    weight, $\max_k w_k$, for different values of
    $\mu_{\log\alpha^{-1}}$. In every case
    $\sigma_{\log\alpha^{-1}}=0.5$ and $K_{\max}=15$. The reference choice
    $\mu_{\log\alpha^{-1}}=-1.5$ suppresses prior accumulation near a
    single component carrying the entire mixture weight while retaining an
    unequal and sparse allocation among the available components. This
    diagnostic isolates the finite stick-breaking hierarchy before the mass
    tapers and aspect regularization are applied.}
    \label{fig:prior_maximum_weight}
\end{figure}

The concentration hyperprior in the finite stick-breaking construction
controls how readily the prior allocates most of the mixture probability to a
small number of basis functions. To illustrate this dependence separately
from the mass tapers and shape regularization,
Fig.~\ref{fig:prior_maximum_weight} shows the prior distribution of the
largest normalized stick-breaking weight, $\max_k w_k$, for several choices of the parameter
$\mu_{\log\alpha^{-1}}$ in Eq.~\ref{eq:dp_concentration_prior}, keeping
$\sigma_{\log\alpha^{-1}}=0.5$ and $K_{\max}=15$ fixed. Increasing
$\mu_{\log\alpha^{-1}}$ increases the typical value of
$\alpha_{\rm DP}^{-1}$ and favors a more concentrated allocation of the
mixture probability. For choices closer to zero, the prior develops
substantial support near $\max_k w_k=1$, corresponding to an effectively
single-component realization. The reference value
$\mu_{\log\alpha^{-1}}=-1.5$ suppresses this accumulation while retaining a
sparse and unequal allocation of mixture weights.

These prior diagnostics establish the flexibility available to the reference
model before conditioning on the observed catalogue.

\begin{table*}[t]
\centering
\small
\renewcommand{\arraystretch}{1.15}
\caption{Prior ranges used in the reference analysis.}
\label{tab:reference_parameter_ranges}
\begin{tabular}{@{}lll@{}}
\hline
\parbox[t]{0.25\textwidth}{\raggedright Quantity\strut} &
\parbox[t]{0.22\textwidth}{\raggedright Allowed range\strut} &
\parbox[t]{0.45\textwidth}{\raggedright Role\strut} \\
\hline

\parbox[t]{0.25\textwidth}{\raggedright Primary low-mass location
$m_{1,\rm low}$\strut} &
\parbox[t]{0.22\textwidth}{\raggedright $[1.01,3]\,M_\odot$\strut} &
\parbox[t]{0.45\textwidth}{\raggedright Location of the smooth primary
low-mass transition.\strut}
\\[0.6ex]

\parbox[t]{0.25\textwidth}{\raggedright Secondary low-mass location
$m_{2,\rm low}$\strut} &
\parbox[t]{0.22\textwidth}{\raggedright $[1,3]\,M_\odot$\strut} &
\parbox[t]{0.45\textwidth}{\raggedright Location of the smooth secondary
low-mass transition.\strut}
\\[0.6ex]

\parbox[t]{0.25\textwidth}{\raggedright High-mass location
$m_{\rm high}$\strut} &
\parbox[t]{0.22\textwidth}{\raggedright $[100,200]\,M_\odot$\strut} &
\parbox[t]{0.45\textwidth}{\raggedright Location of the smooth high-mass
transition.\strut}
\\[0.6ex]

\parbox[t]{0.25\textwidth}{\raggedright Primary low-mass width
$\Delta m_1$\strut} &
\parbox[t]{0.22\textwidth}{\raggedright $[0.005,1]\,M_\odot$\strut} &
\parbox[t]{0.45\textwidth}{\raggedright Width of the primary low-mass
taper.\strut}
\\[0.6ex]

\parbox[t]{0.25\textwidth}{\raggedright Secondary low-mass width
$\Delta m_2$\strut} &
\parbox[t]{0.22\textwidth}{\raggedright $[0.005,1]\,M_\odot$\strut} &
\parbox[t]{0.45\textwidth}{\raggedright Width of the secondary low-mass
taper.\strut}
\\[0.6ex]

\parbox[t]{0.25\textwidth}{\raggedright High-mass width
$\Delta m_{\rm high}$\strut} &
\parbox[t]{0.22\textwidth}{\raggedright $[5,20]\,M_\odot$\strut} &
\parbox[t]{0.45\textwidth}{\raggedright Width of the primary high-mass
taper.\strut}
\\[0.6ex]

\parbox[t]{0.25\textwidth}{\raggedright Hubble constant $H_0$\strut} &
\parbox[t]{0.22\textwidth}{\raggedright
$[10,130]\,{\rm km\,s^{-1}\,Mpc^{-1}}$\strut} &
\parbox[t]{0.45\textwidth}{\raggedright Inferred jointly with the population
parameters.\strut}
\\[0.6ex]

\parbox[t]{0.25\textwidth}{\raggedright Matter fraction $\Omega_m$\strut} &
\parbox[t]{0.22\textwidth}{\raggedright $[0.05,0.55]$\strut} &
\parbox[t]{0.45\textwidth}{\raggedright Inferred jointly with the population
parameters.\strut}
\\[0.6ex]

\parbox[t]{0.25\textwidth}{\raggedright Madau--Dickinson parameters\strut} &
\parbox[t]{0.22\textwidth}{\raggedright
$\gamma\in[0,6]$, $\kappa\in[0,6]$, $z_p\in[0,4]$\strut} &
\parbox[t]{0.45\textwidth}{\raggedright Control the low-redshift rise,
high-redshift decline, and turnover of the merger-rate density.\strut}
\\[0.6ex]

\hline
\end{tabular}
\end{table*}

\section{Alternative one-dimensional mass parameterization}
\label{app:alternative_parameterizations}

Here, we describe the one-dimensional
mixture model used for comparison in the main text. In this alternative, the flexible nonparametric component is
applied only to the primary mass, while the secondary mass enters through a
single smooth mass-ratio factor. %

We model the primary-mass distribution with a Gaussian mixture in
$\log m_1$,
\begin{equation}
g_1(m_1)
=
\frac{1}{m_1}
\sum_{k=1}^{K_{\max}}
w_k\,
\mathcal{N}\!\left(
\log m_1\mid\mu_k,\sigma_k^2
\right).
\label{eq:alt_primary_mixture}
\end{equation}
The mass ratio is described by a single
power-law envelope,
\begin{equation}
q \equiv \frac{m_2}{m_1},
\qquad
E(q | \beta) = q^\beta .
\label{eq:alt_q_envelope}
\end{equation}

The resulting joint mass model is
\begin{equation}
\begin{split}
p_m^{(1{\rm D}+q)}(m_1,m_2)
\propto{}&
g_1(m_1)\,
E(q | \beta)\,
S_{\rm low}(m_1)\,
S_{\rm high}(m_1)\, \\
&\times S_{\rm low}(m_2)\,
\mathbf{1}(m_2<m_1).
\end{split}
\label{eq:alt_primary_q_mass_model}
\end{equation}
Here $S_{\rm low}$ and $S_{\rm high}$ are the same smooth low- and
high-mass taper functions used in the reference model. %

\section{Hierarchical inference details}
\label{app:stats}

This appendix gives the mathematical formulation of the hierarchical analysis
summarized in Sec.~\ref{sec:inference}. The reference analysis samples the
single-event parameters together with the population and cosmological
hyperparameters. We first write the corresponding posterior, then
describe the bounded Gaussian-mixture approximations used for the
single-event posteriors and the injection-based selection correction. 

\paragraph{Posterior distribution.}

Let $\theta_{D,i}$ denote the detector-frame parameters of event $i$, and let
$d_i$ denote its strain data. The single-event analysis provides the posterior
density
\begin{equation}
p_{\rm PE}(\theta_{D,i}\mid d_i)
\propto
p(d_i\mid\theta_{D,i})\,
\pi_{\rm PE}(\theta_{D,i}),
\label{eq:pe_posterior_app}
\end{equation}
where $\pi_{\rm PE}$ is the prior adopted in parameter estimation. For a
choice of cosmological parameters $\Lambda_c$, detector-frame parameters are
mapped deterministically to source-frame parameters,
\begin{equation}
\theta_i
=
T(\theta_{D,i};\Lambda_c).
\label{eq:detector_source_map_app}
\end{equation}
We denote the Jacobian of the inverse mapping by
\begin{equation}
J(\theta_i;\Lambda_c)
=
\left|
\frac{\partial\theta_{D,i}}{\partial\theta_i}
\right|.
\label{eq:detector_source_jacobian_app}
\end{equation}

After analytically integrating the overall merger-rate amplitude
with a scale-invariant prior, the joint posterior density in
the detector-frame event coordinates and the hyperparameters is
\begin{equation}
\begin{aligned}
& p\!\left(\{\theta_{i}\},\Lambda\mid\{d_i\}\right)
\propto \pi(\Lambda)\,\xi(\Lambda)^{-N_{\rm obs}}
\\
&\quad\times
\prod_{i=1}^{N_{\rm obs}}
\frac{p_{\rm PE}(\theta_{D,i}\mid d_i)}
     {\pi_{\rm PE}(\theta_{D,i})}
\frac{
  p_{\rm pop}\!\left(T(\theta_{D,i};\Lambda_c)\mid\Lambda\right)
}{
  J\!\left(T(\theta_{D,i};\Lambda_c);\Lambda_c\right)
}.
\end{aligned}
\label{eq:hier_like_basic_full}
\end{equation}
The factor $J^{-1}$ converts the source-frame population density
to a density in detector-frame coordinates.
Parameters entering the waveform but not modeled at the population level
retain their single-event prior distribution and therefore cancel between
$p_{\rm pop}$ and $\pi_{\rm PE}$. The factor $\xi(\Lambda)$ accounts for
selection effects and is defined below.

\paragraph{Bounded single-event posterior approximations.}

We evaluate Eq.~\eqref{eq:hier_like_basic_full} using continuous
approximations to the single-event posteriors, following the strategy
introduced in Ref.~\cite{Mancarella:2025uat} and validated in the blinded
mock-data study of Ref.~\cite{Tagliazucchi:2026dpr}. In the present analysis, the posterior approximation is constructed separately
for each event ( labelled by $i$) in the detector-frame coordinates
\begin{equation}
\boldsymbol{y}_i
=
\left(
\mathcal{M}^{\rm det}_i,\,
q_i,\,
 d_{L,i}
\right),
\label{eq:bounded_gmm_raw_coordinates}
\end{equation}

For each event $i$ and coordinate $r$, we determine numerically lower and upper bounds
$a_{ir}$ and $b_{ir}$ that enclose the corresponding posterior samples. Each
coordinate is then mapped to the real line through
\begin{equation}
z_{ir}
=
\log\!\left(
\frac{y_{ir}-a_{ir}}
{b_{ir}-y_{ir}}
\right).
\label{eq:event_bounded_flogit}
\end{equation}
A full-covariance Gaussian mixture is fitted in the transformed coordinates
$\boldsymbol{z}_i$,
\begin{equation}
\widehat{p}_i(\boldsymbol{z}\mid d_i)
=
\sum_{\ell=1}^{K_i}
\omega_{i\ell}\,
\mathcal{N}\!\left(
\boldsymbol{z}
\mid
\boldsymbol{\mu}_{i\ell},
\boldsymbol{\Sigma}_{i\ell}
\right).
\label{eq:event_gmm_surrogate}
\end{equation}
The number of components $K_i$ is chosen separately for each event by
minimizing the Bayesian Information Criterion over the fitted candidates.
The fitted weights, means, covariances, and event-specific coordinate bounds
are then stored and used in the hierarchical analysis.
During the inference, the inverse
transformation is applied to recover
the masses and
luminosity distance before evaluating the hierarchical posterior. This
construction restricts the posterior approximation to the event-specific
coordinate ranges covered by the posterior samples.

\paragraph{Selection effects.}

The detection efficiency is the population-averaged probability of passing
the search selection,
\begin{equation}
\xi(\Lambda)
=
\int
\mathrm{d}\theta_D\,
P_{\rm det}(\theta_D)\,
\frac{
p_{\rm pop}\!\left(
T(\theta_D;\Lambda_c)
\mid\Lambda
\right)
}{
J\!\left(
T(\theta_D;\Lambda_c);
\Lambda_c
\right)
}.
\label{eq:selection_integral_app}
\end{equation}
We estimate this integral using simulated injections
\cite{Tiwari:2017ndi,Farr:2019rap}. If $N_{\rm draw}$ injections are generated
from $\pi_{\rm inj}(\theta_D)$ and $\mathcal{D}_{\rm inj}$ denotes those
passing the same selection threshold used for the event catalog, then
\begin{equation}
\widehat{\xi}(\Lambda)
=
\frac{1}{N_{\rm draw}}
\sum_{k\in\mathcal{D}_{\rm inj}}
\frac{
p_{\rm pop}\!\left(
T(\theta_{D,k};\Lambda_c)
\mid\Lambda
\right)
}{
\pi_{\rm inj}(\theta_{D,k})\,
J\!\left(
T(\theta_{D,k};\Lambda_c);
\Lambda_c
\right)
}.
\label{eq:alpha_inj_app}
\end{equation}

\paragraph{Monte Carlo uncertainty.}

For the reference latent-variable evaluation, the event contributions are
computed from continuous posterior surrogates and do not contain
posterior-sample Monte Carlo sums. The stochastic likelihood error considered
here therefore comes from the finite injection set. Following
Refs.~\cite{Farr:2019rap,Essick:2022ojx,Talbot:2023pex,Heinzel:2025ogf}, its
contribution to the variance of the estimated log likelihood is
\begin{equation}
\sigma^2_{\ln\widehat{\mathcal{L}}}(\Lambda)
\simeq
N_{\rm obs}^2
\frac{
\sigma^2_{\widehat{\xi}}(\Lambda)
}{
\widehat{\xi}(\Lambda)^2
}.
\label{eq:loglik_variance_full_app}
\end{equation}
Following
Refs.~\cite{Talbot:2023pex,Heinzel:2025ogf,LIGOScientific:2025pvj}, we ensure that
\begin{equation}
\sigma^2_{\ln\widehat{\mathcal{L}}}(\Lambda)<1
\label{eq:loglik_variance_requirement_app}
\end{equation}

\paragraph{Posterior-sample marginalization as a robustness check.}

As a separate check, we also evaluate the commonly used form in which the
single-event parameters are integrated out. The hyperposterior is then
\begin{equation}
\begin{split}
p(\Lambda\mid\{d_i\})
\propto
\pi(\Lambda)\,
\xi(\Lambda)^{-N_{\rm obs}}
\prod_{i=1}^{N_{\rm obs}}
\int
\mathrm{d}\theta_{D,i}\,
p(d_i\mid\theta_{D,i})
\\
\times
\frac{
p_{\rm pop}\!\left(
T(\theta_{D,i};\Lambda_c)
\mid\Lambda
\right)
}{
J\!\left(
T(\theta_{D,i};\Lambda_c);
\Lambda_c
\right)
}.
\end{split}
\label{eq:hier_like_basic_app}
\end{equation}
Given posterior samples
$\{\theta_{D,ij}\}_{j=1}^{n_i}$ generated with the reference prior, the
event integral is estimated as
\begin{equation}
\widehat{\mathcal{L}}_i(\Lambda)
=
\frac{1}{n_i}
\sum_{j=1}^{n_i}
\frac{
p_{\rm pop}\!\left( T(\theta_{D,ij};\Lambda_c) \mid\Lambda \right) }{ \pi_{\rm PE}(\theta_{D,ij})\, J\!\left( T(\theta_{D,ij};\Lambda_c); \Lambda_c \right) }.
\label{eq:reweight_event_app_integral}
\end{equation}
In this alternative evaluation, finite posterior and injection samples both
contribute to the stochastic error. The corresponding approximation is
\begin{equation}
\sigma^2_{\ln\widehat{\mathcal{L}}}(\Lambda)
\simeq
\sum_{i=1}^{N_{\rm obs}}
\frac{
\sigma^2_{\widehat{\mathcal{L}}_i}(\Lambda)
}{
\widehat{\mathcal{L}}_i(\Lambda)^2
}
+
N_{\rm obs}^2
\frac{
\sigma^2_{\widehat{\xi}}(\Lambda)
}{
\widehat{\xi}(\Lambda)^2
}.
\label{eq:loglik_variance_marginal_app}
\end{equation}
We apply the same requirement
$\sigma^2_{\ln\widehat{\mathcal{L}}}<1$ when performing this robustness
check . %

\section{Feature-finder algorithm}
\label{app:feature-finder}

Here, we present the algorithm used for feature finding and examine sensitivity to analysis choices.
The framework is described in detail in the companion paper~\cite{plandscape-paper} and released in the open--source package \texttt{posterior-landscape}~\cite{plandscape}.

The %
framework identifies population structure
from posterior realizations of a reconstructed probability density. Its input
is the density itself, so the same procedure can be applied to different
population parameterizations. 
The input to the feature finder is the normalized density
$p_{\ln}(m_1,m_2)$, evaluated on a grid covering the physical domain
$m_2<m_1$. The analysis uses $(\ell_1,\ell_2)\equiv(\ln m_1,\ln m_2)$ as spatial coordinates.
We first construct
a reference landscape by taking the posterior median at each grid position. The analysis then proceeds in two
stages: select reference features on this median landscape, and search for
compatible counterparts in individual posterior realizations. This produces
both a catalogue of feature identities and draw-dependent measurements of
their locations, shapes, and strengths.

\paragraph{Smoothing the density.}

Small fluctuations in a gridded density can produce narrow maxima, minima, and
branches that do not represent resolved population structure. We control this
effect by repeating the search after smoothing the density over several
length scales.
Smoothing replaces the density at a given location with a Gaussian averaging kernel. In simple terms, the density in each cell is replaced
by a weighted average of nearby cells, with the weight decreasing as a
Gaussian function of their log-mass distance. Near the physical boundary, the average is corrected for
the fraction of the kernel lying inside the allowed domain. The resulting
field is normalized again. %

The
reference analysis uses Gaussian widths
$\in\{0,\,0.025,\,0.05\}$, 
where $s$ is the standard deviation of the kernel along each logarithmic mass
coordinate and $s=0$ leaves the normalized density unchanged. We denote the
resulting field by $F_s$. 
A candidate is retained
only if a compatible feature of the same class is recovered at least at two
of the three scales.

\paragraph{Locating peaks, pits, ridges, and valleys.}
The density can be viewed as the height of a landscape over the mass plane.
Its maxima and minima locate concentrations and depressions, while its passes
connect neighboring regions. A \emph{saddle} is such a pass: the density rises
along one direction through it and falls along the crossing direction. The
algorithm finds these connections by adding neighboring grid cells in order
of their density. The four associated feature classes are defined as follows.

\begin{itemize}

\item \emph{Peaks} are local density maxima surrounded by lower density. To find peaks, the
algorithm sorts the grid cells from highest to lowest density and adds them
one at a time. A cell that has no previously added neighbour starts a new
high-density region and marks a local maximum. As cells of progressively lower
density are added, this region expands. Eventually it touches a region growing
from another maximum. The density at which the two regions first meet defines
the saddle, or pass, separating the two peaks. When two regions meet, it records the connecting
pass. The higher maximum continues to represent the joined region; the lower
maximum is paired with the pass at which its region joined the higher one.
This pairing defines the reference level against which the lower peak's
contrast is measured.

\item \emph{Pits} are local density minima. Pits are found by reversing the preceding search: the complementary procedure starts
at low density and progressively raises the level. Regions grow from the
minima until they connect at a pass. The deeper minimum continues to represent
the joined depression, and the shallower minimum is paired with that pass.

\item \emph{Ridges} are crest curves traced uphill from the passes to
neighboring maxima. At a saddle where
two high-density regions meet, the algorithm follows the locally increasing
density from the saddle toward each associated peak. Each path from the saddle
to a peak defines one ridge branch. A ridge therefore has two defining
landmarks: the saddle at which it starts and the peak at which it ends.
Individual ridge branches are retained as geometric diagnostics, but are
excluded from the scientific feature catalogue discussed here because their
correspondence across posterior realizations is less stable and they do not
enter the cosmological analysis.

\item \emph{Valleys} are trough curves traced downhill from the passes to
neighboring minima. They describe the trajectory of an extended density
deficit. Starting from the recorded connections, the algorithm follows the
local downhill paths and refines the curves using the density gradient and
curvature across the trough. Their endpoints preserve the pass--minimum
association, while their paths follow the extended depression.

\end{itemize}

\paragraph{Persistence and automatic threshold selection.}
The connection levels above quantify how strongly each extremum is separated
from its surroundings. For a non-global peak or pit, its \emph{persistence} is
\begin{equation}
    \Pi_{\rm peak}=F_{\rm peak}-F_{\rm pass},
    \qquad
    \Pi_{\rm pit}=F_{\rm pass}-F_{\rm pit},
    \label{eq:app_finder_persistence}
\end{equation}
with all densities evaluated on the same field $F_s$. A small persistence
corresponds to a shallow local variation; a large persistence indicates that
a substantial change in density level is required to connect that extremum
to a higher peak or deeper pit. 

An automatic threshold for retaining features is selected separately for each of the three
reference-density smoothing scales through the following procedure.

\begin{enumerate}

\item The algorithm calculates the persistence of every finite peak and pit,
excluding the global maximum and global minimum.

\item Peak and pit persistence values are placed in one common list and sorted
from smallest to largest. Values below the numerical resolution of the density
field are discarded.

\item For every adjacent pair, the algorithm calculates the logarithmic gap
\begin{equation}
G_i
=
\ln\!\left(\frac{\Pi_{i+1}}{\Pi_i}\right).
\end{equation}
A large value of $G_i$ identifies a separation between a group of
low-persistence fluctuations and a group of more persistent structures.

\item The final gap, between the two largest finite persistence values, is
excluded. Selecting this gap would allow the strongest finite feature alone to
set the catalogue resolution.

\item Among the remaining gaps, the algorithm selects the largest one. This gap must satisfy
\begin{equation}
G_i\geq g_{\min},
\end{equation}
The reference choice is $g_{\min}=1$, corresponding to a ratio of at least
$e$ between the bounding persistence values.

\item When this condition is satisfied, the threshold is placed at the
geometric mean of the two persistence values surrounding the selected gap,
\begin{equation}
T_s
=
\sqrt{\Pi_i\Pi_{i+1}}.
\end{equation}
All finite peaks and pits with
$\Pi\geq T_s$ are retained.

\end{enumerate}

When no eligible separation meets the requirement, the algorithm uses
\begin{equation}
    T_s=\max\!\left[
      \Pi_{\rm floor},\,
      \frac14 Q_{0.10}\!\left(\{\Pi_{(i)}\}_{i=1}^{N}\right)
    \right],
    \label{eq:app_finder_fallback_threshold}
\end{equation}
where $Q_{0.10}$ is the tenth percentile. With fewer than three eligible
values, it uses $T_s=\Pi_{\rm floor}$. An explicitly supplied numerical
threshold can replace this automatic choice. Peaks and pits with persistence
at least $T_s$ are retained at that scale, and the associated connection
hierarchy determines the eligible curve branches. The global extrema remain
eligible at this stage independently of the cut.

The parameter $g_{\min}$ controls acceptance of a separation in the
persistence spectrum. Increasing it can switch the algorithm from the gap
rule to the fallback, with a resulting cutoff that may be lower. %
These thresholds specify catalogue selection.

\paragraph{Reference features across smoothing scales.}
The median landscape is analyzed separately at each smoothing scale, including a separate automatic
threshold calculation. Candidates at the first scale form the starting
catalogue. A feature enters the reference catalogue if compatible
counterparts occur at at least two of the three scales. %
Matching compares feature types and nearby locations. %

\paragraph{Slope changes.}
A separate search identifies extended changes of logarithmic slope, called
\emph{shoulders} in the implementation. We define
\begin{equation}
    t=\frac{\ell_1+\ell_2}{2},
    \,
    r=\frac{\ell_1-\ell_2}{2},
    \,
    \alpha_t=\left.\frac{\partial\ln F_s}{\partial t}\right|_r,
    \,
    \kappa_t=\left.\frac{\partial\alpha_t}{\partial t}\right|_r .
    \label{eq:app_finder_directional_slope}
\end{equation}
Increasing $t$ at fixed $r$ rescales both component masses by the same factor
while keeping their ratio fixed. Along each such profile, the finder searches
for a local flattening followed by renewed steepening, identified by a local
maximum of $\alpha_t$ bracketed by the corresponding changes in $\kappa_t$.
Candidate transitions must exceed the local derivative-fluctuation floor. The criterion $\alpha_t<2$ at
the flattening ensures that the corresponding density per $dm_1\,dm_2$ remains
decreasing along the common-mass direction.

For a bounded transition, the logarithmic slopes $\alpha_{\rm pre}$ and
$\alpha_{\rm post}$ are measured over the intervals immediately preceding and
following the flattening, as defined by the neighboring extrema and
zero-crossing of $\kappa_t$. Its strength is then
\begin{equation}
    \Delta\alpha=\alpha_{\rm pre}-\alpha_{\rm post}.
\end{equation}

Compatible transitions on neighboring mass-ratio profiles are linked into an
extended slope-change curve. When both the onset and end of the transition
can be identified across enough neighboring profiles, they define a finite
region whose centroid and projections can be measured. 
 For a transition, let \(t_T^{(s)}\) and
\(r_T^{(s)}\) denote the separate medians of the coordinates \(t\)
and \(r\), defined above, along the recovered transition curve in
posterior draw \(s\). The front point
\((x_{1,T}^{(s)},x_{2,T}^{(s)})\) is the component-mass pair with
these \((t,r)\) coordinates. Its common mass scale and mass ratio
therefore satisfy
\(\ln m_{\parallel,T}^{(s)}=t_T^{(s)}\) and
\(\ln q_T^{(s)}=-2r_T^{(s)}\).
These draw-level points supply the landscape contours, location
table, and \(H_0\) diagnostics. %

\paragraph{Finite regions, prominence, and valley depth.}
For a concentration, the relative prominence is
\begin{equation}
    A_{\rm P}
    =
    1-\frac{F_{\rm base}}{F_{\rm peak}},
    \label{eq:app_finder_peak_prominence}
\end{equation}
where $F_{\rm peak}$ is the density at the maximum and $F_{\rm base}$ is the
density at its paired saddle. This is the quantity reported as feature
strength in the main text. To define a finite region for the concentration,
we use the connected region containing the peak above
\begin{equation}
    F_{1/2}=\frac{F_{\rm peak}+F_{\rm base}}{2}.
\end{equation}
Density-weighted centroids and marginal projections are measured within this
region.

For a valley, the finder samples profiles perpendicular to the trough
trajectory and locates the first resolved density maximum on either side.
Denoting these side densities by $F_{\rm L}(u)$ and $F_{\rm R}(u)$, and the
density on the valley trajectory by $F_{\rm V}(u)$, the local transverse
log-density contrast is
\begin{equation}
    c(u)=
    \ln\!\left[
       \frac{\sqrt{F_{\rm L}(u)F_{\rm R}(u)}}{F_{\rm V}(u)}
    \right].
    \label{eq:app_finder_valley_local_contrast}
\end{equation}
The feature-level contrast is the median over valid transverse sections,
\begin{equation}
    \Delta\ln p_\perp
    =
    \operatorname{median}_{u\,{\rm valid}} c(u),
    \qquad
    A_{\rm V}=1-\exp(-\Delta\ln p_\perp).
    \label{eq:app_finder_valley_depth}
\end{equation}
Locally,
\begin{equation}
    \exp[-c(u)]
    =
    \frac{F_{\rm V}(u)}
         {\sqrt{F_{\rm L}(u)F_{\rm R}(u)}} ,
\end{equation}
so that the main-text expression
$A_{\rm V}=1-p_{\rm V}/\sqrt{p_{\rm L}p_{\rm R}}$
is the corresponding relative-depth form of the same transverse density
contrast.  The valley centroid uses the reconstructed deficit over the full
valid local measurement region in each draw. %

\paragraph{Posterior recovery and conditional measurements.}
Once the reference identities are fixed, each posterior density draw is
analyzed and its candidates are matched to those identities. For peaks,
pits, ridges, and valleys, this step uses the first analysis scale, $s=0$
here, and the fixed numerical cutoff $T_0$ obtained from the first-scale
median landscape. %
Matches are assigned between candidates of the same type. %
The tracker keeps the fixed reference identities throughout the posterior
analysis and records unmatched realizations explicitly.

Let $Z_f^{(s)}=1$ denote an accepted match for feature $f$ in posterior draw
$s$. The
morphology-recovery probability is
\begin{equation}
    P_{{\rm morph},f}
    =
    \frac{1}{N_s}\sum_{s=1}^{N_s} Z_f^{(s)}.
\end{equation}
For equally weighted draws this is their recovered fraction. Package
summaries retain tracked identities with $P_{\rm morph}\geq0.05$ in the
reference configuration. The final scientific catalogue additionally applies
the morphology choices described above. %

Recovery of a location and availability of a finite strength measurement are
tracked independently. 
Posterior medians and
intervals are calculated from the valid draws for the relevant quantity,
retaining their posterior weights.

\paragraph{Independent marginals and comparisons between configurations.}
The one-dimensional catalogues are constructed independently from the
$m_1$ and $m_2$ marginals. Their density amplitudes are Gaussian-smoothed
in logarithmic mass at widths $0.050$, $0.075$, and $0.100$. Peaks and dips
are found as extrema of the smoothed marginal, and slope changes through its
logarithmic slope and curvature. Candidates must recur at two or more scales.
The same multiscale search is performed on individual marginal posterior
draws. %

After the one- and two-dimensional catalogues have been constructed, the
framework compares compatible morphology classes using the overlap of their
projected logarithmic-mass intervals. %
The fraction of projected feature weight contained in an independently
identified marginal interval quantifies its capture. Capture fractions and
comparisons of mass scales are evaluated on draws where both measurements
are valid. 

For comparisons between population-model or finder configurations, the
catalogues are matched by morphology type and proximity in logarithmic mass,
using peak locations or the pass and endpoint positions of curves. 

An unmatched entry denotes the absence of a compatible
catalogue counterpart under that configuration's selection rules.

\section{Simulations}
\label{app:mass_model_simulations}

This appendix describes in detail an injection-recovery test that validates the framework.
We simulate a full CBC population whose source-frame component-mass distribution is constructed to be compatible with the population constraints inferred by the LVK Collaboration from the GWTC-5.0 catalog~\cite{LIGOScientific:2026ctl}. 
This contains a primary-mass spectrum with smooth low- and high-mass tapers, a neutron-star--black-hole mass gap, a three-regime
power-law with two multiplicative Gaussian excesses associated with the main features reconstructed by the LVK population analysis, and a conditional secondary-mass distribution with a primary-mass-dependent pairing exponent.

We use a modified version of the \texttt{GWMockCat} \texttt{python} package~\cite{Farah:2023vsc} to generate mock data and injections sets. 
We consider an 
LVK O5-like observing scenario, with $1000$ detected events, all with $\text{SNR}>8$. %

This is an end-to-end
test: the mock observations are processed through the same population
reconstruction and feature-finding stages used for the real catalogue, the
joint mass distribution is reconstructed, and the feature finder is then applied to the reconstructed
posterior.  The comparison therefore tests both recovery of the overall
density and recovery of its morphology.

\subsection{Model}

Here, we define the source-frame component-mass model used to generate the simulated population. 

The source-frame component masses are ordered as \(m_1\ge m_2\) and modeled hierarchically as
\begin{equation}
\begin{split}
& p(m_1,m_2)= p_{\rm S}(m_1),p(m_2\mid m_1),
\\
& m_{\rm NS,min}<m_2\le m_1<m_{\rm BH,max},
\end{split}
\label{eq:joint-factorization}
\end{equation}
with zero density outside this domain. The model combines compact boundary tapers, a neutron-star--black-hole depletion, three smoothly connected power-law regimes, two multiplicative Gaussian enhancements in \(p_{\rm S}\), and a primary-mass-dependent pairing law; optional localized structure may additionally be included in the \((m_1,m_2)\) plane.

Boundary and transition functions are defined via the cubic smooth step

$$
h(u)=
\begin{cases}
0,&u\le0,\\
u^2(3-2u),&0<u<1,\\
1,&u\ge1,
\end{cases}
$$

and, for \(j\in\{1,2\}\),
\begin{align}
&S_{\rm low}^{(j)}(m)=h \left(\frac{m-m_{\rm NS,min}}{\delta_{\rm low}^{(j)}}\right),
\,
\\
&S_{\rm high}^{(j)}(m)=h \left(\frac{m_{\rm BH,max}-m}{\delta_{\rm high}^{(j)}}\right).
\label{eq:compact-tapers}
\end{align}
Thus \(S_{\rm low}^{(j)}=0\) for \(m\le m_{\rm NS,min}\) and \(S_{\rm high}^{(j)}=0\) for \(m\ge m_{\rm BH,max}\); \(j=1,2\) denote the primary and conditional-secondary distributions, respectively. Smooth transitions elsewhere use
\(s(m;m_0,\delta)=[1+\exp(-(m-m_0)/\delta)]^{-1}\).

\paragraph{Primary mass spectrum. }

Define
\(H(m;m_0,\eta)=[1+(m_0/m)^\eta]^{-1}\),
\(L(m;m_0,\eta)=[1+(m/m_0)^\eta]^{-1}\), and

$$
N_{\rm gap}(m)=
1-A_{\rm gap}H(m;m_{\rm NS,max},\eta_{\rm NS})
L(m;m_{\rm BH,min},\eta_{\rm BH}).
$$

The three-regime backbone is
\begin{equation}
\begin{aligned}
B_1(m)&=m^{\alpha_1},\\
B_2(m)&=m_{\rm BH,min}^{\alpha_1-\alpha_2}m^{\alpha_2},\\
B_3(m)&=m_{\rm BH,min}^{\alpha_1-\alpha_2}
m_{\rm br}^{\alpha_2-\alpha_3}m^{\alpha_3},\\
B_{12}(m)&=[1-s(m;m_{\rm BH,min},\delta_{\rm BH})]B_1(m) \\
&+s(m;m_{\rm BH,min},\delta_{\rm BH})B_2(m),\\
B(m)&=[1-s(m;m_{\rm br},\delta_{\rm br})]B_{12}(m) \\
&+s(m;m_{\rm br},\delta_{\rm br})B_3(m).
\end{aligned}
\label{eq:backbone}
\end{equation}
The prefactors make the corresponding piecewise power law continuous at \(m_{\rm BH,min}\) and \(m_{\rm br}\).

The Gaussian density normalized on \([a,b]\) is denoted
\(\mathcal N_{[a,b]}(m\mid\mu,\sigma) \). 
With \(a=m_{\rm NS,min}\) and \(b=m_{\rm BH,max}\),

$$
P_{\rm peak}(m)=
1+c_1\mathcal N_{[a,b]}(m\mid\mu_1,\sigma_1)
+c_2\mathcal N_{[a,b]}(m\mid\mu_2,\sigma_2).
$$

The full primary spectrum is
\begin{equation}
p_{\rm S}(m) \propto
S_{\rm low}^{(1)}(m)S_{\rm high}^{(1)}(m)
N_{\rm gap}(m)P_{\rm peak}(m)B(m).
\label{eq:primary-spectrum}
\end{equation}

\paragraph{Secondary mass spectrum.}

The pairing exponent varies with primary mass as
$$
\beta(m_1)=
\beta_{\rm NS}
+(\beta_{\rm BH}-\beta_{\rm NS})
s(m_1;m_\beta,\delta_\beta),
$$
so positive \(\beta\) favors comparable masses.

For fixed \(m_1\), the normalized secondary distribution is
\begin{equation}
p(m_2\mid m_1) \propto
S_{\rm low}^{(2)}(m_2)S_{\rm high}^{(2)}(m_2)
\left(\frac{m_2}{m_1}\right)^{\beta(m_1)}.
\label{eq:secondary-spectrum}
\end{equation}
Together with Eq.~\eqref{eq:joint-factorization}, this defines a normalized joint density per \(dm_1\,dm_2\). Unlike a symmetric construction proportional to \(p_{\rm S}(m_1)p_{\rm S}(m_2)\), features in the primary spectrum are therefore not automatically duplicated along the secondary-mass axis.

\paragraph{Fiducial values. }

The primary-spectrum parameters, calibrated against the LVK GWTC-5.0 population analysis, and the baseline pairing parameters are summarized in Table~\ref{tab:fiducial-mass-parameters}.

\begin{table}
\centering
\caption{Fiducial parameters of the simulated mass model.}
\label{tab:fiducial-mass-parameters}
\begin{tabular}{lc}
\hline
Parameter(s) & Fiducial value(s) \\
\hline
$(c_1,c_2)$ & $(125.5580,\,61.6328)$ \\
$(\mu_1,\sigma_1)$ & $(36.3130,\,10.1629)\,M_\odot$ \\
$(\mu_2,\sigma_2)$ & $(8.7445,\,1.3990)\,M_\odot$ \\
$(m_{\rm NS,min},m_{\rm NS,max})$
& $(1.0745,\,2.0340)\,M_\odot$ \\
$(m_{\rm BH,min},m_{\rm BH,max})$ & $(8.7801,\,357.1927)\,M_\odot$ \\ 
$(\alpha_1,\alpha_2,\alpha_3)$ & $(-2.6458,\,-2.1554,\,-3.5193)$ \\
$(m_{\rm br},\delta_{\rm br},\delta_{\rm BH})$
& $(27.3414,\,7.1946,\,1.0547)\,M_\odot$ \\
$(A_{\rm gap},\eta_{\rm NS},\eta_{\rm BH})$
& $(0.8617,\,8.2701,\,19.4166)$ \\
$(\delta_{\rm low}^{(1)},\delta_{\rm high}^{(1)})$
& $(0.6343,\,276.6099)\,M_\odot$ \\
$(\beta_{\rm NS},\beta_{\rm BH},m_\beta,\delta_\beta)$
& $(1.15,\,2.50,\,5.0\,M_\odot,\,2.0\,M_\odot)$ \\
$(\delta_{\rm low}^{(2)},\delta_{\rm high}^{(2)})$
& $(0.17,\,54.0)\,M_\odot$ \\
\hline
\end{tabular}
\end{table}

\subsection{Recovery}

\begin{figure}[t]
\centering
\includegraphics[width=0.45\textwidth]
{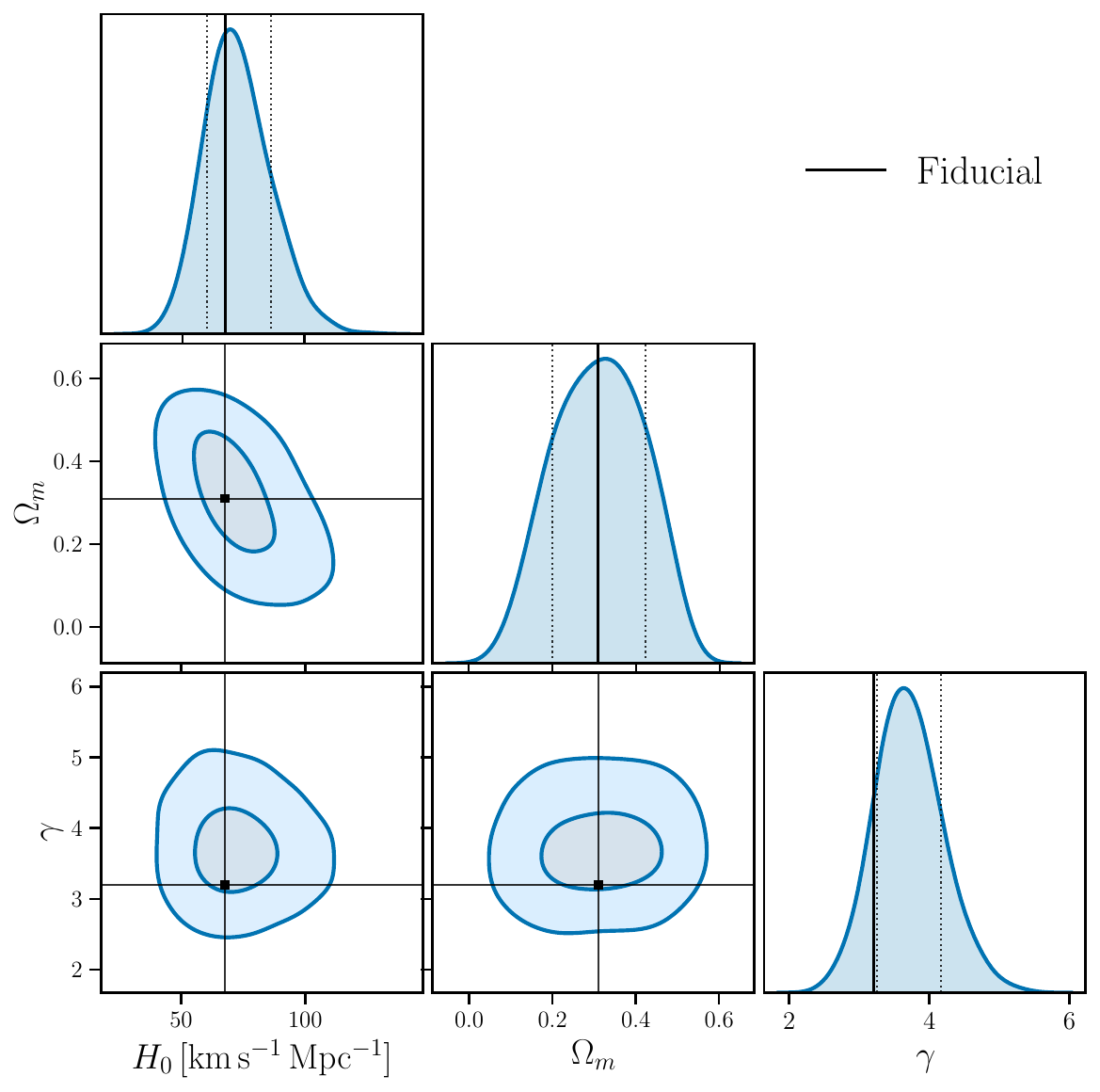}
\caption{Joint and marginal posteriors for $H_0$, $\Omega_{\rm m}$, and $\gamma$. The black vertical lines represents the fiducial values used in the simulation while the dashed lines in the 1D marginals show the 68\% credible intervals.
}
\label{fig:H0_sim}
\end{figure}

We now discuss results of the posterior and feature finder reconstructions.
The injected cosmological parameters $H_0=67.7\,\rm km\,s^{-1}\,Mpc^{-1}$ and $\Omega_{\rm m}=0.31$ are also consistent within our measurement uncertainties. The posterior in the $H_0, \Omega_{\rm m}, \gamma$ plane is shown in Fig.~\ref{fig:H0_sim}.

\begin{figure}[t]
  \centering
  \includegraphics[width=0.48\textwidth]{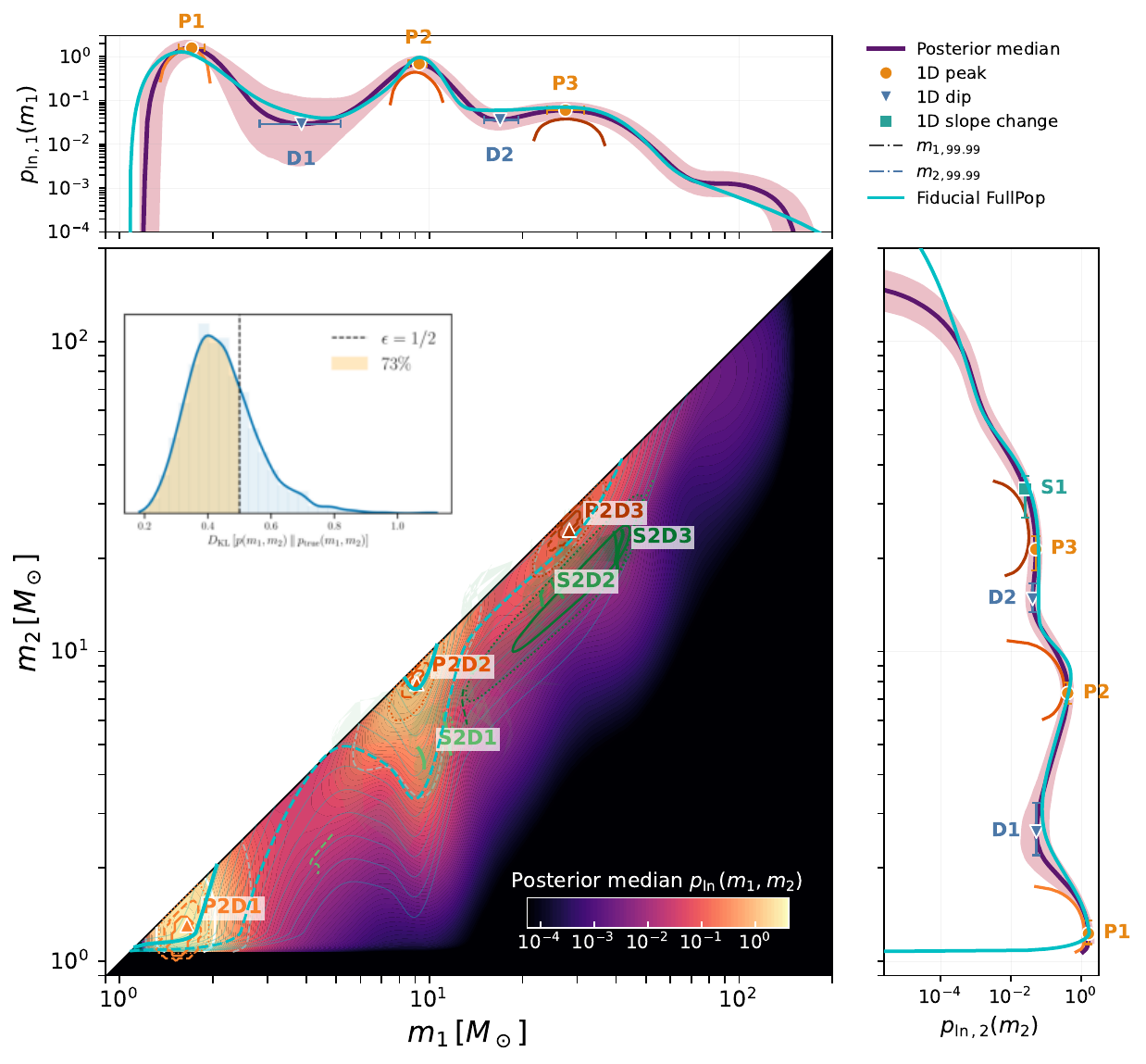}
  \caption{
    End-to-end recovery of the injected mass distribution.  The main
    panel shows the posterior-median joint density per logarithmic mass area,
    $p_{\ln}(m_1,m_2)$.
    The top and right panels show the corresponding marginals.  Purple curves
    and rose bands denote the posterior medians and
    $90\%$ credible regions, while cyan denotes the
    analytic truth.  Fine cyan contours in the joint plane are equally spaced in log density.  Orange marks
    recovered peaks and green marks recovered two-dimensional shoulders.
    The inset shows the distribution of KL divergence between posterior realizations and the injected density.
  }
  \label{fig:fullpop-injection-recovery}
\end{figure}

Figure~\ref{fig:fullpop-injection-recovery} compares the reconstruction and
the feature-finder output with the analytic truth.  The reconstruction
captures the global geometry of the injected distribution in both the
two-dimensional plane and its marginal projections.  The narrow injected
modes are more concentrated near the $m_1=m_2$ boundary than their
non-parametric reconstructions, as expected from finite data and finite
resolution, but their characteristic mass scales are recovered.

We compare
the location of the injected and recovered features using the mass-scale coordinate relevant to spectral-siren
inference,
\begin{equation}
  m_{\parallel}=\sqrt{m_1m_2}.
  \label{eq:mparallel-definition}
\end{equation}
All three injected
values lie within the corresponding 95\% posterior intervals; the first and
third also lie within the 68.3\% intervals.  Each peak has essentially unit
posterior occurrence probability and persists at all three analysis scales.

The finder also identifies three two-dimensional shoulders, S2D1--S2D3.
These were not inserted into the mass model as separate components; they
emerge from the combined shape of the full density.  To assess whether they
reflect structure in the injected distribution rather than artifacts of the
feature-finding procedure, we evaluate an independent directional derivative
on the tabulated fiducial analytic density.  Define
\begin{equation}
  u=\ln\sqrt{m_1m_2},
  \qquad
  q=\frac{m_2}{m_1},
  \label{eq:shoulder-coordinates}
\end{equation}
so that increasing $u$ scales both masses at fixed $q$.  The derivative along
this direction is
\begin{equation}
  D_{\parallel}
  =\frac{\partial}{\partial\ln m_1}
   +\frac{\partial}{\partial\ln m_2}
  =\left.\frac{\partial}{\partial u}\right|_q .
  \label{eq:shoulder-direction}
\end{equation}

Writing $\ell=\ln p_{\ln}$, we define the directional logarithmic slope as
\begin{equation}
  \alpha_{\parallel}(u,q)
  \equiv D_{\parallel}\ell
  =\left.\frac{\partial\ln p_{\ln}}{\partial u}\right|_q ,
\end{equation}
and its steepening as
\begin{equation}
  \kappa_{\parallel}(u,q)
  \equiv -D_{\parallel}\alpha_{\parallel}
  =-D_{\parallel}^{2}\ell .
  \label{eq:shoulder-curvature}
\end{equation}
A positive $\kappa_{\parallel}$ indicates that the density steepens toward
larger masses along a constant-$q$ trajectory.  We evaluate
Eqs.~\eqref{eq:shoulder-direction}--\eqref{eq:shoulder-curvature} directly on the
analytic fiducial density. %

We find that S2D1 coincides with a strong steepening of the fiducial model associated
with the narrow low-mass enhancement and the black-hole turn-on.  S2D3 traces
the principal high-mass steepening produced by the high-mass enhancement
and the smooth change in the high-mass power-law slope.  These are genuine
directional steepenings of the injected total density, although they are
neither separate model components nor obvious by eye.

S2D2 is less straightforward.  It lies near the onset of the same broad
high-mass turnover, but the analytic density has no separate local maximum of
$\kappa_{\parallel}$ between that onset and the stronger S2D3 front.  The
posterior-median density develops an earlier curvature maximum, causing the
catalogue to divide the extended transition into two overlapping branches.
We retain both labels for transparency but interpret them jointly as one
high-mass shoulder complex: S2D3 marks its principal steepening,
whereas S2D2 represents a reconstruction-enhanced onset of
the same transition.  Such subdivisions can be
threshold-dependent; for the real-data analysis we therefore test
stability of the results against the feature-finder settings.

Overall, the simulation recovers the global two-dimensional density and all
three injected modes.  It also places shoulder candidates near the two
strongest directional steepenings of the analytic truth, although the
low-mass shoulder occurs in only 31.7\% of posterior draws.

Finally, we assess the global agreement between the injected and recovered
distributions directly in the two-dimensional mass plane by computing the KL divergence between the fiducial spectrum and posterior realizations. 
The inset of Fig.~\ref{fig:fullpop-injection-recovery} shows the resulting
distribution of KL divergences.  Following Ref.~\cite{Agapito:2026kak}, we
adopt $D_{\mathrm{KL}}\leq\epsilon=1/2$ as a reference tolerance.  Pinsker's inequality then guarantees
$d_{\mathrm{TV}}\leq\sqrt{\epsilon/2}=1/2$, where $d_{\mathrm{TV}}$ is the
largest difference in probability assigned by the two distributions to any
region of the mass plane.  
Approximately $73\%$ of the posterior mass lies within the adopted
$D_{\mathrm{KL}}\leq1/2$ neighbourhood of the injected distribution.
This broad global consistency check complements the recovery
of the injected mass scales and the direct comparison of the
reconstructed and injected densities in
Fig.~\ref{fig:fullpop-injection-recovery}.  We interpret this result as evidence for good
global recovery.

\section{Mass-model learning and robustness diagnostics}
\label{app:mass_model_diagnostics}

In this Appendix we examine what properties of that flexible representation are actually selected by the GWTC--5 data, and how sensitive results are to the model configuration.

In particular, we test whether the reconstruction makes
use of an unnecessarily large number of Gaussian basis functions or drives
their widths toward the minimum scale permitted by the model. 
We next examine the effect of varying some of the baseline configuration choices.

\subsection{Effective mixture complexity and resolution}
\label{app:mixture_complexity}

The raw stick-breaking weights $w_k$ describe the amplitudes of the Gaussian
basis functions before the mass tapers and the ordered physical domain are
applied. Since these restrictions affect different components by different
amounts, $w_k$ cannot in general be interpreted as the fraction of the physical
population carried by component $k$. Individual components also have no unique
astrophysical interpretation: the same structure in the complete density may
be represented by one broad component or by several overlapping narrower
components in different posterior realizations. We therefore use the
component-level quantities below only as diagnostics of the effective
complexity and resolution of the mixture, while all population features are
identified from the complete reconstructed density.

For component $k$, let $\mathcal I_k$ denote its integral after applying the
mass tapers and restricting the density to the ordered region $m_2<m_1$. We
define the corresponding physical component fraction as
\begin{equation}
    \widetilde w_k
    =
    \frac{w_k\mathcal I_k}
    {\displaystyle\sum_{\ell=1}^{K_{\max}}w_\ell\mathcal I_\ell},
    \qquad
    \sum_{k=1}^{K_{\max}}\widetilde w_k=1.
    \label{eq:app_physical_component_fraction}
\end{equation}

We characterize the allocation of physical probability with the largest
component fraction,
\begin{equation}
    \widetilde w_{\max}
    =
    \max_k\widetilde w_k,
\end{equation}
and with the effective number of contributing components,
\begin{equation}
    K_{\rm eff}
    =
    \left(
        \sum_{k=1}^{K_{\max}}\widetilde w_k^2
    \right)^{-1}.
    \label{eq:app_effective_component_number}
\end{equation}
The latter is the inverse participation ratio. It approaches unity when a
single component contains almost all of the physical probability and equals
$N$ when that probability is shared equally among $N$ components. It therefore
provides a continuous measure of effective mixture complexity. %

The data also determine the characteristic width scale used by the mixture.
For each mass coordinate we define the physical-fraction-weighted width
\begin{equation}
    \overline{\sigma}_j
    =
    \sum_{k=1}^{K_{\max}}
    \widetilde w_k\,\sigma_{jk},
    \qquad
    j\in\{1,2\},
    \label{eq:app_weighted_component_width}
\end{equation}
where $\sigma_{jk}$ is the logarithmic width of component $k$ in coordinate
$j$. These weighted averages describe the typical kernel width carrying
appreciable population probability. %

\begin{figure*}[t]
    \centering
    \includegraphics[width=\textwidth]
    {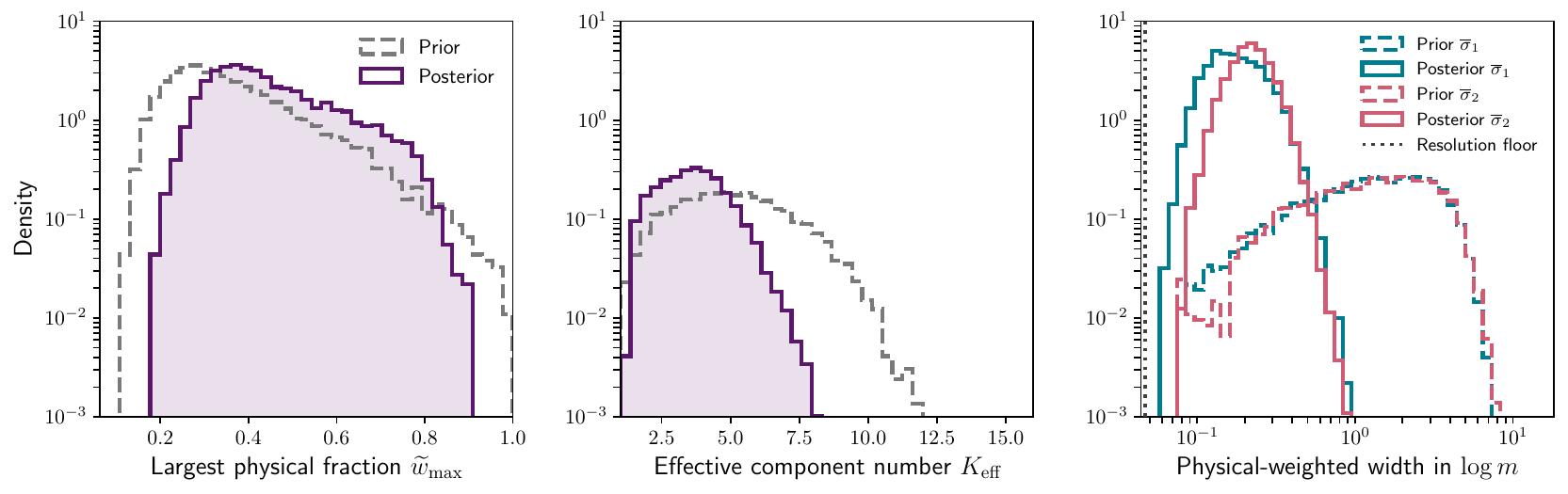}
    \caption{Prior and posterior diagnostics of the effective complexity and
    characteristic width of the two-dimensional mass mixture. From left to
    right, the panels show the largest physical component fraction
    $\widetilde w_{\max}$, the effective number of components
    $K_{\rm eff}$, and the physical-fraction-weighted logarithmic widths
    $\overline{\sigma}_1$ and $\overline{\sigma}_2$. Physical component
    fractions include the effects of the mass tapers and the restriction to
    the ordered domain $m_2<m_1$. The prior distributions are induced by the
    complete reference configuration, while the posterior distributions are
    conditioned on the GWTC--5 observations. Vertical markers in the width
    panel indicate the minimum event-resolution-based scales $L_1$ and $L_2$.
    The posterior favours smaller physical-fraction-weighted widths
than the prior, with both weighted-width distributions remaining
well above the component floors.}
    \label{fig:app_prior_posterior_complexity_widths}
\end{figure*}

Figure~\ref{fig:app_prior_posterior_complexity_widths} compares the induced
prior and GWTC--5 posterior distributions of these quantities. The data produce
a moderate but clear change in effective mixture complexity. The median largest
physical component fraction increases from $0.34$ under the prior to $0.42$ in
the posterior, while the median effective number of components decreases from
$K_{\rm eff}=5.0$ to $3.6$. The reconstruction therefore uses only a subset of
the $K_{\max}=15$ basis functions available in the reference model rather than
populating the full component dictionary.

The substantially stronger update concerns the characteristic width scale.
The prior medians of $\overline{\sigma}_1$ and
$\overline{\sigma}_2$ are both close to $2.4$, whereas the posterior medians
decrease to $0.21$ and $0.24$, respectively. Since the mixture coordinates are
logarithmic, a width $\sigma$ corresponds approximately to a fractional mass
scale $e^\sigma-1$. The posterior values therefore correspond to characteristic
fractional scales of approximately $23\%$ and $27\%$. For orientation, these
amount to roughly $2$--$3\,M_\odot$ at $10\,M_\odot$ and
$7$--$8\,M_\odot$ at $30\,M_\odot$. The comparison shows that the catalogue
contains substantial information about the scale on which the mass density
varies: the posterior selects kernels far narrower than those typically drawn
from the prior.

At the same time, this narrowing does not drive the model to its imposed
resolution limit. The event-level posterior information sets minimum
logarithmic scales of approximately
\begin{equation}
    L_1\simeq L_2\simeq0.05,
\end{equation}
corresponding to a fractional mass scale of about $5\%$. The physical-fraction-weighted widths remains above these floors rather than
accumulating against them. The localized structures reconstructed in the mass
plane are therefore supported on scales selected by the data, rather than being
generated by components saturating the finest resolution permitted by the
model.

Taken together, the complexity and width diagnostics show that GWTC--5
substantially updates the flexible mass prior in a controlled manner. The data
favor a relatively small effective number of contributing basis functions and
select considerably finer mass scales than are typical under the prior, while
leaving a clear separation from the imposed minimum resolution. This behavior
provides an important check that the recovered population structure is neither
the consequence of activating the full mixture dictionary nor of pushing the
basis functions to the narrowest scale allowed by construction.

\subsection{Variations around the baseline configuration}
\label{app:baseline_variations}

We next test whether the population structures and cosmological conclusions
obtained with the reference configuration depend sensitively on the technical
choices controlling the flexible mass reconstruction. We repeat the inference
under variations of the mixture-weight prior, component-shape regularization,
minimum resolved scale, maximum number of available Gaussian components,
data-adaptive mass-ratio bounds, and likelihood treatment. Each test changes
the indicated element of the analysis while retaining the reference choices
for the remaining settings.

The variations and the aspect of the inference that each is designed to probe
are summarized in Table~\ref{tab:population_robustness_variations}. The set
includes both more restrictive choices, which reduce the flexibility or
resolution of the mixture, and less restrictive choices, which provide the
model with additional freedom. We also include substantially smaller and
larger component dictionaries than the reference value $K_{\max}=15$. %

\begin{table*}[t]
\centering
\small
\renewcommand{\arraystretch}{1.18}
\caption{
Variations around the reference configuration. All settings not explicitly
listed are kept at their baseline values given in
Table~\ref{tab:reference_population_settings}. The tests modify the
stick-breaking concentration prior, the aspect regularization, the minimum
resolved component width, the maximum available mixture complexity, the
lower mass-ratio-center bound, or the treatment of likelihood uncertainty.
The variants are used as controlled sensitivity tests and are not
model-averaged.}
\label{tab:population_robustness_variations}
\begin{tabular}{@{}lll@{}}
\hline
\parbox[t]{0.21\textwidth}{\raggedright Variation\strut} &
\parbox[t]{0.36\textwidth}{\raggedright Change relative to the reference configuration\strut} &
\parbox[t]{0.35\textwidth}{\raggedright Purpose\strut} \\
\hline

\parbox[t]{0.21\textwidth}{\raggedright
$\mu_{\log\alpha^{-1}}=-1.0$\strut} &
\parbox[t]{0.36\textwidth}{\raggedright
Shift the mean of the $\log\alpha^{-1}$ prior from $-1.5$ to $-1.0$.\strut} &
\parbox[t]{0.35\textwidth}{\raggedright
Tests sensitivity to the prior controlling how strongly the mixture
probability can concentrate into a small number of components.\strut}
\\[0.7ex]

\parbox[t]{0.21\textwidth}{\raggedright
Weaker aspect prior\strut} &
\parbox[t]{0.36\textwidth}{\raggedright
Broaden and weaken the soft penalty against strongly anisotropic Gaussian
components.\strut} &
\parbox[t]{0.35\textwidth}{\raggedright
Tests whether the inferred morphology relies on suppressing highly elongated
basis functions.\strut}
\\[0.7ex]

\parbox[t]{0.21\textwidth}{\raggedright
Weaker aspect prior + finer $L_{\rm small}$\strut} &
\parbox[t]{0.36\textwidth}{\raggedright
Combine the weaker aspect regularization with a smaller minimum resolved
component-width scale.\strut} &
\parbox[t]{0.35\textwidth}{\raggedright
Provides additional freedom simultaneously in component shape and small-scale
resolution and therefore constitutes one of the least restrictive
reconstructions considered.\strut}
\\[0.7ex]

\parbox[t]{0.21\textwidth}{\raggedright
Stronger aspect prior\strut} &
\parbox[t]{0.36\textwidth}{\raggedright
Narrow and strengthen the soft penalty against strongly anisotropic Gaussian
components.\strut} &
\parbox[t]{0.35\textwidth}{\raggedright
Tests whether the inferred structures persist when elongated components are
more strongly disfavored.\strut}
\\[0.7ex]

\parbox[t]{0.21\textwidth}{\raggedright
Finer $L_{\rm small}$\strut} &
\parbox[t]{0.36\textwidth}{\raggedright
Decrease the event-resolution-based minimum component-width scale.\strut} &
\parbox[t]{0.35\textwidth}{\raggedright
Allows the mixture to resolve finer population structure and tests whether
additional narrow features are introduced.\strut}
\\[0.7ex]

\parbox[t]{0.21\textwidth}{\raggedright
Coarser $L_{\rm small}$\strut} &
\parbox[t]{0.36\textwidth}{\raggedright
Increase the event-resolution-based minimum component-width scale.\strut} &
\parbox[t]{0.35\textwidth}{\raggedright
Tests whether the inferred features and cosmological result survive stronger
smoothing of the mass distribution.\strut}
\\[0.7ex]

\parbox[t]{0.21\textwidth}{\raggedright
$K_{\max}=10$\strut} &
\parbox[t]{0.36\textwidth}{\raggedright
Reduce the maximum number of available Gaussian components from $15$ to
$10$.\strut} &
\parbox[t]{0.35\textwidth}{\raggedright
Checks whether the reference reconstruction requires a component dictionary
as large as the baseline one.\strut}
\\[0.7ex]

\parbox[t]{0.21\textwidth}{\raggedright
$K_{\max}=30$\strut} &
\parbox[t]{0.36\textwidth}{\raggedright
Increase the maximum number of available components from $15$ to $30$.\strut} &
\parbox[t]{0.35\textwidth}{\raggedright
Tests whether the reconstructed population is limited by the baseline
mixture truncation.\strut}
\\[0.7ex]

\parbox[t]{0.21\textwidth}{\raggedright
$K_{\max}=100$\strut} &
\parbox[t]{0.36\textwidth}{\raggedright
Increase the maximum number of available components from $15$ to $100$.\strut} &
\parbox[t]{0.35\textwidth}{\raggedright
Provides a deliberately overcomplete component dictionary and tests the
stability of the inference far from the baseline truncation.\strut}
\\[0.7ex]

\parbox[t]{0.21\textwidth}{\raggedright
Lower $q$-bound width $10^{-4}$\strut} &
\parbox[t]{0.36\textwidth}{\raggedright
Extend the data-adaptive lower support of the component-center mass ratio
substantially farther into the low-$q$ tail, while retaining the reference
upper bound.\strut} &
\parbox[t]{0.35\textwidth}{\raggedright
Tests sensitivity to the treatment of the broad low-mass-ratio posterior
tails used to construct the component-center support.\strut}
\\[0.7ex]

\parbox[t]{0.21\textwidth}{\raggedright
Marginal likelihood,
$\sigma^2_{\log\mathcal L}<1$\strut} &
\parbox[t]{0.36\textwidth}{\raggedright
Use the alternative marginal-likelihood treatment with the estimated
log-likelihood variance restricted below unity.\strut} &
\parbox[t]{0.35\textwidth}{\raggedright
Tests sensitivity to the treatment and numerical control of likelihood
uncertainty.\strut}
\\[0.7ex]

\hline
\end{tabular}
\end{table*}

\begin{figure*}[t]
    \centering
    \includegraphics[width=\textwidth]
    {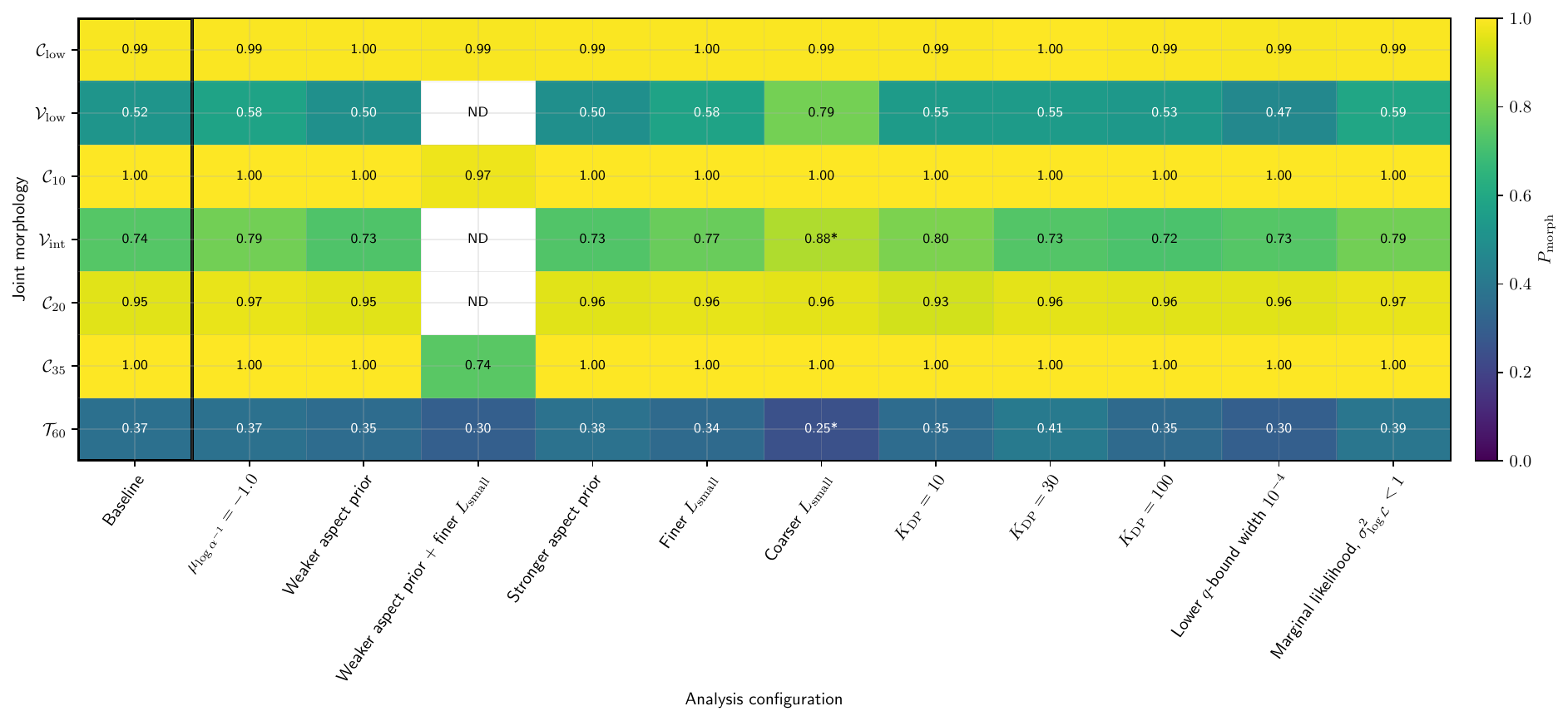}
    \caption{
    Stability of the principal joint mass-plane morphologies under variations
    of the reference analysis. Rows correspond to the seven structures
    interpreted in the main text and columns to the reference configuration
    and 11 robustness variations. Each detected cell reports the corresponding
    $P_{\rm morph}$; the baseline column is outlined. ``ND'' indicates that no
    compatible counterpart of the baseline feature is identified in that
    configuration and should not be interpreted as $P_{\rm morph}=0$. An
    asterisk marks a compatibility-based split relation, for which the plotted
    cell gives the recovery probability of the matched branch. The variants
    are independent sensitivity tests and the values are not averaged across
    columns.}
    \label{fig:feature_morphology_robustness}
\end{figure*}

The most direct test of the scientific interpretation is whether the joint
morphologies identified in the reference analysis continue to be recovered
when these choices are changed. We therefore cross-match the seven structures
discussed in the main text against the independently reconstructed catalogue
of every robustness run and record their morphology-recovery probability
$P_{\rm morph}$. %
The resulting morphology matrix is shown in
Fig.~\ref{fig:feature_morphology_robustness}.

The principal concentrations are remarkably stable. The low-mass
concentration $\mathcal C_{\rm low}$ is recovered in all 12 configurations
with
$0.991\lesssim P_{\rm morph}\lesssim0.996$, while
$\mathcal C_{10}$ is recovered in all configurations with
$0.974\lesssim P_{\rm morph}\lesssim0.999$.
The near-equal-mass $\mathcal C_{35}$ concentration is likewise identified in
all 12 analyses. Its morphology probability remains essentially unity in all
but the combined weaker-aspect/finer-resolution test, where it decreases to
$P_{\rm morph}=0.744$. Thus the existence of the joint
$\sim35\,M_\odot$ concentration is robust, although its posterior morphology
support becomes weaker in the most permissive combined
shape--resolution variation.

The weaker intermediate structures show the expected greater sensitivity.
The off-diagonal concentration $\mathcal C_{20}$ is recovered in 11 of the
12 configurations, with
$0.932\leq P_{\rm morph}\leq0.966$ whenever a compatible feature is
identified. The only missing cross-match occurs for the combined weaker-aspect
prior and finer $L_{\rm small}$ analysis. The same configuration does not
yield compatible counterparts for either $\mathcal V_{\rm low}$ or
$\mathcal V_{\rm int}$. In their remaining 11 configurations, the two valleys
span
$0.466\leq P_{\rm morph}\leq0.792$ and
$0.716\leq P_{\rm morph}\leq0.882$, respectively. The greater sensitivity of
these structures is therefore consistent with their lower morphology
probabilities already found in the baseline reconstruction, rather than with
a general instability of the dominant mass-plane concentrations.

The high-mass transition $\mathcal T_{60}$ behaves differently. A compatible
high-mass slope-change family is identified in all 12 configurations, but its
morphology probability remains low in every case,
$0.249\leq P_{\rm morph}\leq0.411$. %
 Thus the joint morphology has low posterior recovery throughout
the tested model choices. This comparison concerns recovery, not the
strength of the conditional \(H_0\)--front-scale association discussed in the main text.

\begin{figure}[t]
\centering
\includegraphics[width=0.45\textwidth]
{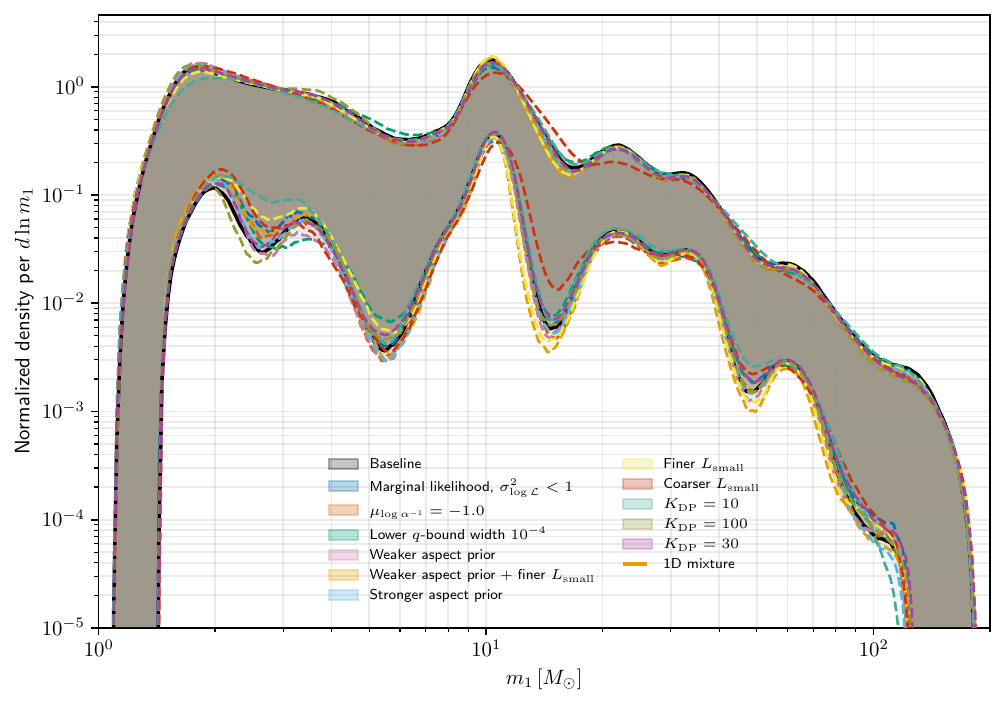}
\caption{
Robustness of the projected primary-mass distribution, normalized per
$d\ln m_1$. The shaded baseline region shows the posterior uncertainty of the
reference joint model, while the alternative curves show the corresponding
reconstructions for the variations in
Table~\ref{tab:population_robustness_variations}. The principal structures of the projected spectrum are
preserved across the population-model variations.}
\label{fig:pm1_robustness}
\end{figure}

As a complementary global diagnostic, we compare the projected primary-mass
spectrum across the same population-model variations. This projection does not
test the two-dimensional morphology directly, but shows whether the overall
mass spectrum from which the familiar one-dimensional features are inferred
changes appreciably under the robustness choices. The baseline posterior band
and the alternative reconstructions are shown in
Fig.~\ref{fig:pm1_robustness}. The flexible one-dimensional model is also
shown for context. %

The projected spectrum is broadly stable across the tests. In particular, the
low-mass support, the pronounced scale near $10\,M_\odot$, the subsequent
intermediate-mass modulation, and the broad high-mass decline remain present
under substantial changes in the allowed mixture complexity, component
resolution, shape regularization, and location support. The variations are
most visible where the baseline posterior is already broad. This projection
therefore gives a complementary check that the main one-dimensional
population scales are not generated by a single regularization prescription,
while the morphology matrix above provides the more direct test of their
joint two-dimensional interpretation.

\begin{figure}[t]
\centering
\includegraphics[width=0.50\textwidth]
{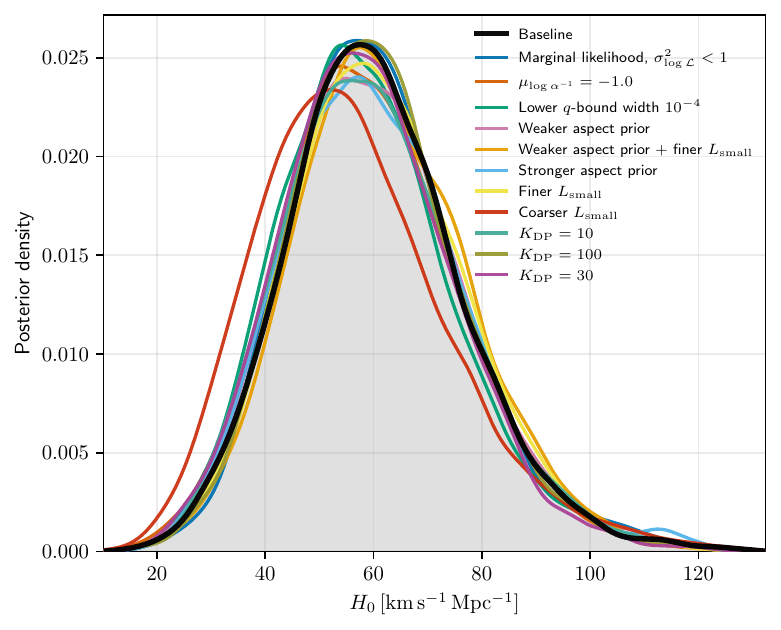}
\caption{
Marginal posterior density of $H_0$ for the baseline joint mass model and the
11 robustness variations listed in
Table~\ref{tab:population_robustness_variations}. The reference posterior is
shown in black; alternative configurations use the same color convention as
the population-robustness comparison. The strong overlap demonstrates that
the cosmological result is stable against the tested choices controlling the
mixture representation and likelihood treatment.}
\label{fig:H0_robustness}
\end{figure}

Finally, we test whether the same modelling choices propagate appreciably into
the cosmological inference. We compare the marginal $H_0$ posterior obtained
with the baseline analysis and each of the 11 robustness configurations. This
comparison, containing only variations around the joint reference model, is
shown in Fig.~\ref{fig:H0_robustness}.

The $H_0$ posteriors remain strongly overlapping across the tested
configurations. Their medians span
$54.4$--$60.5\,{\rm km\,s^{-1}\,Mpc^{-1}}$, compared with
$59.5\,{\rm km\,s^{-1}\,Mpc^{-1}}$ for the reference analysis, a variation
that is small relative to the width of the individual posteriors. The largest
visible displacement is obtained for the coarser $L_{\rm small}$ analysis,
whose median decreases to
$54.4\,{\rm km\,s^{-1}\,Mpc^{-1}}$. This configuration imposes the strongest
smoothing among the tested resolution choices and also produces one of the
largest visible modifications of the reconstructed primary-mass spectrum.
The coincidence is qualitatively consistent with the spectral-siren
sensitivity to resolved mass-spectrum structure. %

Overall, the controlled variations support the robustness claims used in the
main text. The dominant concentrations $\mathcal C_{\rm low}$,
$\mathcal C_{10}$, and $\mathcal C_{35}$ remain identifiable throughout the
tested model space, while the weaker $\mathcal C_{20}$ concentration and the
two valleys show greater sensitivity in precisely the configurations that
provide additional small-scale and shape freedom.  The \(\mathcal T_{60}\) family retains low joint morphology recovery under these alternatives. At the same time, the global
primary-mass reconstruction and the $H_0$ posterior remain broadly stable.

\subsection{Sensitivity to feature-finder choices}
\label{app:feature_finder_variations}

We additionally test the sensitivity of the catalogue to choices internal to the
feature finder, keeping the underlying population posteriors fixed. Starting
from the reference settings described in Appendix~\ref{app:feature-finder}, we repeat the comparison across
all population-model variations with a lower persistence-gap requirement,
$g_{\min}=0.5$, a higher requirement, $g_{\min}=2$, and a coarser set of
two-dimensional smoothing scales, $s=\{0,0.05,0.10\}$. These tests probe the
stability of the catalogue construction separately from the robustness of the
inferred population density.

The principal concentrations are largely insensitive to these choices. The
most informative exception occurs for the population-model variation combining
a weaker aspect prior with finer $L_{\rm small}$. This is a more
permissive shape-and-resolution configuration, since it weakens the preferred
component geometry while allowing smaller mixture widths. Under the reference
finder, this reconstruction does not yield catalogue counterparts for
$\mathcal C_{20}$, $\mathcal V_{\rm int}$, or $\mathcal V_{\rm low}$.
Lowering the gap requirement to $g_{\min}=0.5$ leaves this result unchanged,
as the same automatic persistence gap and numerical cutoff are selected.
Increasing it to $g_{\min}=2$, however, rejects that gap and activates the
fallback calibration, producing a lower persistence cutoff. A
compatible $\mathcal C_{20}$ concentration is then recovered with
$P_{\rm morph}=0.933$ and median $A_{\rm P}=0.538$, while compatible
$\mathcal V_{\rm int}$ and $\mathcal V_{\rm low}$ structures are recovered
with $P_{\rm morph}=0.784$ and $0.649$, respectively.

This behaviour shows that the absence of these weaker structures from the
reference catalogue in this particular population reconstruction depends on
the adopted persistence calibration. The alternative smoothing test leads to
the same broad feature census, with only limited changes in individual
cross-configuration associations. We therefore regard the dominant
concentrations as robust to the feature-finder choices explored here, while the
intermediate structures and valleys require more caution: their catalogue
inclusion can depend jointly on population-model flexibility and on the
threshold used to resolve weak morphology. Such sensitivity does not by itself
distinguish genuine low-contrast population structure from features associated
with finite-catalogue fluctuations, and motivates further tests with additional
data and simulated catalogues.

\section{Calibration of the feature--\texorpdfstring{$H_0$}{H0}
information diagnostics}
\label{app:h0_information_calibration}

 This appendix defines the information statistics, their validated
nearest-neighbour estimation, conditional-randomization calibration, and
Holm correction. %

\subsection{Information statistics and estimation}

 For each feature \(f\), we define
  
\begin{equation}
 X_{1,f}=\ln m_{1,f}^{\rm (1D)},\quad
 U_f=\ln m_{\parallel,f},\quad V_f=\ln q_f.
 \label{eq:h0_information_variables}
\end{equation}
 
 Here \(X_{1,f}\) describes the independently reconstructed primary-mass
feature, while \((U_f,V_f)\) describes the joint-feature position:
the centroid for concentrations and valleys, or the front point for
\(\mathcal T_{60}\). For the latter, \(U_T=t_T\) and \(V_T=-2r_T\).

For a feature representation \(Y\), its mutual information with \(H_0\) is
\begin{equation}
I(H_0;Y)
=
\int
p(H_0,Y)
\log_2\!\left[
\frac{p(H_0,Y)}{p(H_0)p(Y)}
\right]
\,dH_0\,dY .
\label{eq:h0_mutual_information}
\end{equation}
We evaluate
\begin{equation}
\begin{split}
&I_{X_1,f}=I(H_0;X_{1,f}),
\\
&I_{\parallel,f}=I(H_0;U_f),
\\
&I_{{\rm 2D},f}=I\!\left(H_0;(U_f,V_f)\right).
\end{split}
\label{eq:h0_information_statistics}
\end{equation}
The first two quantities are alternative, non-nested summaries of the same
feature, so their difference is not interpreted as a separate information
contribution.

To isolate the dependence carried by the mass-ratio direction after fixing
the common mass scale, we also evaluate
\begin{equation}
\begin{split}
& I_{q\mid\parallel,f}
= 
I(H_0;V_f\mid U_f) \\
& = 
\int
p(H_0,U_f,V_f)
\log_2\!\left[
\frac{p(H_0,V_f\mid U_f)}
     {p(H_0\mid U_f)p(V_f\mid U_f)}
\right]
\,dH_0\,dU_f\,dV_f .
\end{split}
\label{eq:h0_orthogonal_cmi}
\end{equation}

For each feature, all four quantities are evaluated on the same common-valid
posterior sample.
 We estimate \(I_{\parallel,f}\) and \(I_{q\mid\parallel,f}\) directly
and report the total through the information chain rule,
  
\begin{equation}
 \widehat I_{{\rm 2D},f}
 =\widehat I_{\parallel,f}+\widehat I_{q\mid\parallel,f}.
 \label{eq:h0_information_sum}
\end{equation}

 We use the KSG-1 mutual-information estimator introduced in Ref.~\cite{Kraskov2004}
and the direct conditional-information estimator of Ref.~\cite{FrenzelPompe2007}
with \(k=5\).
The primary analysis retains all common-valid draws.
All estimates are expressed in bits.

\subsection{ Significance of residual mass-ratio dependence}

 This subsection describes the significance test used
in Sec.~\ref{sec:joint_features_h0}. For each feature, we ask whether
the observed mass-ratio dependence on \(H_0\) exceeds what could
arise by chance after accounting for the common mass scale.

 The residual test targets \(H_0\perp V_f\mid U_f\), allowing both
variables to depend on the common scale. We use local randomization of
\(V_f\) at fixed \((H_0,U_f)\), following Ref.~\cite{H0UpdateRunge2018}.
Each draw receives a donor value from its five nearest neighbours in
\(U_f\), including itself. Draws and candidate donors are visited in random
order, preferring unused donors; reuse is allowed when all five have been
used. This approximates the conditional null while preserving local
dependence on \(U_f\).

For each feature we generate \(B=1999\) null realizations and compare
their raw conditional-information estimates with the observed estimate:
  
\begin{equation}
 p_f=\frac{1+\sum_{b=1}^{B}
 \mathbf 1\!\left[\widehat I_{q\mid\parallel,f}^{(b)}
 \geq\widehat I_{q\mid\parallel,f}^{\rm obs}\right]}{B+1}.
 \label{eq:h0_conditional_randomization_p}
\end{equation}

 Here \(p_f\) is the individual, unadjusted \(p\)-value
for feature \(f\). It measures how often local permutation produces
residual information at least as large as observed. A small value
indicates that the observed residual is unusual under the assumption
of no residual dependence. The next subsection applies Holm's
correction to these five \(p_f\) values, giving the adjusted
\(p_{\rm Holm}\) reported in Table~\ref{tab:H0_feature_information}.

\subsection{Family-wise calibration of the conditional-information tests}

 The test family consists of the five higher-mass structures
examined in Sec.~\ref{sec:joint_features_h0}:
$\mathcal C_{10}$, $\mathcal C_{35}$, $\mathcal C_{20}$,
$\mathcal V_{\rm int}$, and $\mathcal T_{60}$.
We apply Holm's step-down correction~\cite{H0UpdateHolm1979}.
For ordered raw values \(p_{(1)}\leq\cdots\leq p_{(5)}\),
  
\begin{equation}
 p_{{\rm Holm},(j)}
 =\min\!\left\{1,\max_{1\leq i\leq j}
                  \left[(6-i)p_{(i)}\right]\right\}.
 \label{eq:residual_cmi_holm}
\end{equation}
 
 The adjusted values are then returned to their feature order.
Holm's correction does not require independence between the five tests.
We use a \(0.05\) threshold. The minimum raw \(p\)-value is \(0.0005\);
both \(\mathcal C_{20}\) and \(\mathcal V_{\rm int}\) have
\(p_{\rm Holm}=0.0025\).

Estimator stability is assessed separately with \(200\) random subsets
containing \(80\%\) of each feature's common-valid draws.
The \(5\)th--\(95\)th percentiles of the raw conditional estimates give
the whiskers in Fig.~\ref{fig:H0_information_by_representation}.
These estimates may be negative; the ranges are not posterior credible
intervals or significance tests.

We checked the false-positive calibration using synthetic Gaussian
chains with zero residual information by construction. We generated
\(500\) datasets with sample sizes and serial dependence fitted to
the posterior chains, and another \(500\) with stronger serial
dependence. Because the null hypothesis is true in these simulations,
every rejection is a false positive: a test calibrated at \(5\%\)
should produce about \(25\) rejections per \(500\) datasets.
We obtained \(18/500\) (\(3.6\%\)) and \(25/500\) (\(5.0\%\)),
respectively, consistent with the intended false-positive rate
within simulation uncertainty. %

\subsection{Recovery dependence of the high-mass transition}
\label{app:h0_recovery_dependence}

For \(\mathcal T_{60}\), a continuous location is available only when the
slope-change morphology is recovered. We therefore test separately whether
its binary recovery indicator
\begin{equation}
Z_{60}^{(s)}
=
\begin{cases}
1, & \mathcal T_{60}\ \text{is recovered in realization }s,\\
0, & \text{otherwise}
\end{cases}
\label{eq:t60_recovery_indicator}
\end{equation}
depends on \(H_0\).

The recovered and unrecovered draws define two samples of \(H_0\). Because
these samples need not be Gaussian and can differ through their ordering
rather than only their means, we compare them using the Mann--Whitney rank
statistic. If \(H_{0,i}^{(1)}\) is a recovered draw and
\(H_{0,j}^{(0)}\) an unrecovered draw, we define
\begin{equation}
U
=
\sum_{i=1}^{n_1}
\sum_{j=1}^{n_0}
\left[
\mathbf 1\!\left(H_{0,i}^{(1)}>H_{0,j}^{(0)}\right)
+
\frac{1}{2}
\mathbf 1\!\left(H_{0,i}^{(1)}=H_{0,j}^{(0)}\right)
\right],
\label{eq:t60_mann_whitney}
\end{equation}
where \(n_1\) and \(n_0\) are the numbers of recovered and unrecovered draws.
Thus, \(U/(n_1n_0)\) is the fraction of recovered--unrecovered pairs in which
the recovered draw has the larger \(H_0\), with half weight assigned to ties.

We report the corresponding rank-biserial coefficient
\begin{equation}
r_{\rm rb}
=
\frac{2U}{n_1n_0}-1.
\label{eq:t60_rank_biserial}
\end{equation}
This coefficient provides a signed and normalized effect size for the
association between a continuous variable and a binary indicator without
assuming a parametric form for either \(H_0\) distribution. It ranges from
\(-1\) to \(1\); with the convention above, \(r_{\rm rb}<0\) indicates that
recovered draws preferentially occupy lower values of \(H_0\).

 We calibrate its absolute magnitude with its own \(B=2000\)
within-chain circular shifts of \(H_0\) relative to the binary recovery
indicator. In chain \(c\),
  
\begin{equation}
 H_{0,c,t}^{(b)}=H_{0,c,(t+\delta_{bc})\bmod n_c},
 \label{eq:h0_circular_shift}
\end{equation}
 
 where \(n_c\) is the chain length and \(\delta_{bc}\) is uniform
over non-zero shifts. This retains each chain's cyclic ordering while
changing its alignment with recovery. It tests unconditional recovery
dependence, separately from the conditional-information randomization.
The two-sided probability is
\begin{equation}
p_{\rm circ}
=
\frac{
1+\displaystyle\sum_{b=1}^{B}
\mathbf 1\!\left[
\left|r_{\rm rb}^{(b)}\right|
\geq
\left|r_{\rm rb}^{\rm obs}\right|
\right]
}{
B+1
}.
\label{eq:t60_rank_biserial_null}
\end{equation}

This is a separate two-sided recovery test and is not included in the
five-feature conditional-information family entering  \(p_{\rm Holm}\).

We obtain
\begin{equation}
r_{\rm rb}(H_0,Z_{60})=-0.199,
\qquad
p_{\rm circ}=5.0\times10^{-4},
\label{eq:t60_recovery_h0_result}
\end{equation}
showing that recovery is associated with lower \(H_0\). The corresponding
posterior medians are
\begin{align}
&H_0^{Z_{60}=1}
=
55.87\,{\rm km\,s^{-1}\,Mpc^{-1}},
\qquad
\\
&H_0^{Z_{60}=0}
=
61.52\,{\rm km\,s^{-1}\,Mpc^{-1}}.
\label{eq:t60_recovery_h0_medians}
\end{align}
As a descriptive check, dividing the complete posterior into five
equal-posterior-mass \(H_0\) bins gives
\begin{equation}
P(Z_{60}=1\mid H_0\ {\rm bin})
=
0.479,\ 0.419,\ 0.373,\ 0.304,\ 0.262,
\label{eq:t60_recovery_h0_bins}
\end{equation}
from the lowest to the highest \(H_0\) bin.

These quantities describe recovery under the fixed catalogue identities,
matching rules, and analysis settings.

\end{document}